\documentclass[
reprint,
aps,
superscriptaddress]
{revtex4-2}
\usepackage{graphicx} 
\usepackage{mathtools}
\usepackage{amsfonts}
\usepackage{graphicx}
\usepackage[caption=false]{subfig}
\usepackage{hyperref}
\usepackage[margin=0.8in]{geometry}
\usepackage{float}
\usepackage{amsthm}
\hypersetup{
	linkcolor=black,
	colorlinks=true,
	urlcolor=blue,
	filecolor=black,
	citecolor=blue,
	pdftitle={Current Enabled Optical Response}
}
\usepackage{bm}
\usepackage[dvipsnames]{xcolor}
\newcommand{\nt}[1]{\textcolor{blue}{#1}}

\date{March 2025}
\begin{document}
\title{Current-Enabled Optical Conductivity of Collective Modes in Unconventional Superconductors}
\author{Gerrit Niederhoff}
\affiliation{Institute for Theoretical Physics, ETH Zürich, 8093 Zürich, Switzerland}
\affiliation{Department of Physics, University of Tokyo, 7-3-1 Hongo, Tokyo 113-0033, Japan}
\author{Ryusei Kataoka}
\affiliation{Department of Physics, University of Tokyo, 7-3-1 Hongo, Tokyo 113-0033, Japan}
\author{Kazuaki Takasan}
\affiliation{Department of Physics, University of Tokyo, 7-3-1 Hongo, Tokyo 113-0033, Japan}
\author{Naoto Tsuji}
\affiliation{Department of Physics, University of Tokyo, 7-3-1 Hongo, Tokyo 113-0033, Japan}
\affiliation{RIKEN Center for Emergent Matter Science (CEMS), 2-1 Hirosawa, Wako, Saitama 351-0198, Japan}
\begin{abstract}
We theoretically investigate the current-enabled linear optical conductivity of collective modes 
in superconductors with unconventional pairing symmetries.
After deriving general formulas for the optical conductivity of a superconductor featuring multiple pairing channels and bands using the path integral formalism, we apply these formulas to several models.
Using a model of competing $s$- and $d$-wave pairing interactions, we find that several known collective modes generate peaks in the optical conductivity upon injection of a supercurrent. This includes single- and multiband versions of Bardasis-Schrieffer modes, mixed-symmetry Bardasis-Schrieffer modes, and Leggett modes.
Using a model for interband $p$-wave superconductivity with Rashba spin-orbit coupling, we find that in such a system Bardasis-Schrieffer modes are optically active even without introducing a supercurrent. In a $p+ip$ chiral ground state, these modes turn out to produce peaks in the longitudinal and transverse optical conductivity. Other collective modes belonging to the chiral $p+ip$ order parameter turn out to be unaffected by the spin-orbit coupling but contribute to the optical response when a supercurrent is introduced. These results promise new avenues for the observation of collective modes in a variety of superconducting systems, including multiband superconductors and superconductors that feature multiple pairing channels or multi-component order parameters, such as chiral $p$- or $d$-wave superconductors.
\end{abstract}

\maketitle

\section{Introduction}

Determining the pairing symmetry and other properties of potentially unconventional superconductors remains a difficult task for many materials \cite{sigristGL, unconventionalReview, recentFeSC-review, kagomePairingSymmetryTheory, UTe2-pairingsymmetry, strontiumSC}.
One useful indicator of the nature of the superconducting ground state is the spectrum of collective modes. Although every superconductor can in principle host an amplitude and phase modes, usually referred to as the Higgs 
\cite{Anderson1958, Higgs1964, tsujisan-higgs-review} 
and Nambu-Goldstone (NG) \cite{Nambu1961, goldstoneMode} modes respectively, other collective modes can also be found in certain specific systems. One example is the Leggett mode \cite{leggett-original-paper}, a relative phase oscillation between superconducting gaps, typically found in multiband superconductors \footnote{In the present work, we are interested in spatially uniform oscillations of the relative phase of the superconducting order parameters. There are, however, also situations where a stable state with a topological winding of the relative phase in space can exist, which is called a phase soliton \cite{phaseSolitonTanaka, phaseSolitonMesoscopic, phaseSolitonTopology}.}.
Another collective excitation can occur when there are several competing superconducting pairing channels. In those systems, where the ground state belongs to one pairing channel but other attractive channels exist, it is possible to find well-defined excitations below the quasi-particle gap. These can be identified as fluctuations of the order parameter belonging to the sub-dominant pairing channel, and are referred to as the Bardasis-Schrieffer (BS) modes \cite{bsmode-original}.

One issue with studying these collective modes is that they often do not couple linearly to light. A useful tool is Raman spectroscopy, which has been used to study the Higgs \cite{Sooryakumar1980},
Leggett \cite{Blumberg2007},
and BS modes \cite{bs-Raman}.
The Higgs mode has also been detected using pump-probe spectroscopy \cite{higgsPumpPropeSpectroscopy, Katsumi2018} and third harmonic generation \cite{higgsTHG, Matsunaga2017, Chu2020}.
All of these probes rely on nonlinear optical effects. In recent years, however, there has been several progress in predicting and observing signatures of collective modes of superconductors in the linear optical response. For example, the BS mode with a different parity than the ground state has been predicted to respond to light linearly in locally non-centrosymmetric compounds such as CeRh$_2$As$_2$ \cite{linearBS}. The Leggett mode has been predicted to become visible in the linear optical response in certain multiband superconductors
\cite{leggettOpticalResponse, nagashima-san-leggett, Nagashima_PDW}. 
Furthermore, the Higgs mode has been shown to become optically active in the presence of a supercurrent \cite{higgs_moving_condensate} (see also Refs.~\cite{Kubo2024, Wang2024}), which has been observed in an experiment \cite{higgsExperiment}.
This idea of injecting a supercurrent to induce a non-trivial optical response has also been considered in studying the effect of quasiparticle excitations in the optical conductivity \cite{supercurrent_crowley, supercurrent_papaj}, where the supercurrent mainly plays a role of breaking inversion symmetry to allow the necessary optical transitions \cite{supercurrent_papaj}.

In the present work, we investigate the effect of such an injected current on the optical response of collective modes in unconventional superconductors. First, the path integral formalism in imaginary time is used to derive the optical conductivity of general one- or multiband superconductors, with an arbitrary number of pairing channels. These resulting formulas are then applied to different models. Specifically, a model with competing $s$-wave and $d$-wave pairings will be used \cite{s-vs-d-BSmode}, which is a situation possibly relevant to iron-based superconductors \cite{recentFeSC-review}. Upon tuning the $d$-wave interaction, the model undergoes a quantum phase transition from an $s$-wave to a $d$-wave ground state, with a mixed-symmetry $s+id$ state in between. The previous study has found a well-defined Bardasis-Schrieffer mode below the gap in the $s$-wave phase, and a mixed-symmetry Bardasis-Schrieffer (MSBS) mode in the $s+id$-wave phase of this model \cite{s-vs-d-BSmode}, making it an excellent candidate for our purposes.

\begin{figure}[t]
    \centering
    \includegraphics[width=\linewidth]{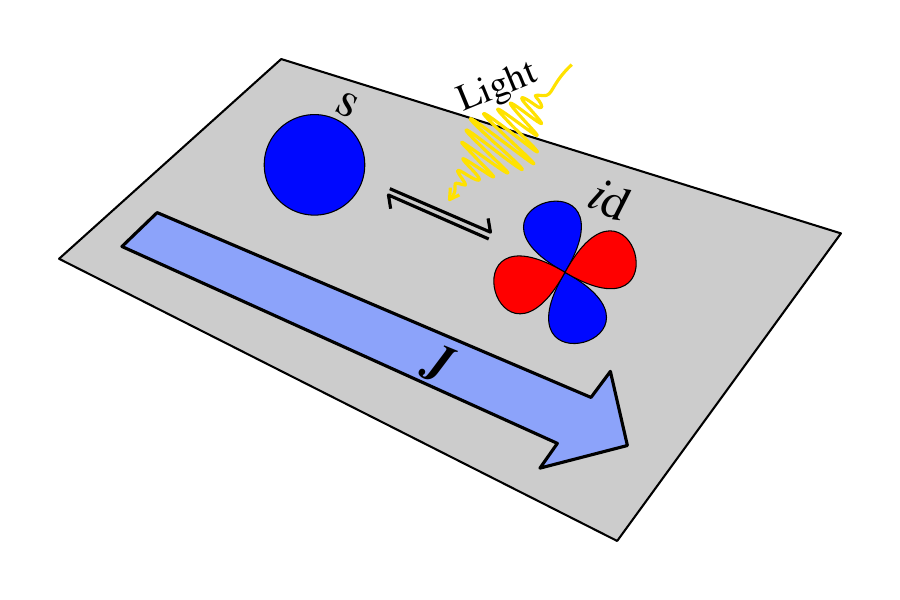}
    \caption{
    Schematic representation of current-enabled optical response of collective modes in unconventional superconductors. One can, for example, excite a mixed-symmetry Bardasis-Schrieffer (MSBS) mode linearly by light in an $s+id$-wave superconductor in the presence of externally injected supercurrent $\bm{J}$.
    }
    \label{cartoon}
\end{figure}

Using the formulas we derived, we find that this predicted BS/MSBS mode generates a peak in the optical conductivity when a supercurrent is injected \nt{(Fig.~\ref{cartoon})}.
We then extend this to a two-band model, and find two BS modes as well as a Leggett mode in the resulting spectrum. Again, we find that all of these modes result in sharp peaks in the optical conductivity, though their relative intensity depends strongly on the chosen parameters.

Finally, we study a model of $p$-wave interband-pairing superconductors with a Rashba-type spin-orbit coupling (SOC). It has been pointed out that such a system features Lifshitz invariants \cite{LandauLifshitz, edelsteinGL, Lifshitz_Rashba, nagashima-san-leggett}, which are terms in the Ginzburg-Landau energy that are linear in the spatial gradient. Since a connection between these terms and the optical response of collective modes has been made \cite{nagashima-san-leggett}, the Rashba system is a promising candidate for optically active collective modes, even without an injected current. 
We study this model across a quantum phase transition between two different $p$-wave ground states (chiral and non-chiral). Applying our formulas for the optical response to this system, we indeed find peaks due to the BS modes, without applying a supercurrent. In the chiral $p+ip$-wave ground state, we see these peaks not only in the longitudinal but also in the transverse optical conductivity. We also observe that even in this already non-centrosymmetric system, an injected supercurrent still has a qualitative effect, causing the appearance of another peak in the longitudinal optical conductivity, which is associated with a relative phase oscillation of the components of the chiral order parameter. This mode has been referred to as a generalized clapping mode \cite{clappingModes, trsbModes}.

This paper is organized as follows: Sec. \ref{effectiveActionSection} contains the derivation of the formulas for the optical conductivity. In Sec. \ref{s_vs_d_section} these formulas are applied to a superconductor featuring $s$- and $d$-wave pairing, first with one band then with two. As a function of increasing strength of $d$-wave pairing, collective mode spectra and optical conductivities are calculated. Finally, in Sec. \ref{rashba}, the formalism is slightly extended and applied to a Rashba system featuring interband pairing. Once again, two pairing channels are chosen and collective mode spectra as well as optical conductivities are calculated for different interaction strengths in the two channels.

\section{Effective action for the optical conductivity}\label{effectiveActionSection}
In this section, we review the derivation of the optical conductivity from the path integral formalism \cite{AltlandSimons, manyParticlePhysics} and adapt previous calculations to derive formulas for the optical conductivity of a generic single-band or multiband superconductor featuring multiple pairing channels. Fluctuations of order parameters and the corresponding collective modes are introduced in a similar manner as in Ref. \cite{leggettOpticalResponse}, but adapted to general unconventional pairing symmetries.

\subsection{One-band systems}

In the path-integral formalism,
the fermionic imaginary-time action for a generic one-band superconductor is given by
	\begin{align}
    \begin{split}
		S[\bar{\psi},\psi]
		&=\int_0^\beta\text{d}\tau
		\left(
		\sum_{\bm{k}\sigma}\bar{\psi}_{\bm{k}\sigma}\partial_\tau\psi_{\bm{k}\sigma}
		-H[\bar{\psi},\psi]\right),\\
		H&=H_0+H_\text{int},\\
		H_0&=\sum_{\bm{k}\sigma}\bar{\psi}_{\bm{k}\sigma}\xi_{\bm{k}}\psi_{\bm{k}\sigma},\\
		H_\text{int}
		&=
		\sum_{\bm{kk'}}
		V_{\bm{kk'}}
		\bar{\psi}_{\bm{k+q}\uparrow}
		\bar{\psi}_{\bm{-k+q}\downarrow}
		\psi_{\bm{-k'+q}\downarrow}
		\psi_{\bm{k'+q}\uparrow},
        \end{split}
	\end{align}
	with a general interaction $V_{\bm{kk'}}$ and dispersion $\xi_{\bm{k}}$. 
    Here $\psi_{\bm k\sigma}$ are fermionic Grassmann variables with momentum $\bm k$ and spin $\sigma$, $\beta$ is the inverse temperature, and
    Cooper pairs are explicitly given a total momentum of $2\bm{q}$, which is considered as an external parameter and is not summed over. This is a way of simulating the effect of an injected supercurrent \cite{supercurrent_papaj}. 
    In the present work, we assume a separable form of the interaction,
    \begin{align}
        V_{\bm{kk'}}&=\sum_\mu V^\mu\varphi_\mu(\bm{k})\varphi_\mu(\bm{k'}),
		\label{interactiondecomposition}
    \end{align}
    where $\varphi_\mu(\bm k)$ is the basis function for each pairing channel $\mu$.
    In order to obtain a mean-field description, 
    we introduce Hubbard-Stratonovich fields $b_{\bm{k}}$ and $b^*_{\bm{k'}}$, which allows us to rewrite the action as
    \begin{align}
		\begin{split}
			S[\bar{\psi},\psi,b^*,b]
			&=\int_0^\beta\text{d}\tau
			\biggm(
			\sum_{\bm{k}\sigma}\bar{\psi}_{\bm{k}\sigma}
			(\partial_\tau+\xi_{\bm{k}})
			\psi_{\bm{k}\sigma}
			\\
			-\sum_{\bm{kk'}}
			V_{\bm{kk'}}
			&\big[
			b^*_{\bm{k}}b_{\bm{k'}}
			+b^*_{\bm{k}}
			\psi_{{\bm{-k'+q}}\downarrow}
			\psi_{{\bm{k'+q}}\uparrow}
            \\
			&+b_{\bm{k'}}
			\bar{\psi}_{\bm{k+q}\uparrow}
			\bar{\psi}_{{\bm{-k+q}}\downarrow}
			]\biggm),
		\end{split}
        \label{hubbardStratonovichAction}
	\end{align}
    where the quartic interaction term was replaced by an interaction with the auxiliary fields $b$. 
    We define the gap function,
    \begin{align}
    \Delta_{\bm{k}}=-\sum_{\bm{k'}}V_{\bm{kk'}}b_{\bm{k'}}=\sum_\mu\varphi_\mu(\bm{k})\Delta^\mu,
    \label{gap-decomposition}
    \end{align}
    where the decomposition of $\Delta_{\bm{k}}$ into its components $\Delta^\mu$ follows from the form of the interaction [Eq.~(\ref{interactiondecomposition})].
    Inserting this into Eq.~(\ref{hubbardStratonovichAction}),
    the action becomes:
    \begin{align}
		\begin{split}
			S &=
			\int_0^\beta\text{d}\tau
			\biggm(
			-\sum_{\bm{k}}\Psi^\dagger_{\bm{k}}
			G^{-1}_{\bm{k},\tau}\Psi_{\bm{k}}
			-\sum_\mu\frac{|\Delta_\mu|^2}{V^\mu}
			\biggm),
			\end{split}\\
			G^{-1}_{\bm{k}\tau}
			&=-\partial_\tau-\mathcal{H}_\text{BdG}
            ,\quad
            \Psi_{\bm{k}} =\begin{pmatrix}
                \psi_{\bm{k+q}\uparrow}\\
                \bar{\psi}_{\bm{-k+q}\downarrow}
            \end{pmatrix},
	\end{align}
    where the Bogoliubov-de Gennes (BdG) Hamiltonian takes the form of
	\begin{align}
		\mathcal{H}_\text{BdG}&=\begin{pmatrix}
			\xi_{\bm{k+q}}&\Delta_{\bm{k}}\\
			\Delta^*_{\bm{k}}&\xi_{\bm{-k+q}}
		\end{pmatrix},
	\end{align}
	whose eigenvalues may be written as
	\begin{align}
		E^\pm_{\bm{k}}&=\xi_{\bm{k}}'\pm\delta_{\bm{k}}\label{bdg-eigenvalues}
	\end{align}
	with
	\begin{align}
		\begin{split}
			\xi_{\bm{k}}'&=\frac{1}{2}(\xi_{\bm{k+q}}-\xi_{\bm{-k+q}})\\
			\delta_{\bm{k}}&=\sqrt{\bar{\xi}_{\bm k}^2+|\Delta_{\bm{k}}|^2}\\
			\bar{\xi}_{\bm{k}}&=\frac12(\xi_{\bm{k+q}}+\xi_{\bm{-k+q}})
		\end{split}\label{defineEnergyTerms}.
	\end{align}
    The vector potential $\bm{A}$ is now introduced via the minimal coupling
	$\bm{k}\mapsto\bm{k}\mp e\bm{A}$ for particles and holes, respectively. 
    We expand the Green's function to the first order in $\bm A$.
    The second-order term contributes only to the diamagnetic optical conductivity \cite{pairwaveVertexCorrection}, which does not contribute to the real part of the optical conductivity and will be neglected here.
    
    In order to investigate fluctuations of the order parameters, we write the gap components as
    \begin{align}
        \Delta^\mu_p &= \Delta^\mu_0+\Delta^{\mu x}_p+i\Delta^{\mu y}_p,
    \end{align}
    where $\Delta^\mu_0$ is the static, homogeneous equilibrium value determined from the gap equation (see Appendix \ref{gapAppendix}) and $p=(i\Omega,\bm{p})$ is a 4-momentum with a bosonic Matsubara frequency.
    In the saddle-point approximation, $\Delta_0^\mu$ is assumed to be fixed and not integrated over.
    The terms containing only $\Delta_0^\mu$ therefore contribute as a global factor to the action.
    The inverse Green's function $G^{-1}$ then splits into the inverse equilibrium Green's function $\mathcal{G}$ and the self-energy corrections $\Sigma$. In the momentum-frequency representation with $k=(i\omega,\bm{k})$, they are given as
    \begin{align}
		G^{-1}_{k,p}&=\mathcal{G}_k^{-1}-\Sigma^e_{k,p}-\Sigma^\Delta_{k,p},\\
		\Sigma^e_{k,p} &=v_i(\bm{k})A_i(p),\quad 
		\Sigma^\Delta_{k,p}=\sum_{\nu\mu}\Delta^\nu_\mu(p)\varphi^\mu_{\bm{k}}\tau_\nu,
        \label{selfEnergy}
        \\
		v_i(\bm{k})&=
		\begin{pmatrix}
			\partial_i\xi_{\bm{k+q}}&0\\
			0&\partial_i\xi_{\bm{k-q}}\\
		\end{pmatrix}
		=
		v_i^0\tau^0+v_i^3\tau^3\label{bareVertex},
    \end{align}
    where $\mathcal G$ is the equilibrium Green's function, $\Sigma^e$ and $\Sigma^\Delta$ are the corrections due to the vector potential and gap function, respectively, and $v_i(\bm k)$ is the group velocity. For one-band systems, the equilibrium Green's function $\mathcal{G}$ can be calculated explicitly:
    \begin{align}
        \mathcal{G}^{-1}_k &=\frac{
			(i\omega-\xi_{\bm{k}}')\tau_0
			+\bar{\xi}_{\bm{k}}\tau_3
			+\Delta_{\bm{k}}^R\tau_1-\Delta_{\bm{k}}^I\tau_2}
		{(i\omega-\xi_{\bm{k}}')^2-\delta_{\bm{k}}^2},\label{explicitGreensfunction}
	\end{align}
    where $\Delta^{R/I}$ denotes the real/imaginary component of the equilibrium gap function, 
    and $\tau_\mu$ denote the Pauli matrices for $\mu=1,2,3$ and the $2\times2$ unit matrix for $\mu=0$.
    This makes the action
    \begin{align}
		\begin{split}
			S &= -\beta\sum_{\mu}\frac{|\Delta^0_\mu|^2}{V^\mu}
			-\beta\sum_{\mu\nu p}\frac{\Delta^{\nu}_\mu(-p)\Delta^{\nu}_\mu(p)}{V^\mu}
			\\
			&\quad-\beta\sum_k\Psi^\dagger_k\mathcal{G}^{-1}_{k}\Psi_k
			+\beta\sum_{k,p}\Psi^\dagger_{k+p}\Sigma^\Delta_{k,p}\Psi_k\\
            &\quad
			+\beta\sum_{k,p}\Psi^\dagger_{k+p}\Sigma^e_{k,p}\Psi_k.
		\end{split}
        \label{eq: action}
	\end{align}
Integrating out the fermions (see, for example, Ref.~\cite{AltlandSimons}) leaves us with an effective action,
whose components that depend on the fluctuations of the order parameters and the vector potential $\bm{A}$ are, with $\Sigma=\Sigma^e+\Sigma^\Delta$:
\begin{align}
		S_\text{eff}^\text{FL} &=
		-\beta\sum_{\mu\nu p}\frac{\Delta^{\nu}_\mu(-p)\Delta^{\nu}_\mu(p)}{V^\mu}
		+\sum_{l=1}^\infty\frac{\text{Tr}(\mathcal{G}\Sigma)^l}{l}.\label{effectiveActionFluctuations}
	\end{align}
    
\begin{figure}[t]
		\centering
		\includegraphics[width=\linewidth]{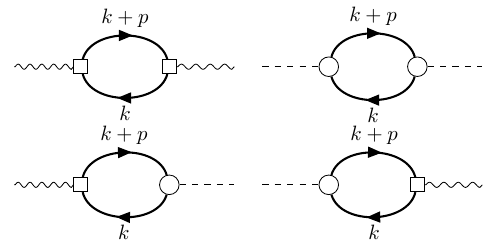}
		\caption{
        Feynman diagrams for each term in the effective action $S_\text{eff}^\text{FL}[\Delta^{\nu\mu};\bm{A}]$ (see Eq.~(\ref{effectiveActionFluctuations})). The dashed (wavy) lines represent the fluctuations $\Delta^\nu_\mu$ (the vector potential $\bm{A}$) with the corresponding vertex $\varphi^\mu\tau^\nu$ ($v^i$).
        }
		\label{feynman1}
	\end{figure}

\begin{figure}[t]
        \centering
        \includegraphics[width=\linewidth]{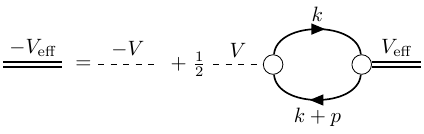}
        \caption{Diagrammatic representation for the effective coupling $V_\text{eff}$ [Eq.~(\ref{eq: V_eff})].}
        \label{feynmanEffectiveV}
    \end{figure}

\begin{figure}[t]
        \centering
        \includegraphics[width=\linewidth]{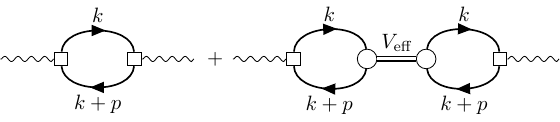}
        \caption{Diagrammatic representation for the components of the effective action [Eq.~(\ref{eq: S_eff^EM})] that contribute to the real part of the optical conductivity.}
        \label{feynmanConductivity}
    \end{figure}

    Here the $l=2$ term is the lowest relevant order. All the possible terms contained in the $l=2$ term can be expressed in terms of Feynman diagrams, as depicted in Fig.~\ref{feynman1}. 
    Explicitly, this reads (with summation over repeated indices):
	\begin{align}
		\begin{split}
			S_\text{eff}^{\rm FL}=
			&\beta\sum_{p}\biggm(\frac{\Delta^{\mu\nu}_{-p}\Delta^{\mu\nu}_p}{-V^\mu}
			+\frac12\Delta^{\mu\nu}_p\Delta^{\mu'\nu'}_{-p}\Pi^{\mu\mu'}_{\nu\nu'}(p)\\
            +&\frac{e^2}{2} A^a_pA^b_{-p}\Phi^{ab}(p)
			-\frac e2 \Delta^{\mu\nu}_{-p}A^a_{p}Q_a^{\mu\nu}(-p)\\
			-&\frac e2 \Delta^{\mu\nu}_{-p}A^a_{-p}Q_a^{\mu\nu}(p)
			\biggm),
		\end{split}\label{effectiveActionOneBand}\\
		\Phi^{ab}(p)
		&=\frac{1}{\beta}\sum_{k}
		\text{Tr}\left[
		v_a(\bm{k})\mathcal{G}_{k+p}
		v_b(\bm{k})\mathcal{G}_k
		\right],\label{phiGeneral}\\
		\Pi^{\mu\mu'}_{\nu\nu'}(p)
		&=\frac{1}{\beta}\sum_{k}
		\varphi^\mu_{\bm{k}}\varphi^{\mu'}_{\bm{k}}
		\text{Tr}
		\left[
		\tau^\nu\mathcal{G}_{k+p}
		\tau^{\nu'}\mathcal{G}_k
		\right],\label{piGeneral}\\
		Q^{\mu\nu}_a(p)
		&=\frac{1}{\beta}\sum_{k} \varphi^{\mu}_{\bm{k}}
		\text{Tr}\left[
		v_a(\bm{k})\mathcal{G}_{k+p}
		\tau^\mu\mathcal{G}_k
		\right].\label{qGeneral}
	\end{align}
    The integrals (\ref{phiGeneral})-(\ref{qGeneral}) can be simplified for one-band systems, by writing the Green's function in terms of the Pauli matrices (see Appendix \ref{simplifying}),
    \begin{align}
		\Phi^{ab}&=-4I[v^3_a|\Delta|^2v^3_b],\label{phiGeneralSimplified}\\
		\begin{split}
		\Pi^{\mu\mu'}
		&=I\biggm[
		\varphi^\mu
		\biggm\{
		\begin{pmatrix}
			-4(\bar{\xi})^2
			&2i(i\Omega)\bar{\xi}\\
			-2i(i\Omega)\bar{\xi}&
			-4(\bar{\xi})^2
		\end{pmatrix}
        \\
        &\qquad\qquad-
        \begin{pmatrix}
			4(\Delta^I)^2
			&4\Delta^I\Delta^R\\
			4\Delta^I\Delta^R
			&4(\Delta^R)^2
		\end{pmatrix}
		\biggm\}
		\varphi^{\mu'}\biggm],
		\end{split}
		\label{piGeneralSimplified}\\
		\bm{Q^\mu_a}&=I\left[v^3_a\begin{pmatrix}
			4\Delta^R\bar{\xi}-2i(i\Omega)\Delta^I\\
			-4\Delta^I\bar{\xi}-2i(i\Omega)\Delta^R
		\end{pmatrix}\varphi^\mu\right],
		\label{qvecGeneralSimplified}
	\end{align}
	where the functional $I$ is defined as
	\begin{align}
		I[f]&=\sum_{\bm{k}}\frac{\varphi_d(\bm{k})
        [n_F(E^+)-n_F(E^-)]
        }
		{\delta_{\bm{k}}
			\left[(i\Omega)^2-4\delta_{\bm{k}}^2\right]
		}\label{bigIntegral},
	\end{align}
    where $n_F(E)$ is the Fermi-Dirac-distribution.
    The terms in the effective action that depend on the vector potential $\bm{A}$ can be again found by performing a gaussian integral, which defines a new effective coupling-vertex between photons. Diagrammatically, this step is shown in Fig.~\ref{feynmanEffectiveV}, and one obtains the effective coupling,
	\begin{align}
		\left[V_\text{eff}^{-1}(i\Omega)\right]^{\mu\mu'}_{\nu\nu'}&=
		\frac{\delta^{\mu\mu'}\delta_{\nu\nu'}}{V^\mu}
		- \frac{1}{2}\Pi^{\mu\mu'}_{\nu\nu'}(i\Omega).
        \label{eq: V_eff}
	\end{align}
	The spectrum of the collective modes in the system is given by the condition,
	\begin{align}
		\det V_\text{eff}^{-1} \overset{!}{=}0.\label{modecondition}
	\end{align}
	The final form of the effective action, shown diagrammatically in Fig.~\ref{feynmanConductivity}, is
	\begin{align}
		\begin{split}
			S_\text{eff}^\text{EM}
			&=
			\frac{\beta e^2}{2}\sum_{p}\sum_{ab}A_p^aA_{-p}^b\\\times&\left(
			\Phi^{ab}(p)+\frac12\bm{Q}_a^\text{T}(p)
			\left[V_\text{eff}^{-1}(p)\right]
			\bm{Q}_b(-p)
			\right).
		\end{split}
        \label{eq: S_eff^EM}
	\end{align}
    The optical conductivity follows from this by differentiating twice with respect to $\bm{A}$ \cite{AltlandSimons}. In the long wave-length limit ($\bm{p}\approx0$) and after performing the analytical continuation $i\omega\mapsto\omega+i\eta$, one obtains:
	\begin{align}
		\sigma^{ab}(\omega) &=
		\frac{ie^2}{2\hbar^2\omega}\Phi^{ab}(\omega)+
		\frac{ie^2}{4\hbar^2\omega}\bm{Q_a^\text{T}}(\omega)
		\left[V_\text{eff}(\omega)\right]\bm{Q_b}(-\omega).\label{conductivityFormula}
	\end{align}
    It should be noted that including the fluctuations in this way leads to a gauge-invariant optical response, but the compressibility sum rule may be violated \cite{pathIntegralGaugeInvariance}, which would be fixed by calculating the vertex corrections to explicitly preserve the Ward-Takahashi identity, as has been done for conventional order parameters, for example, in Refs.~\cite{pairwaveVertexCorrection} and \cite{supercurrent_response}. 
    Since we are only interested in the optical properties, the path integral approach is sufficient in our case.
    
    \subsection{Multiband systems}\label{multiBandTheory}
    
    In the case of multiband superconductors, the Hamiltonian takes the form of
    \begin{align}
		H_\text{kin} &= \sum_{\alpha\bm{k}}\xi^\alpha_{\bm{k}}c^\dagger_{\bm{k}\alpha}c_{\bm{k}\alpha},\\
		H_\text{int}
		&=
		\sum_{\bm{kk'}\alpha\beta\gamma\rho}
		V^{\alpha\beta\gamma\rho}_{\bm{kk'}}
		c^{\dagger}_{\bm{k+q},\alpha}c^{\dagger}_{{\bm{-k+q},\beta}}
		c_{{\bm{-k'+q},\gamma}} c_{{\bm{k'+q},\rho}},\label{generalMultibandInteraction}
	\end{align}
    where $\alpha,\gamma,\beta,\rho$ are band indices, and $c^\dagger_{\bm{k},\alpha}$ is the creation operator for electrons with momentum $\bm{k}$ in band $\alpha$ (spin indices are omitted).
    In order to study non-trivial multiband effects, it is sufficient to include interband pair scatterings, but neglect interband pairings. Such an assumption is reasonable when the separation between the bands is larger than the superconducting gap energy. In Eq.~(\ref{generalMultibandInteraction}), this means setting $\beta=\alpha$ and $\rho=\gamma$, thus restricting Cooper pairs to be in each band individually. Once again a separable interaction potential will be assumed, now of the form of
    \begin{align}
		V^{\alpha\gamma}_{\bm{kk'}}
		&=
		\sum_\mu
		V^\mu_{\alpha\gamma}\varphi^\mu_{\bm{k}\alpha}(\varphi^\mu_{\bm{k'}\gamma})^*,\label{multibandMatrix}
	\end{align}
    where $V^\mu_{\alpha\gamma}$ is a hermitian matrix describing 
    intraband pairing and interband pair scattering, and $\varphi_{\bm k \alpha}^\mu$ are the basis functions for each pairing channel. After the Hubbard-Stratonovich transformation (which is analogous to the one-band case), the mean-field term of the action can be written as
    \begin{align}
        -\beta\sum_{\bm{kk'}\alpha\gamma}V^{\alpha\gamma}_{\bm{kk'}}(b^\alpha_{\bm{k}})^*b^\gamma_{\bm{k'}}
        &=
        -\beta
        \sum_{\alpha\gamma\mu}
        \Delta^\mu_\alpha\left[V^\mu\right]^{-1}_{\alpha\gamma}(\Delta^\mu_\gamma)^*,
    \end{align}
    where
    \begin{align}
        \Delta^\alpha_{\bm{k}}&=\sum_\mu\Delta^\mu_\alpha\varphi^\mu_\alpha(\bm{k}).
    \end{align}
    The advantage of considering no interband pairing is that $\mathcal{H}_\text{BdG}$ becomes completely decoupled between the bands. This means that the relevant polarization bubbles can all be calculated using Eqs.~(\ref{phiGeneralSimplified})-(\ref{qvecGeneralSimplified}) for each band.
    So, for the simplified case considered here, we instead get a slightly modified version of Eq.~(\ref{effectiveActionOneBand}) with additional sums over the bands,
    \begin{align}
        \begin{split}
			S_\text{eff}^{FL}=
			&\beta\sum_{p}\biggm(
            \left[V^\mu\right]^{-1}_{\alpha\gamma}
            \Delta^{\mu\nu}_{\alpha;\bar{p}}\Delta^{\mu\nu}_{\gamma;p}
            \\
			+\frac12
            &\Delta^{\mu\nu}_{\alpha;\bar{p}}\Delta^{\mu'\nu'}_{\alpha;p}\Pi^{\mu\mu'}_{\nu\nu';\alpha}(p)
            +\frac{e^2}{2} A^a_{\bar{p}}A^b_{p}\Phi^{ab}_\alpha(p)\\
			-\frac e2 &\Delta^{\mu\nu}_{\alpha;\bar{p}}A^a_{p}Q_{a;\alpha}^{\mu\nu}(-p)
			-\frac e2 \Delta^{\mu\nu}_{\alpha;p}A^a_{\bar{p}}Q_{a;\alpha}^{\mu\nu}(p)
			\biggm),
		\end{split}\label{effectiveActionMultiBand}
    \end{align}
    where all the indices (repeated or not) are summed over and $\bar{p}=-p$. 
    This gives the optical conductivity with a modified effective coupling $V_\text{eff}$,
    \begin{align}
    \begin{split}
		\sigma_{ab}(\omega)&= \frac{ie^2}{2\hbar^2\omega}\sum_\alpha\Phi_{ab}^\alpha(\omega)\\
		+\frac{ie^2}{4\hbar^2\omega}
        \sum_{\alpha\gamma}&\bm{{Q^{\alpha}_{a}}^T}(\omega)
		\left[V_\text{eff}^{\alpha\gamma}(\omega)\right]^{-1}
		{\bm{Q_{b}}^{\gamma}}(-\omega),
    \end{split}\\
		\left[V_{\text{eff }\nu\nu'}^{\alpha\gamma\mu\mu'}
        (\omega)\right]^{-1} &= 
			\delta^\mu_{\mu'}\delta^{\nu}_{\nu'}
			\left[V^\mu\right]^{-1}_{\alpha\gamma}
			-\frac12 \delta^{\alpha\gamma}\Pi^{\mu\mu'}_{\nu\nu',\alpha}(\omega),
	\end{align}
    where the off-diagonal elements of $V^\mu$ are responsible for mixing different bands, while the off-diagonal elements of $\Pi$ are responsible for mixing different pairing channels.
The same condition Eq. \ref{modecondition} applies here as in the single-band case.   
\section{Competition between \textit{s}- and \textit{d}-wave pairings}\label{s_vs_d_section}

\begin{figure}[h]
    \centering
    \includegraphics[width=\linewidth]{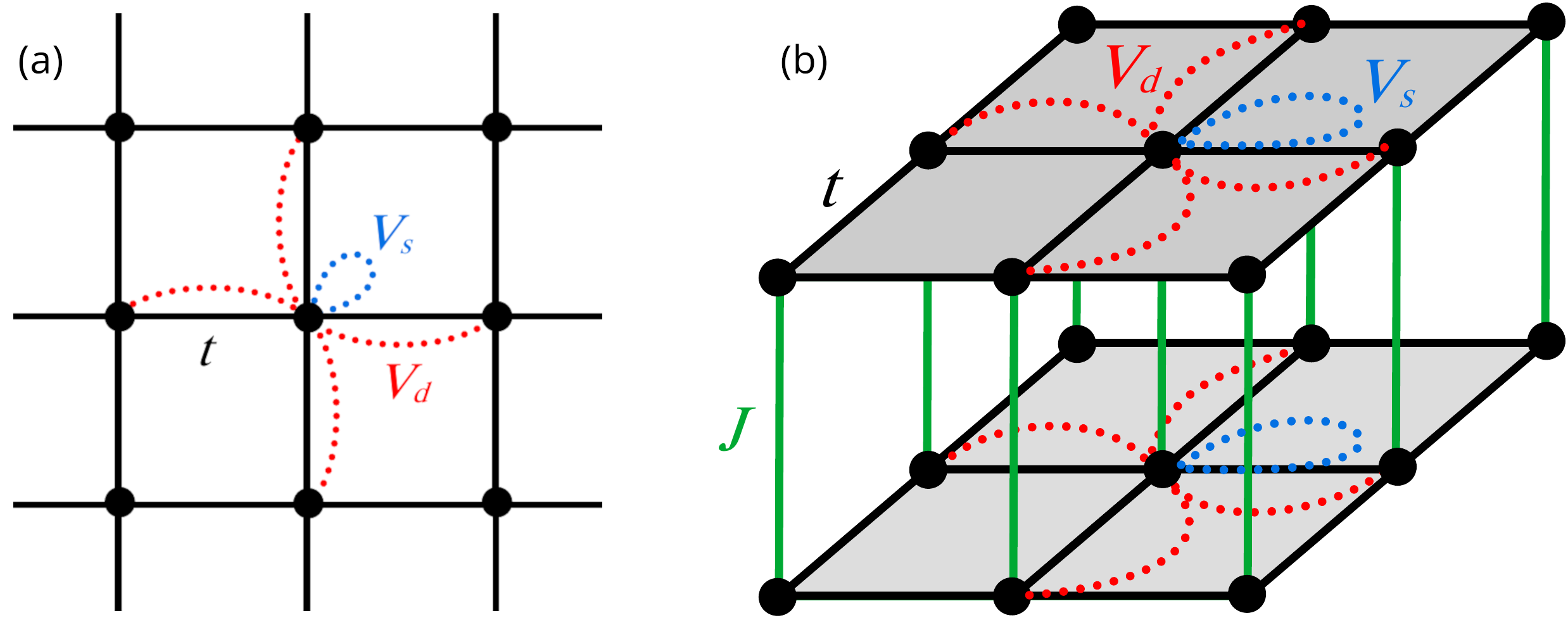}
    \caption{Schematics of the two tight-binding Hamiltonians considered for the competing $s$- and $d$-wave interactions: (a) The single layer model with one band and (b) the bilayer model with two bands. The dotted lines in blue and red illustrate attractive interactions in the $s$- and $d$-wave channels, while the solid lines in black and green represent intralayer and interlayer single-particle hoppings, respectively.}
    \label{hubbardSchematics}
\end{figure}

In this section, the derived formulas will be applied to superconductors featuring both $s$- and $d$-wave pairings. Given that a non-trivial optical response is possible only in systems that break the Galilei invariance \cite{pairwaveVertexCorrection}, we first study this situation in a one-band (single-layer) model on the square lattice (Fig.~\ref{hubbardSchematics}a). Subsequently, the results are extended to a bilayer model (Fig.~\ref{hubbardSchematics}b) to investigate how a second band affects the results.

\subsection{One-band model}
\label{sec: one-band model}

\begin{figure}[t]
			\centering
				\includegraphics[width=\linewidth]{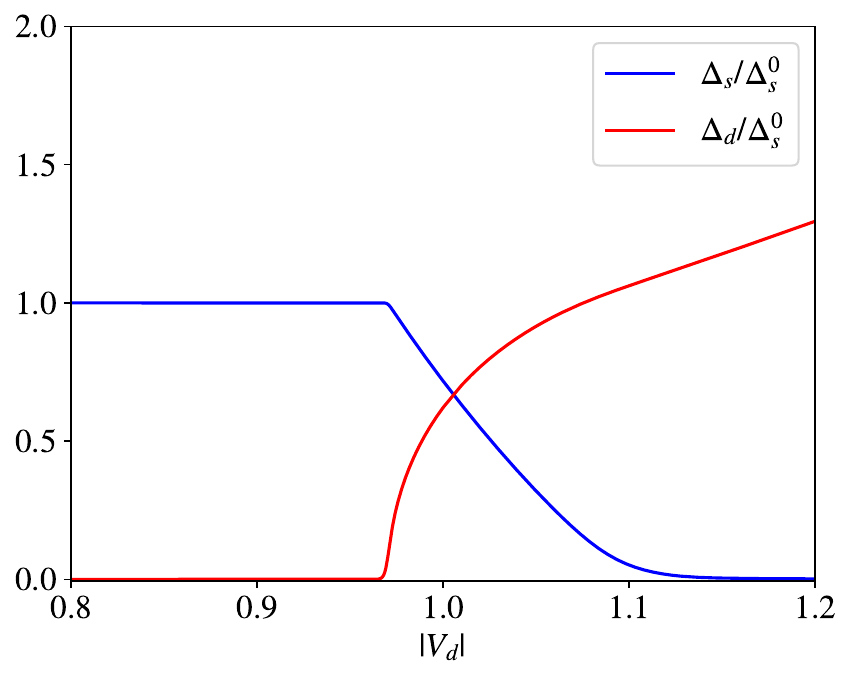}
				\caption{
                Order-parameter components $\Delta_s$ and $\Delta_d$ as a function of $|V_d|$ for the one-band model with the $s$- and $d$-wave pairing interactions on the square lattice. The parameters are $\mu=0$, $V_s=-1$, and $t=1$.
                }
				\label{sd-transition}
    \end{figure}

 On the square lattice with the hopping $t$ and chemical potential $\mu$, the free electron dispersion is
\begin{align}
		\xi_{\bm{k}}&=-2t(\cos k_x+\cos k_y)-\mu.
	\end{align}
A nearest-neighbor interaction $V_d$ results in an extended $s$-wave, $d$-wave, as well as odd-parity pairing terms in momentum space. Neglecting for simplicity all but the $s$-wave and $d$-wave terms, the interaction can be written as
\begin{align}
		\begin{split}
		V_{\bm{k,k'}}&= V_s + V_d\varphi_d(\bm{k})\varphi_d(\bm{k'}),\\
		\varphi_d(\bm{k})&=\frac{1}{\sqrt[]{2}}(\cos k_x-\cos k_y).
		\end{split}\label{s-vs-d-decomposition}
	\end{align}
With such an interaction, the only stable mixed symmetry state is of the form of $s+id$ \cite{s-vs-d-mixedSymmetry}. Taking that into account, the self-consistency condition for the gap components $\Delta^s$ and $\Delta^d$ (See Eq.~(\ref{gap-decomposition})) can be written as
\begin{align}
		\begin{split}
		\Delta_s&=
		V_s\sum_{\bm{k'}}\frac{\Delta_s}{2\delta_{\bm{k}}}n_F^\pm,\\
		\Delta_d &= V_d\sum_{\bm{k'}}\frac{\Delta_d\varphi_d^2(\bm{k})}{2\delta_{\bm{k}}}n_F^\pm,
		\end{split}\label{self-consistency}
	\end{align}
    where $n_F^\pm=n_F(E^+)-n_F(E^-)$.
    The numerical solution for these equations are shown in Fig.~\ref{sd-transition}. This plot is consistent with previous findings \cite{s-vs-d-BSmode}, where a continuum-model was used instead of a lattice. Calculations are always done at zero Temperature in the present work. To confirm the fact that the mixed symmetry state becomes the true ground state, we checked that it has a lower free energy than the pure $s$-wave ground state, see Appendix \ref{freeEnergy}.
    \begin{figure}[t]
				\includegraphics[width=\linewidth]{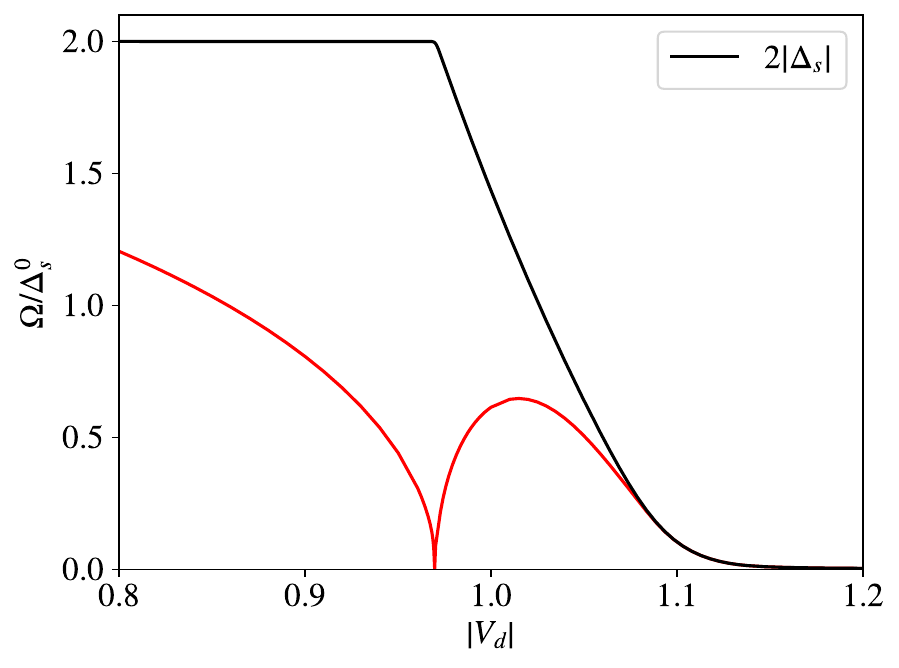}
			\caption{
            Collective-mode spectrum for the one-band model with the $s$- and $d$-wave pairing interactions on the square lattice. The NG mode (not shown here) is always present at $\Omega=0$. Another collective mode (red line) is the BS mode in the $s$-wave phase, which becomes the MSBS mode in the $s+id$-phase. The black line shows the $s$-wave component of the gap function. Again, the parameters are $\mu=0, V_s=-1,t=1$.
            }
            \label{BSmode-spectrum}
		\end{figure}
        
	 In order to determine the full collective-mode spectrum, Eq. (\ref{modecondition}) has to be solved for $\Omega$. In Appendix \ref{simplifying}, this condition is simplified into more explicit mode equations for $\bm{q}=0$. The equations relevant to the BS Mode are given in Eqs.~(\ref{swave_BSmode}) and (\ref{s+idwave_MSBSmode}). The spectrum obtained from solving these equations numerically in the respective phase is shown in Fig.~\ref{BSmode-spectrum}.
     In agreement with the previous findings \cite{s-vs-d-BSmode}, for such systems a BS mode is found in the $s$-wave phase, which softens at the transition to the $s+id$-wave ground state. In the $s+id$-wave phase, the mode consists of $s-id$-fluctuations, meaning out-of-phase amplitude fluctuations of the two order parameter components, which has been referred to as a mixed-symmetry Bardasis-Schrieffer (MSBS) mode \cite{s-vs-d-BSmode}, that stays below the gap throughout the phase.
     After the transition to a pure $d$-wave ground state the quasiparticle gap features nodes, and consequently there can be no collective excitations below the gap. \cite{s-vs-d-BSmode}

     We extract the character of the modes (i.e., whether the mode corresponds to the amplitude or phase fluctuation, etc.) from the eigenvectors of the effective coupling matrix (Eq.~(\ref{eq: V_eff})). For the case of the one-band model with the $s$- and $d$-wave pairing interactions, this approach is further verified by real-time simulations of the order-parameters dynamics driven by an interaction quench (see Appendix \ref{dynamicsAppendix}). From the quench dynamics, we can also see how each mode decays in time, which is related to the stability of the mode. For example, the Higgs mode in $s$-wave superconductors is known to decay in a power law as $1/\sqrt{t}$ after a quench \cite{VolkovKogan1973}. This is because the Higgs mode sits right at the bottom ($2\Delta$) of the quasiparticle excitation continuum. In contrast, the BS and MSBS modes exist well below the gap. In fact, we find that these modes do not decay after a quench in the long-time limit. We can also show the frequency of each collective mode and the damping behavior analytically based on the linearized equation-of-motion approach \cite{Anderson1958, Tsuji2015} (Appendix \ref{dynamicsAppendix}).
     
	\begin{figure}[t]
		\includegraphics[width=\linewidth]{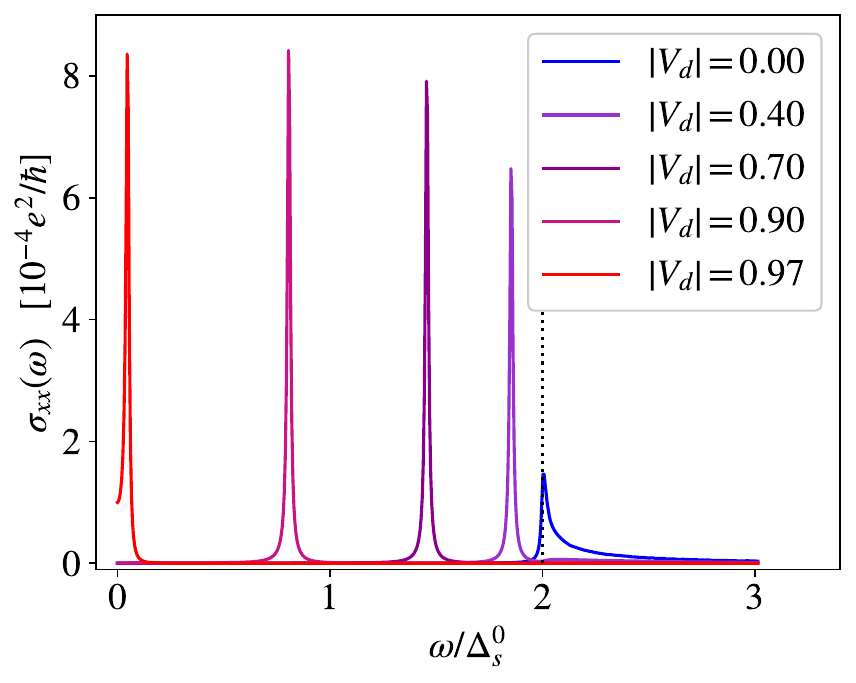}
		\caption{
        Optical conductivity in the $s$-wave phase of the one-band model with the $s$- and $d$-wave pairing interactions in the presence of supercurrent. The dotted line marks the minimal quasiparticle gap. The parameters are $\mu=0$, $V_s=-1$, $t=1$, and $\bm{q}=0.001\bm{\hat{e}}_x$.}
		\label{conductivities_swave}
	\end{figure}

     The optical conductivity of the system given by Eq.~(\ref{conductivityFormula}) consists of the quasiparticle response and the collective-mode response, neither of which is gauge-invariant on its own. The sum of both, however, yields a gauge-invariant optical response \cite{pathIntegralGaugeInvariance}. The conductivity clearly becomes non-zero only once a current is applied ($v^3_a\neq0$). For the numerical calculations, we take $\bm{q}=0.001\bm{e}_x$, and consequently only $\sigma_{xx}$ is non-trivial. Note that all the contributions to the optical conductivity are proportional to $\bm{q}^2$. This means that the absolute value of $\bm{q}$ can be chosen arbitrarily as long as it is small enough not to affect the value of the gap.
     
    \begin{figure}[t]
		\includegraphics[width=\linewidth]{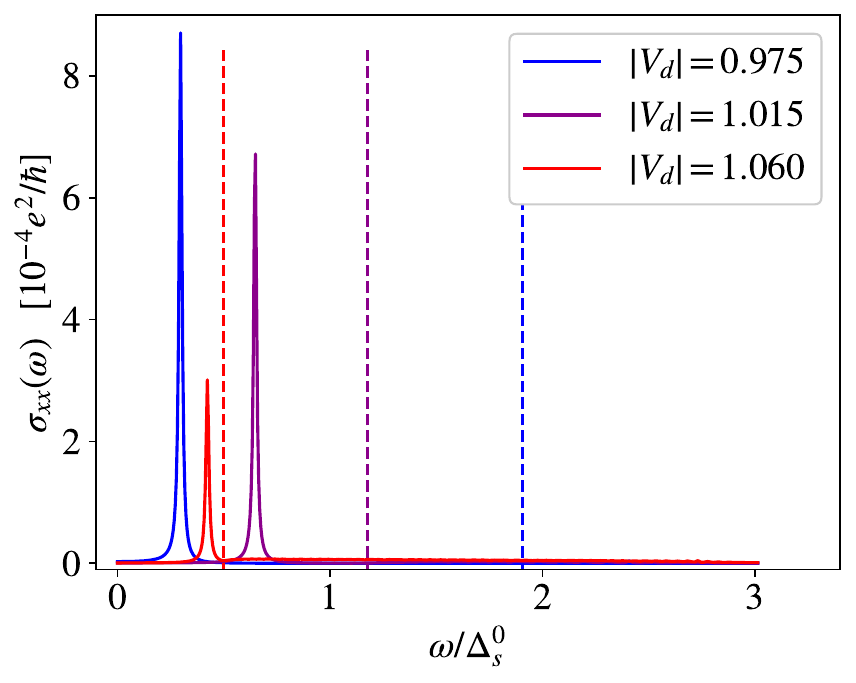}
		\caption{
        Optical conductivity in the $s+id$-wave phase of the one-band model with the $s$- and $d$-wave pairing interactions in the presence of supercurrent. The dotted lines in each color mark the minimal quasiparticle gap for each value of $V_d$. The parameters are $\mu=0$, $V_s=-1$, $t=1$, and $\bm{q}=0.001\bm{\hat{e}}_x$.}
		\label{conductivities_s+idwave}
	\end{figure}
      Figure \ref{conductivities_swave} shows the numerical evaluation of Eq.~(\ref{conductivityFormula}) for several values of $V_d$ in the $s$-wave phase.
	  For $V_d=0$, there is simply a current-induced quasi-particle response at $\Omega=2\Delta$, which has previously been investigated with different theoretical methods \cite{supercurrent_crowley, supercurrent_papaj}. As the $d$-wave pairing interaction is turned on, the peak becomes stronger and moves to lower energies. By manually suppressing individual components, it can be checked that this happens entirely due to the BS mode of the subdominant $d$-wave pairing. 
      The "bare" quasiparticle response $\Phi^{ab}$ is unaffected by $V_d$ as long as one remains in the $s$-wave phase. The corresponding peak at $2\Delta$ is however strongly suppressed by a corresponding negative peak due to the BS mode.
	  The results for the optical conductivity in the mixed-symmetry (i.e., $s+id$-wave) phase are shown in Fig.~\ref{conductivities_s+idwave}. As predicted by the spectrum in Fig.~\ref{BSmode-spectrum}, the peak due to the collective mode first moves to larger frequencies as $|V_d|$ is increased before moving back down again, always staying below $2\Delta_s$. 
      Just like in the $s$-wave phase, the collective mode also suppresses the response of the quasiparticles at $2\Delta$.
      Furthermore, on this side of the transition the optical response also becomes weaker when moving further from the $s/s+id$-transition point.

\subsection{Two-band model}\label{s_vs_d_twoband}

To explore multiband effects 
on the optical spectrum of the collective modes,
we employ a simple extension of the one-band model used in the previous section.
Instead of the two-dimensional single-layer model, we consider a bilayer model with a weaker single-particle interband hopping $J<t$. This corresponds to simply having two copies of the same cosine-dispersion band, placed at $\pm J-\mu$:
\begin{align}
    \xi^\pm_{\bm{k}}&= -2t(\cos k_x +\cos k_y) -\mu \pm J.
\end{align}
We furthermore choose the pairing matrices [Eq.~(\ref{multibandMatrix})] as
\begin{align}
    V^\mu_{\alpha\gamma}&= V^\mu 
    \begin{pmatrix}
    1&0.1\\
    0.1&1\\
    \end{pmatrix}
\end{align}
to introduce interband effects to the model.
\subsubsection{The symmetric case}\label{symmMultiBand}

\begin{figure}[t]
	\centering
	\includegraphics[width=\linewidth]{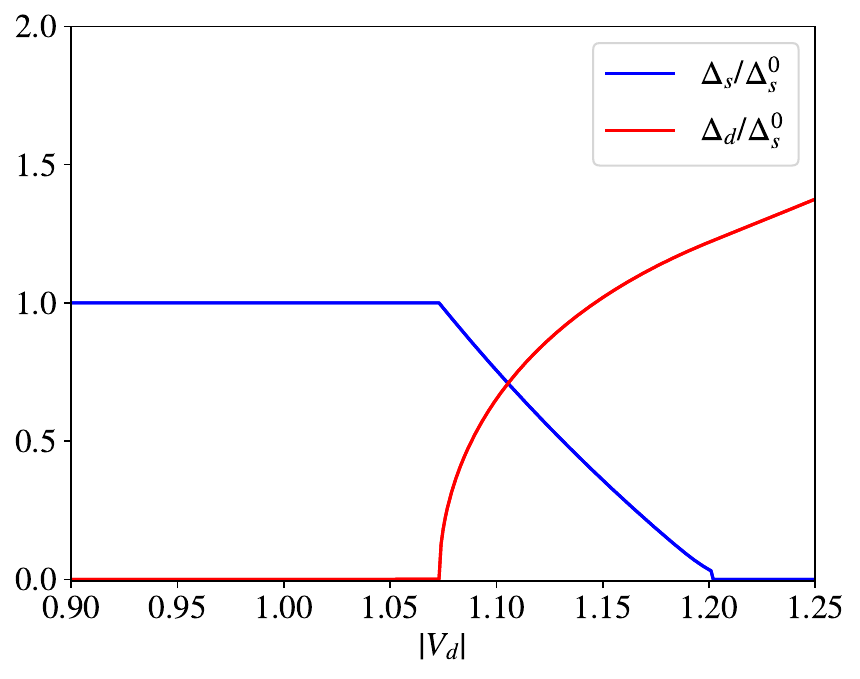}
	\caption{
    Order-parameter components $\Delta_s$ and $\Delta_d$ as a function of $|V_d|$ for the symmetric bilayer model with the $s$- and $d$-wave pairing interactions. The parameters are $\mu=0$, $V_s=-1$, $t=1$, and $\bm{q}=0.001\bm{\hat{e}}_x$.
    }
    \label{multibandPhasetransitionSymmetric}
\end{figure}

When $\mu=0$, the bands are centered at $\pm J$, and the multiband gap equation (See Appendix \ref{gapAppendix}) yields the same gap values in each band. The resulting quantum phase transition that takes place while tuning $V_d$ can be seen in Fig.~\ref{multibandPhasetransitionSymmetric}. The small discontinuity at small $\Delta_s$ is a result of the gap equation being solved in the presence of a supercurrent. Again, we confirmed the transition with a free energy comparison, see Appendix \ref{freeEnergy}.
\begin{figure}[h]
\centering
	\includegraphics[width=\linewidth]{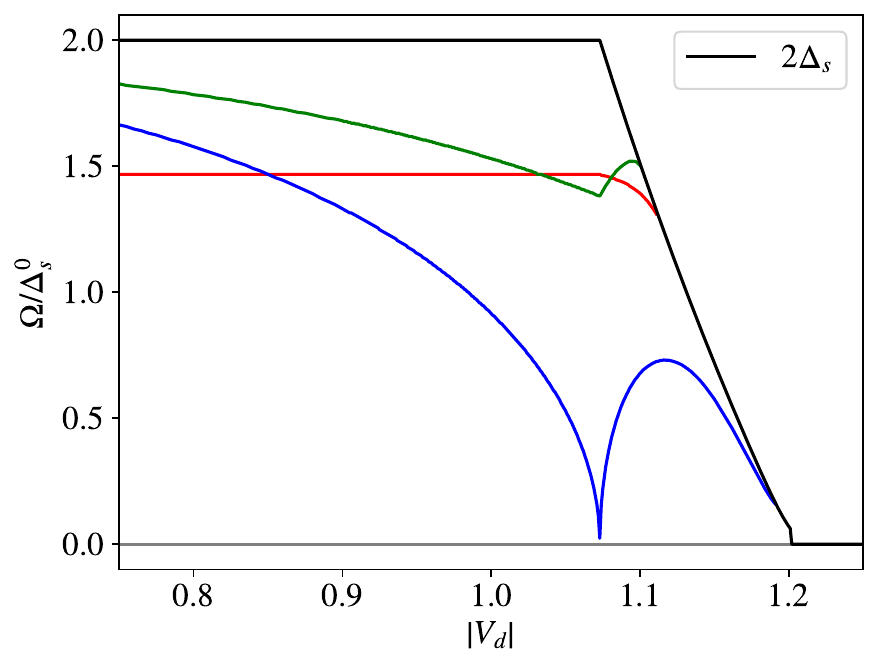}
	\caption{
    Collective-mode spectrum for the symmetric bilayer model with the $s$- and $d$-wave pairing interactions.
    The four sub-gap modes are an out-of-phase BS/MSBS mode (green), a Leggett mode (red), an in-phase BS/MSBS mode (blue), and the NG mode (gray).
    The black line shows the $s$-wave component of the gap function. 
    The parameters are $\mu=0$, $V_s=-1$, $t=1$, and $\bm{q}=0.001\bm{\hat{e}}_x$.
    }
	\label{multiBandSpectrumSymmetric}
\end{figure}
The inclusion of the interband pair scattering is sufficient to generate a more diverse spectrum of collective modes across this phase transition.
This multiband spectrum is shown in Fig.~\ref{multiBandSpectrumSymmetric}. Note that the two BS modes are not degenerate, and do not simply correspond to the $id$-fluctuations in each band. In fact, there is one BS mode where the $id$-component fluctuates in phase for the two bands, and another one which fluctuates out of phase.
This is the same as in the $s+id$-phase, except that the modes are now MSBS modes instead of BS modes. The main qualitative difference between the phases is that the frequency of the Leggett mode also depends on the value of $V_d$ in the mixed phase.

\begin{figure}[h]
			\centering
            \subfloat[
                \label{multibandConductivitiesSymmetric_swave}
            ]{
                \includegraphics[width=0.97\linewidth]{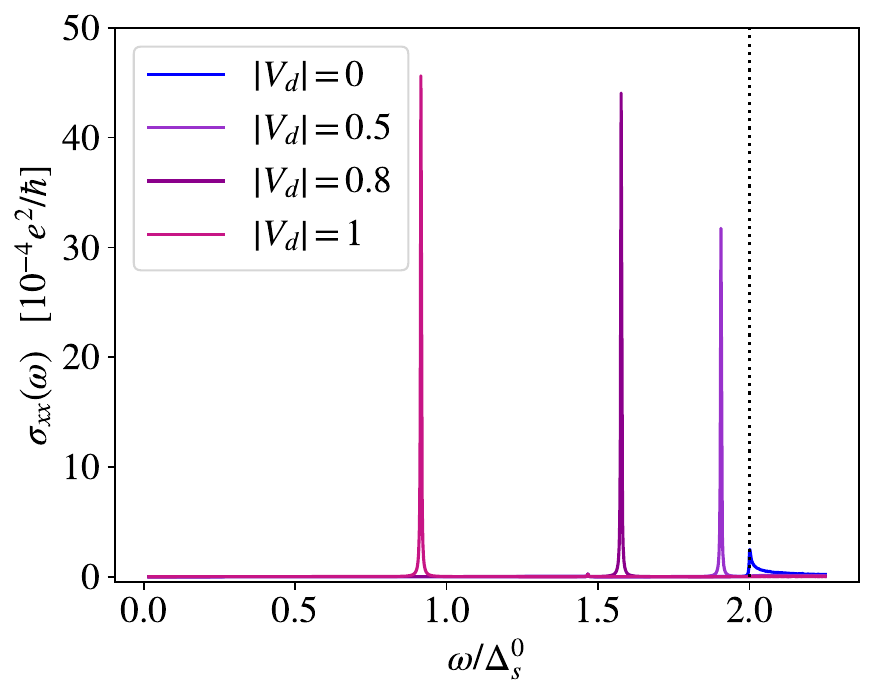}
            }\\
            \subfloat[
                \label{multibandConductivitiesSymmetric_swave_zoomed}
            ]
            {
                \includegraphics[width=0.97\linewidth]{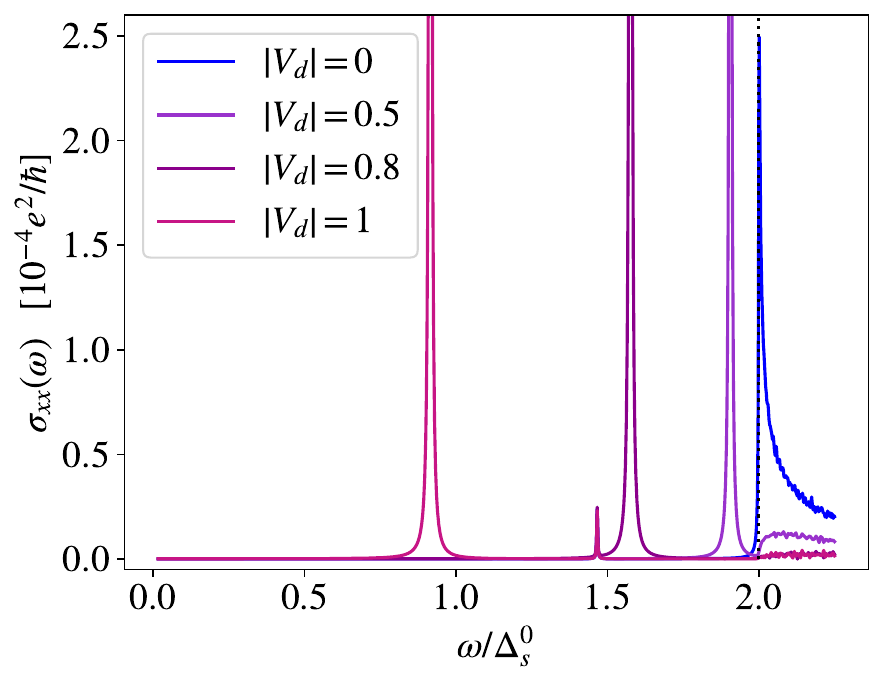}
            }
			\caption{
            Optical conductivity in the $s$-wave phase of the symmetric bilayer model with the $s$- and $d$-wave pairing interactions in the presence of supercurrent. (a) and (b) are the same plots with different scales.
            The Leggett-mode peak appears at the same position ($\omega\sim 1.47\Delta_s^0$) for each value of $V_d$, but is barely visible next to the BS-mode peak (the largest peak).
            The parameters are $\mu=0$, $V_s=-1$, $t=1$, and $\bm{q}=0.001\bm{\hat{e}}_x$.
            }
			\label{symmetricMultibandConductivity}
\end{figure}

The optical response is shown in Fig.~\ref{symmetricMultibandConductivity}, again for several values of $V_d$. A very large peak is generated by the in-phase BS-mode, which moves to lower frequencies for larger $|V_d|$ exactly as the spectrum in Fig. \ref{multiBandSpectrumSymmetric} predicts. Just like in the one-band model, the quasiparticle peak at $2\Delta$ becomes suppressed by this collective mode.
The Leggett mode also results in a smaller peak in the optical spectrum, which can barely be seen in Fig. \ref{multibandConductivitiesSymmetric_swave}. Figure \ref{multibandConductivitiesSymmetric_swave_zoomed} shows the same data on a finer scale. There, it can be seen that the Leggett-mode results in a peak which is the same for different values of $V_d$. The out-of-phase BS mode does not appear in the linear optical response. Physically, this corresponds to the intuitive fact that the electromagnetic field couples to both bands in the same way. Since the gap in both bands is also the same, it is impossible for light to excite fluctuations in opposite directions in the two bands.

\begin{figure}[h]
	\includegraphics[width=\linewidth]
    {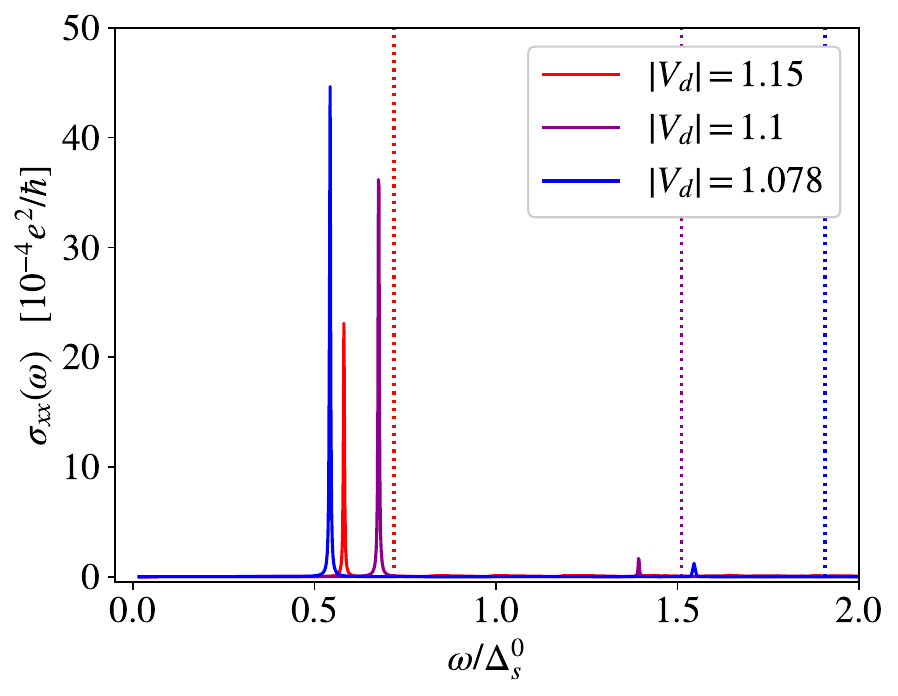}
	\caption{
    Optical conductivity in the $s+id$-phase of the symmetric bilayer model with the $s$- and $d$-wave pairing interactions in the presence of supercurrent. The dotted lines show the minimal quasiparticle gap for each value of $|V_d|$. The parameters are $\mu=0$, $V_s=-1$, $t=1$, and $\bm{q}=0.001\bm{\hat{e}}_x$.
    }
	\label{multibandConductivitiesSymmetric_s+id-wave}
\end{figure}

Once the system is in the $s+id$-wave ground state, the frequency of the Leggett mode depends on $V_d$ as well, and the optical response becomes stronger compared to the $s$-wave phase as shown in Fig.~\ref{multibandConductivitiesSymmetric_s+id-wave}. 
Again, the minimal quasiparticle gap is shown for each data set by the dotted line, and it becomes smaller as $|V_d|$ is increased, since the system moves towards a pure $d$-wave state with nodes in the gap. Just as in the $s$-wave phase, we only see a peak due to the in-phase MSBS mode, and the out-of-phase MSBS mode does not contribute to the optical conductivity. Furthermore, the peak at the quasiparticle gap is still suppressed.
For $V_d=-1.15$, the Leggett mode no longer lies below the gap, and there is only one visible peak.

\subsubsection{The asymmetric case}\label{asymmMultiband}

\begin{figure}[h]
			\centering
            \includegraphics[width=0.97\linewidth]{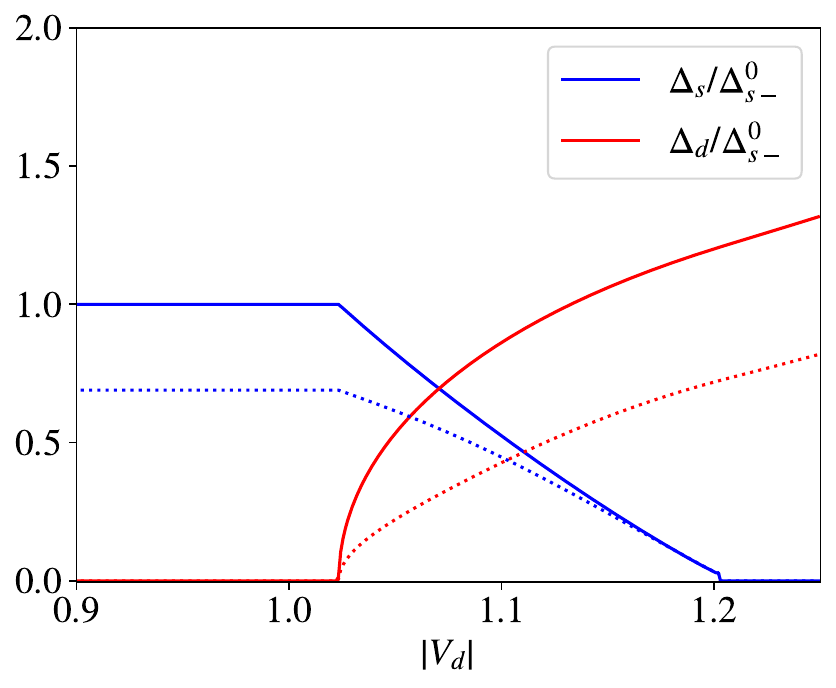}
			\caption{
            Order-parameter components as a function of $|V_d|$ for the asymmetric bilayer model with the $s$- and $d$-wave pairing interactions. The solid and dotted lines correspond to $\Delta_{s-}$ and $\Delta_{s+}$ (i.e., the gap in the lower and upper band), respectively. The parameters are $\mu=-0.2$, $V_s=-1$, $t=1$, and $\bm{q}=0.001\bm{\hat{e}}_x$.
            }
            \label{multibandPhasediagram}
\end{figure}

The situation changes qualitatively when the gaps for the two bands are different. Besides manually changing the interaction strength between the two bands, an easy way to achieve this is to set $\mu\neq0$, which naturally leads to different gaps, $\Delta_{\mu-}$ and $\Delta_{\mu+}$. The phase diagram for $\mu=-0.2$ with the usual interband scattering [Eq.~(\ref{multibandMatrix})] is shown in Fig.~\ref{multibandPhasediagram}.
The most notable difference to the symmetric case discussed in the last section is that the order parameter now takes different values in the two bands. 

We plot the collective-mode spectrum across the transition in Fig.~\ref{multigapSpectrum(zoom)_μ=-0.2,J=0.2,supp=0.1}.
\begin{figure}[h]
    \centering
    \includegraphics[width=\linewidth]{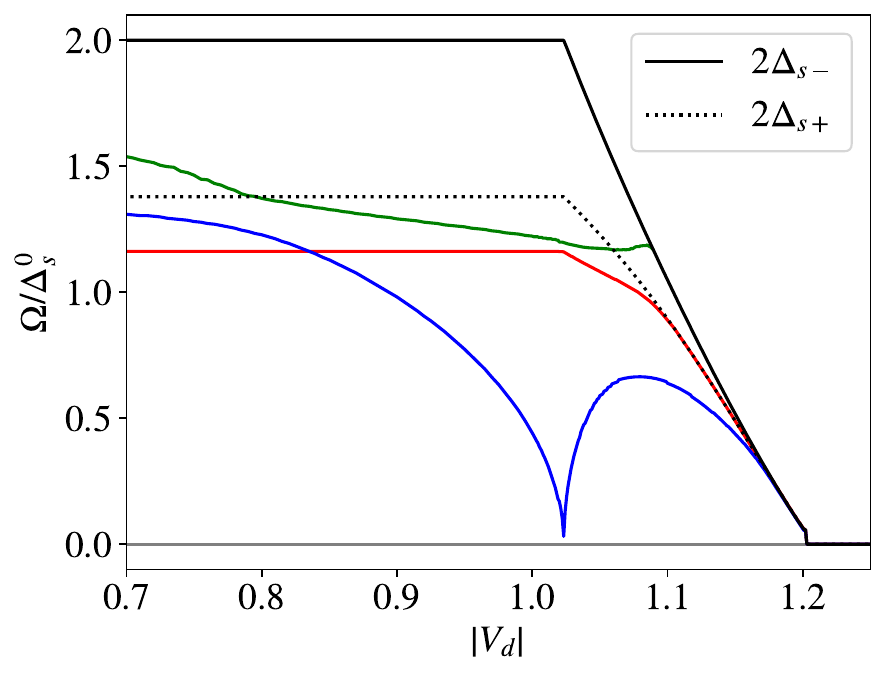}
    \caption{
    Collective-mode spectrum for the asymmetric bilayer model with the $s$- and $d$-wave pairing interactions.
    The four sub-gap modes are an out-of-phase BS/MSBS mode (green), a Leggett mode (red), an in-phase BS/MSBS mode (blue), and the NG mode (gray). The gaps in the lower ($-$) and upper $(+)$ bands are shown by the black solid and dotted lines, respectively.
    The parameters are $\mu=-0.2$, $V_s=-1$, $t=1$, and $\bm{q}=0.001\bm{\hat{e}}_x$.
    }
    \label{multigapSpectrum(zoom)_μ=-0.2,J=0.2,supp=0.1}
\end{figure}
The lower BS mode again becomes gapless at the transition point to the $s+id$ ground state.
\begin{figure}[h]
			\centering
            \subfloat[
                \label{multibandConductivitiesμ=-0.2_swave}
            ]{
                \includegraphics[width=0.97\linewidth]{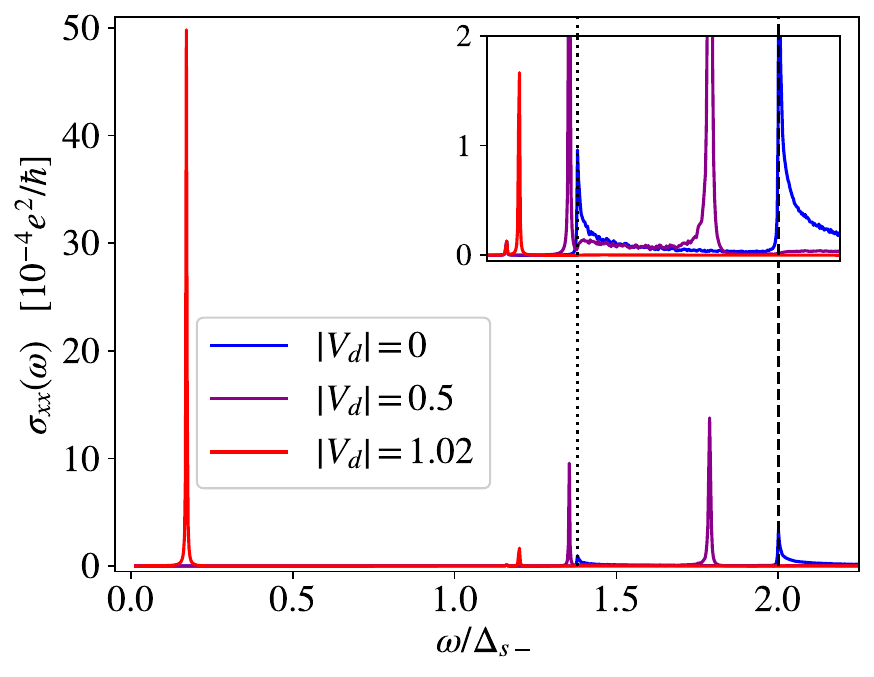}
            }
			\\
            \subfloat[
                \label{multibandConductivities,μ=-0.2,s+id}
            ]{
                \includegraphics[width=0.97\linewidth]{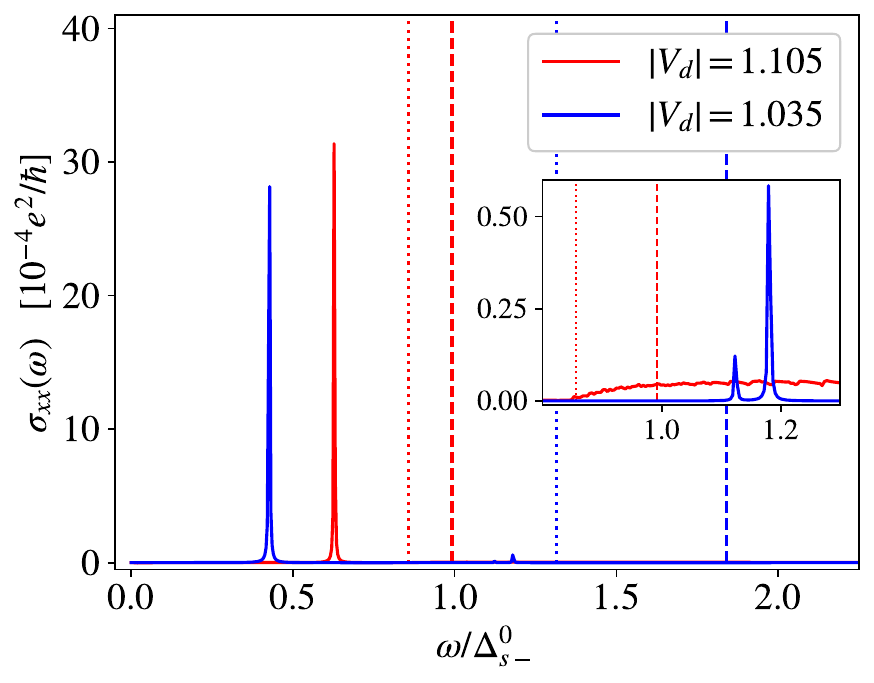}
            }
			\caption{
				Optical conductivity for the asymmetric bilayer model with the $s$- and $d$-wave pairing interactions in the presence of supercurrent (a) in the pure $s$-wave phase and (b) in the $s+id$ phase.
            The inset in each panel shows an enlarged view of the same data (with the $x$-axis shared in (a)). Both of the plots use $\mu=-0.2$, $V_s=-1$, $t=1$, and $\bm{q}=0.001\bm{\hat{e}}_x$.
			}
		\end{figure}
Figure \ref{multibandConductivitiesμ=-0.2_swave} shows how, as $|V_d|$ is varied, the peaks at each of the two gaps become much stronger and move to smaller frequencies, now being associated with the two BS modes rather than quasi-particle excitations. The peak due to the Leggett mode, which can only be seen in the inset of Fig.~\ref{multibandConductivitiesμ=-0.2_swave}, does not change. As the frequency of the lower BS mode approaches zero, the intensity of the corresponding peak becomes larger, while the other one becomes smaller.
After the transition to the $s+id$ ground state, the optical spectrum shown in Fig.~\ref{multibandConductivities,μ=-0.2,s+id} exhibits three peaks, for $V_d=-1.035$.
As expected from Fig.~\ref{multigapSpectrum(zoom)_μ=-0.2,J=0.2,supp=0.1}, the upper two peaks, corresponding to the out of phase MSBS and Leggett mode, eventually disappear from the sub-gap spectrum. This can be seen in Fig.~\ref{multibandConductivities,μ=-0.2,s+id}, for $V_d=1.105$. Here, only a broad peak above the gap remains of these dampened collective modes, which is visible only in the inset.

\section{Rashba System with Interband Pairing}\label{rashba}

It has been pointed out that the linear optical response of collective modes is related to Lifshitz invariants in the Ginzburg-Landau free energy for superconductors \cite{nagashima-san-leggett}. Inspired by this, we use the derived formulas to investigate optical responses in a specific system which has been shown to feature such invariants, namely a superconductor with Rashba-type spin-orbit coupling (SOC) and interband pairing \cite{Lifshitz_Rashba}. We will now introduce the main features of this model, before applying the derived formulas to investigate the collective-mode spectrum and the resulting optical conductivity.
\subsection{Model}
	\subsubsection{Non-interacting Hamiltonian}
        A single-particle Hamiltonian for a system with the Rashba SOC can be written as
		\begin{align}
			H_0 &= \sum_{\bm{k}\sigma\sigma'}
			(\xi^0_{\bm{k}}+\bm{\gamma_k\sigma}_{\sigma\sigma'})
			c^\dagger_{\bm{k}\sigma}
			c_{\bm{k}\sigma'},
		\end{align}
		where $\xi^0$ is the dispersion of the model without SOC and $\bm{\gamma_k}$ is the Rashba SOC vector.
        Since the Hamiltonian is a $2\times 2$ matrix, it can be explicitly diagonalized by the unitary transformation \cite{Lifshitz_Rashba},
		\begin{align}
			\begin{split}
		\begin{pmatrix}
			a_{\bm{k}+}\\
			a_{\bm{k}-}
		\end{pmatrix}
		&=
		U_{\bm{k}}\begin{pmatrix}
			c_{\bm{k}\uparrow}\\
			c_{\bm{k}\downarrow}
		\end{pmatrix},\\
		U_{\bm{k}}=
		\frac{1}{\sqrt{2}}
		\begin{pmatrix}
			1&t_+(\bm{k})\\
			1&t_-(\bm{k})
		\end{pmatrix}&,\quad
		t_\lambda(\bm{k})=
		\lambda
		\frac{\gamma_x(\bm{k})-i\gamma_y(\bm{k})}{|\bm{\gamma_{\bm{k}}}|},
		\end{split}
		\label{rashbaTransformation}
		\end{align}
		which corresponds to a change from the spin basis ($\sigma=\uparrow, \downarrow$) to the Rashba-band basis $\lambda=\pm$. The eigenvalues take the form of $\xi_\lambda(\bm{k})=\xi^0_{\bm{k}}+\lambda|\bm{\gamma_k}|$.
        
		\subsubsection{Superconductivity in Rashba systems}
        
		A superconducting pairing interaction in the Rashba-bands can generally be written as \cite{Lifshitz_Rashba}
		\begin{align}
			H_\text{int}&= \sum_{\bm{kk'}}
			\sum_{\{\lambda_i\}}
			t_{\lambda_2}(\bm{k})t_{\lambda_3}^*(\bm{k'})
			\widetilde{V}^{\{\lambda_i\}}_{\bm{kk'}}
			a^\dagger_{\bm{k}\lambda_1}
			a^\dagger_{\bm{-k}\lambda_2}
			a_{\bm{-k'}\lambda_3}
			a_{\bm{k'}\lambda_4},
			\label{rashbaInteraction}
		\end{align}
		where the interaction is rescaled with the previously defined phase-factors (see Eq.~(\ref{rashbaTransformation})). The Hubbard-Stratonovich transformation is then performed exactly as before (see Eq.~(\ref{hubbardStratonovichAction})) with the difference being an additional summation over the $\lambda$-indices. Note that the fermionic fields will correspond to the $a$-operators, meaning they are expressed in the diagonal basis of the non-interacting hamiltonian. This gives
		\begin{align}
			\begin{split}
			S 
			&=
			\int_0^\beta\text{d}\tau
			\biggm(
			\sum_{\bm{k}}\sum_\lambda (\partial_\tau-\xi_{\bm{k}\lambda})
			\bar{\psi}_{\bm{k}\lambda}\psi_{\bm{k}\lambda}\\
			&\quad-
			\sum_{\bm{kk'}}\sum_{\{\lambda\}}
			t_{\bm{k}\lambda_2}t^*_{\bm{k}\lambda_3}
			\widetilde{V}^{\{\lambda\}}_{\bm{kk'}}
            \bigg[
			b^*_{\bm{k}\lambda_1\lambda_2}
			b_{\bm{k'}\lambda_3\lambda_4}\\
			&+
			b^*_{\bm{k}\lambda_1\lambda_2}
			\psi_{-\bm{k'}\lambda_3}
			\psi_{\bm{k'}\lambda_4}
			+
			b_{\bm{k'}\lambda_3\lambda_4}
			\bar{\psi}_{\bm{k}\lambda_1}
			\bar{\psi}_{\bm{-k}\lambda_2}
			\bigg]
			\biggm),
			\end{split}
		\end{align}
		where one can define the gap functions, which are now $2\times2$-matrices:
		\begin{align}
			\Delta_{\lambda_1\lambda_2}(\bm{k})&=-t_{\bm{k}\lambda_2}\sum_{\bm{k'}\lambda_3\lambda_4}
			t^*_{\bm{k'}\lambda_3}
			\widetilde{V}^{\{\lambda_i\}}_{\bm{kk'}}b_{\bm{k'}\lambda_3\lambda_4}.
		\end{align}
        Note that when we speak of interband-pairing, this corresponds to the off-diagonal elements of this matrix, connecting the $+$ and the $-$ band with each other. Since the Fermi surfaces of these bands are different, this would likely lead to finite-momentum cooper pairs in the groundstate of a physical system. For our purposes, to separate the effects of the cooper pair momentum and the inversion-breaking of the SOC, we assume that the cooper pairs have zero momentum in the ground state before applying a supercurrent.
		The gap functions obey an equivalent self-consistency equation:
		\begin{align}
			\Delta_{\lambda_1\lambda_2}(\bm{k})
			&=
			t_{\bm{k}\lambda_2}
			\sum_{\bm{k'}\lambda_3\lambda_4}
			t^*_{\bm{k'}\lambda_3}
			\widetilde{V}_{\bm{kk'}}^{\{\lambda\}}
			\text{Tr}\left[\tau^-_{\lambda_3\lambda_4}{G}_{\bm{k'}\lambda_4\lambda_3}\right].
			\label{rashbaGapEquationBare}
		\end{align}
		Here $\tau^-_{\lambda_3\lambda_4}$ and ${G}$ are $4\times4$ matrices written in the frequency-momentum representation as
		\begin{align}
			{G}_{\lambda_3\lambda_4}^{-1}
			&=
			\begin{pmatrix}
				\delta^{\lambda_3}_{\lambda_4}(i\omega-\xi_{\bm{k}\lambda_3})
				&
				\Delta_{\lambda_3\lambda_4}(\bm{k})\\
				\Delta^*_{\lambda_4\lambda_3}(\bm{k})
				&
				\delta^{\lambda_3}_{\lambda_4}(i\omega+\xi_{\bm{-k}\lambda_3})
			\end{pmatrix},
			\\
			\tau^-_{\lambda_3\lambda_4}
			&=
			\begin{pmatrix}
				0&0\\
				\delta^{\lambda_3}_{\lambda_4}&0
			\end{pmatrix}.
		\end{align}
        These matrices act on the Nambu spinors  
        $\Psi^\lambda_k=(\psi^\omega_{\bm{k}\lambda},\bar{\psi}^\omega_{\bm{-k}\lambda})$, which can then be transformed with the unitary transformation as
		\begin{align}
			\Psi_k^\lambda\mapsto\begin{pmatrix}
				e^{i\theta_\lambda/2}&0\\
				0&e^{-i\theta_\lambda/2}
			\end{pmatrix}
			\Psi_k^\lambda,\quad e^{i\theta_\lambda}\equiv t_\lambda(\bm{k}).
		\end{align}
		This yields an action expressed entirely in terms of a rescaled gap function, obeying a simplified self-consistency equation \cite{Lifshitz_Rashba},
		\begin{align}
			\widetilde{\Delta}_{\lambda_1\lambda_2}(\bm{k})
			&\equiv
			t^*_{\bm{k}\lambda_2}
			\Delta_{\lambda_1\lambda_2}(\bm{k}),
			\\
			\widetilde{\Delta}_{\lambda_1\lambda_2}(\bm{k})
			&=
			\sum_{\bm{k'}\lambda_3\lambda_4}
			\widetilde{V}_{\bm{kk'}}^{\{\lambda\}}
			\text{Tr}\left[\tau^-_{\lambda_3\lambda_4}\widetilde{{G}}_{\bm{k'}\lambda_4\lambda_3}\right],
			\label{rashbaGapEquation}
		\end{align}
		where $\widetilde{{G}}$ is given by ${G}$ with $\Delta$ being replaced with $\widetilde{\Delta}$.
		The entries of the gap function $\widetilde{\Delta}$ obey \cite{Lifshitz_Rashba}
		\begin{align}
			\widetilde{\Delta}_{\lambda_1\lambda_2}(\bm{k})
			&=
			\lambda_1\lambda_2
			\widetilde{\Delta}_{\lambda_2\lambda_1}(\bm{-k}).
			\label{rashbaGapSymmetry}
		\end{align}
		To make statements about the parity of the gap, one needs to go back to the original spin basis. An explicit calculation of the change of the basis can be found in Appendix \ref{rashbaGapTransformation}. As a result of this, the singlet component $\Psi$ and the triplet $\bm{d}$-vector can be written as
		\begin{align}
			\psi(\bm{k})
			&=
			-\frac12(
			\widetilde{\Delta}^{++}_{\bm{k}}
			+\widetilde{\Delta}^{--}_{\bm{k}}
			),
            \label{rashbaSingletSpinBasis}
            \\
				{\bm{d(k)}}
				&=
				\bm{d}^\text{intra}(\bm{k})+\bm{d}^\text{inter}(\bm{k})
                \label{rashbaTripletSpinBasis},\\
				\bm{d}^\text{intra}(\bm{k}) &= -\frac{\bm{\hat{\gamma}_k}}2
				(\widetilde{\Delta}^{++}_{\bm{k}}
				-\widetilde{\Delta}^{--}_{\bm{k}}),
                \label{rashbaTripletIntraband}
                \\
            \begin{split}
				\bm{d}^\text{inter}(\bm{k})
				&=
				\frac i2({\bm{\hat{\gamma}_k}}\times
				\bm{\hat{z}})
				(\widetilde{\Delta}^{\text{inter}}_{\bm{k}}
				+\widetilde{\Delta}^{\text{inter}}_{\bm{-k}})
                \\
                &\quad
				+\frac{\bm{\hat{z}}}{2}
				(\widetilde{\Delta}^{\text{inter}}_{\bm{k}}
				-\widetilde{\Delta}^{\text{inter}}_{\bm{-k}}).
			\end{split}
			\label{rashbaTripletInterband}
		\end{align}
		\\
        Here, we have renamed $\widetilde{\Delta}^\text{inter}_{\bm{k}}\equiv \widetilde{\Delta}^{+-}_{\bm{k}}=-\widetilde{\Delta}^{-+}_{\bm{-k}}$.
		Given that the following section will focus on the interband pairing, the important conclusions here are:
		\begin{itemize}
			\item Interband pairings only generate triplet states.
			\item An odd-parity interband pairing yields an out-of-plane $\bm{d}$-vector.
			\item An even-parity interband pairing yields an in-plane $\bm{d}$-vector, orthogonal to the Rashba vector $\bm{\gamma_k}$.
		\end{itemize}
		So the main signature of the interband pairing in a Rashba system is a $\bm{d}$-vector which is not parallel to the SOC vector $\bm{\gamma}$.
		\subsubsection{Parametrizing the interaction}
        
		In order to simplify the path-integral formulation, a separable interaction potential has to be assumed once again. In Ref.~\cite{Lifshitz_Rashba}, an interaction of the form,
		\begin{align}
			\widetilde{V}^{\lambda_1\lambda_2\lambda_3\lambda_4}_{\bm{kk'}}
			&=
			-V_g\delta_{\lambda_1\lambda_2}\delta_{\lambda_3\lambda_4}
            \notag
            \\
            &\quad
			-V_u
			[\hat{\bm{\gamma}}_{\bm{k}}\cdot\hat{\bm{\gamma}}_{\bm{k'}}]
			[{\bm{\tau}}_{\lambda_1\lambda_2}(\bm{k})
			\cdot{\bm{\tau}}_{\lambda_3\lambda_4}(\bm{k'})	],
		\end{align}
		has been used, where $\tau_i(\bm{k})=U(\bm{k})\sigma_iU^\dagger(\bm{k})$ are Pauli matrices transformed into the Rashba basis. This can, in a slightly more general form, be written as
		\begin{align}
			\widetilde{V}^{\lambda_1\lambda_2\lambda_3\lambda_4}_{\bm{kk'}}
			&=
			\sum_\mu V^\mu \varphi^\mu_{\lambda_1\lambda_2}(\bm{k})\varphi^\mu_{\lambda_3\lambda_4}(\bm{k'}),
		\end{align}
		where the basis functions $\varphi^\mu$ are now $2\times2$ matrices, whose entries will be appropriately chosen from the irreps of the point group.
		Putting this parametrization into the gap equation (Eq.~(\ref{rashbaGapEquation})) again leads to a similar parametrization for the gap:
		\begin{align}
			\widetilde{\Delta}_{\lambda_1\lambda_2}(\bm{k})
			&=
			\sum_\mu
			\widetilde{\Delta}^\mu \varphi^\mu_{\lambda_1\lambda_2}(\bm{k})
			\label{rashbaGapParametrization}
		\end{align}
		Now, in order to choose the appropriate basis functions, we have to specify the point group of the system. Taking the square lattice for simplicity, we choose $\mathbb{G}=\textbf{C}_{4v}$ as the point group of the system.
        The Rashba vector for this system takes the form of
		\begin{align}
			\bm{\gamma_k}&=\alpha\begin{pmatrix}
				-\sin(k_y)\\\sin(k_x)
			\end{pmatrix}.\label{rashbaVector}
		\end{align}
		For the interaction,
		we choose to consider the trivial representation $A_1$ and the two-dimensionsal representation $E$, which has also been used in Ref.~\cite{Lifshitz_Rashba}. The corresponding basis matrices are 
		\begin{align}
			\widehat{\varphi}_{A_1}&=
            \frac{1}{\sqrt{2}}\begin{pmatrix}
				0&i\\
				-i&0
			\end{pmatrix},\qquad \widehat{\varphi}_{E_{x,y}} =
			\begin{pmatrix}
				0&\sin(k_{x,y})\\
				\sin(k_{x,y})
			\end{pmatrix}.
		\end{align}
		In this way, the basis matrices fulfill Eq.~(\ref{rashbaGapSymmetry}). The matrix for $A_1$ is chosen imaginary so that it is invariant under the time-reversal symmetry, which in the Rashba basis is given by $\mathcal{T}\Delta_{\lambda_1\lambda_2}=\Delta^*_{\lambda_2\lambda_1}$ \cite{Lifshitz_Rashba}. 
        
		\subsection{Generalization of the effective-action approach}
        In this section we will show how the Rashba system with only interband pairing reduces to two one-band models in the effective action approach.
		Since we will always work with $\widetilde{\Delta}$ in the following, the tilde symbol will be dropped from now on.
		When considering only interband pairings, the effective action turns out to be decoupled into two separate blocks once again, which makes the generalization from the earlier calculations relatively simple. In fact, the self-energy $\Sigma$ (Eq.~(\ref{selfEnergy})) and equilibrium Green's function $\mathcal{G}$ (Eq.~(\ref{explicitGreensfunction})) explicitly take the form of
		\begin{align}
			\Sigma
			&=
			\begin{pmatrix}
				-eA_jv^+_j&0&0&\delta\Delta_{\pm}\\
				0 & -eA_jv^-_j&\delta\Delta_\mp&0\\
				0 & \delta\Delta^*_\mp&-eA_jv^+_j&0\\
				\delta\Delta^*_\pm&0&0&-eA_jv^-_j
			\end{pmatrix},\\
			\mathcal{G}^{-1}
			&=
			\begin{pmatrix}
				i\omega-\xi^+&0&0&\Delta_\pm\\
				0 & i\omega-\xi^-&\Delta_\mp&0\\
				0 & \Delta^*_\mp&i\omega+\xi^+&0\\
				\Delta^*_\pm&0&0&i\omega+\xi^-
			\end{pmatrix}.
		\end{align}
		So one gets two decoupled blocks, $\pm$ and $\mp$. The polarization bubbles for each one of them can be calculated separately according to Eqs.~(\ref{phiGeneralSimplified})-(\ref{qvecGeneralSimplified}), and can be simply added together to get the final result. 
		The quantities corresponding to Eq.~(\ref{defineEnergyTerms}) for the two blocks are:
		\begin{align}
			(\xi'_{\bm{k}})^\pm =\bm{q}\cdot\bm{\nabla}\xi^0_{\bm{k}}
			+
			|\bm{\gamma_k}|&,
			\quad (\xi'_{\bm{k}})^\mp=
			\bm{q}\cdot\bm{\nabla}\xi^0_{\bm{k}}
			-|\bm{\gamma_k}|,
			\\
			\bar{\xi}_{\bm{k}}^\pm =\xi^0_{\bm{k}}
			+\bm{q}\cdot\bm{\nabla}|\bm{\gamma_k}|
			&,
			\quad
			\bar{\xi}_{\bm{k}}^\mp=\xi^0_{\bm{k}}-\bm{q}\cdot\bm{\nabla}|\bm{\gamma_k}|,
			\\
			\delta^\pm_{\bm{k}}=  \sqrt{(\bar{\xi}^\pm_{\bm{k}})^2+|\Delta^\pm_{\bm{k}}|^2}
			&,\quad
			\delta^\mp_{\bm{k}}=  \sqrt{(\bar{\xi}^\mp_{\bm{k}})^2+|\Delta^\mp_{\bm{k}}|^2}.
		\end{align}
		The term $\bm{v}^3$, which again is responsible for any non-trivial optical response, is
		\begin{align}
			\begin{split}
			\bm{v}^3_\pm &=\bm{\nabla}(\bm{q}\cdot\bm{\nabla}\xi^0_{\bm{k}})
			+\bm{\nabla}|\bm{\gamma_k}|,\\
			\bm{v}^3_\mp&=
			\bm{\nabla}(\bm{q}\cdot\bm{\nabla}\xi^0_{\bm{k}})
			-\bm{\nabla}|\bm{\gamma_k}|,
			\end{split}
			\label{rashbaPhotonVertex}
		\end{align}
		where we can see that the supercurrent and the Rashba-SOC both have the contributions. The term due to the supercurrent is even in $\bm{k}$, 
        and so are both of the blocks of the Green's function.
        One should note that $\varphi_{A_1}$ obtains a minus sign in the $\mp$-block. Furthermore, it can be taken real for the calculation of the optical conductivity, so as to be able to use Eq.~(\ref{selfEnergy}). This only means that when identifying the modes, one has to remember that the imaginary (real) fluctuations of $\Delta^{A_1}$ will be found in the component corresponding to $\tau_1$ ($\tau_2$).
        
		With all this in mind, Eqs.~(\ref{phiGeneralSimplified})-(\ref{qvecGeneralSimplified}) can be applied to each block individually and then added up.
        
		\subsection{Results}\label{RashbaResults}
        Here we present the results for the collective mode spectrum and the resulting optical conductivity of the Rashba-system with interband pairing, for the chosen irreps.
		An interaction
		\begin{align}
        \begin{split}
			V^{\lambda_1\lambda_2\lambda_3\lambda_4}_{\bm{kk'}}
			&=
			V_{A_1}\widehat{\varphi}_{A_1}^{\lambda_1\lambda_2}(\bm{k})\widehat{\varphi}_{A_1}^{\lambda_3\lambda_4}(\bm{k'})
            \\
			+
			V_{E}\bigg[
			\widehat{\varphi}_{E_x}^{\lambda_1\lambda_2}(\bm{k})
			&\widehat{\varphi}_{E_x}^{\lambda_3\lambda_4}(\bm{k'})
			+ 
			\widehat{\varphi}_{E_y}^{\lambda_1\lambda_2}(\bm{k})
			\widehat{\varphi}_{E_y}^{\lambda_3\lambda_4}(\bm{k'})
			\bigg]
            \end{split}
		\end{align}
        for $|\xi^0_{\bm{k}}|, |\xi^0_{\bm{k'}}| <\omega_c$
		is used for all the calculations in this subsection, where an energy cutoff $\omega_c$ is introduced. The cutoff does not qualitatively affect results, but makes numerical calculations faster.
		\begin{figure*}
			\centering
            \subfloat[
                \label{s-p-transition}
            ]{
                \includegraphics[width=0.49\textwidth]
                {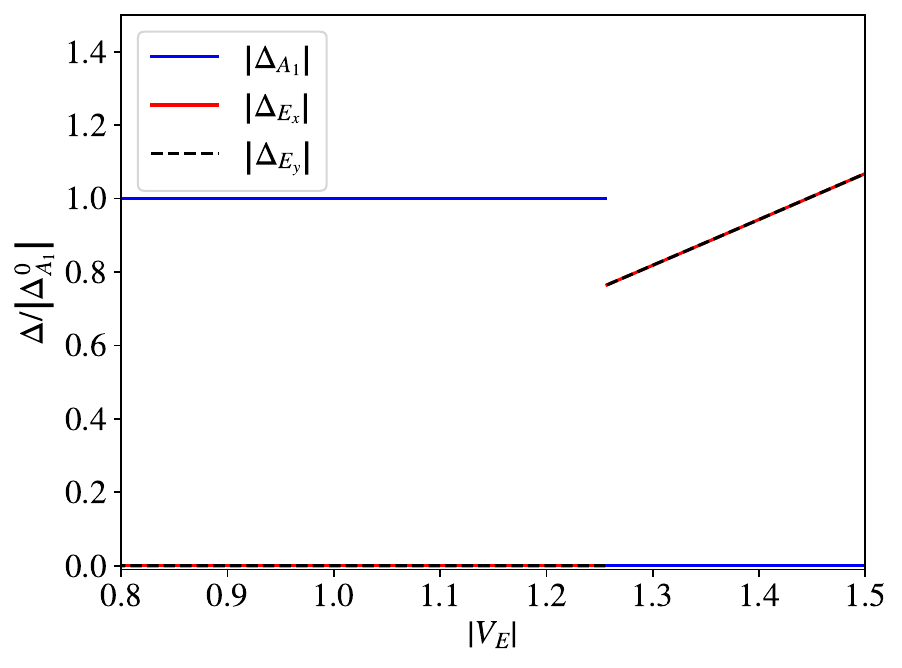}
            }
            \subfloat[
                \label{p-s-transition}
            ]{
                \includegraphics[width=0.49\textwidth]
                {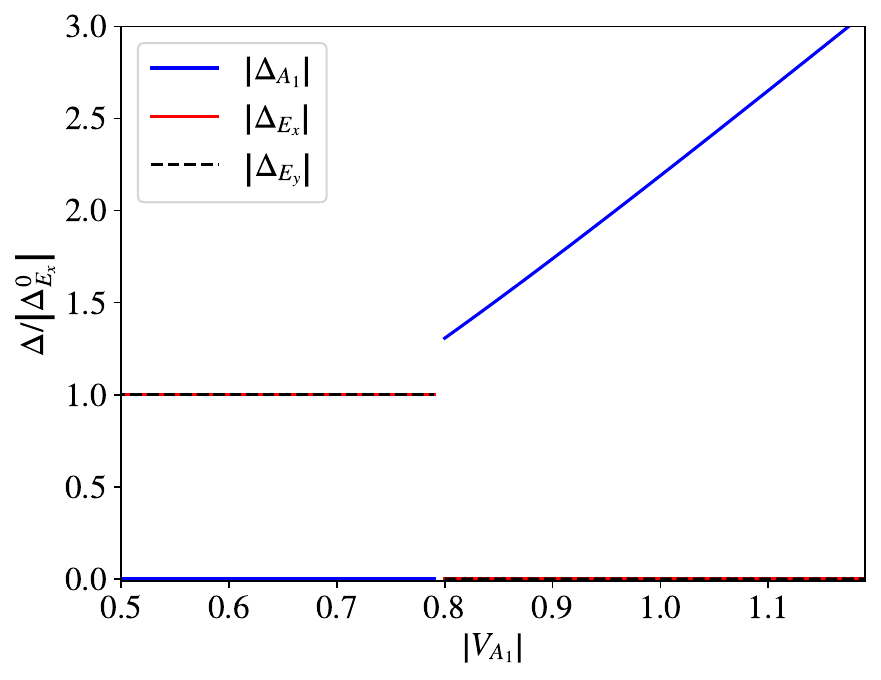}
            }
			\caption{
            Order-parameter components for the Rashba model with the interband pairing (a) for $V_{A_1}=1$ upon increasing $V_E$, and (b) for $V_{E}=1$ upon increasing $V_{A_1}$. The parameters are $\mu=-3$, $t=1$, $\alpha=0.1$, $\bm{q}=0$, and $\omega_c=1$.
            When $\Delta_{E_x}$ and $\Delta_{E_y}$ are nonzero, they have a relative complex phase of $\pi/2$.
            }
		\end{figure*}
		The numerical solution of the gap equation in this system is shown in Fig.~\ref{s-p-transition} for constant $V_{A_1}$ as a function of $V_E$, and in Fig.~\ref{p-s-transition} for constant $V_E$ as a function of $V_{A_1}$. In each case, one can see a discontinuous transition between two phases, the nature of which can be identified from Eq.~(\ref{rashbaTripletInterband}).
        The transition point was determined by comparing the free energy of the different solutions to the self-consistence equation, see Appendix \ref{freeEnergy}.
        The phase in which $V_{A_1}$ is dominant features a real $\bm{d}$-vector that lies in the plane. The phase in which $V_E$ is dominant has a time-reversal symmetry breaking, chiral order parameter of $p+ip$-type.
		Explicitly, the $\bm{d}$-vector takes the form of
		\begin{align}
			\bm{d}\propto
            \begin{cases}
            \begin{pmatrix}
				\sin(k_x)\\
				\sin(k_y)\\
				0
			\end{pmatrix}
            &\text{(real phase)}
			\\
			\begin{pmatrix}
				0\\0\\
				\sin(k_x)
			\end{pmatrix}
			+i
			\begin{pmatrix}
				0\\0\\
				\sin(k_y)
			\end{pmatrix}
            &\text{(chiral phase)}
            \end{cases}.
		\end{align}
		
		\begin{figure*}
			\centering
            \subfloat[
                \label{swave_mode_spectrum}
            ]{
                \includegraphics[width=0.49\textwidth]{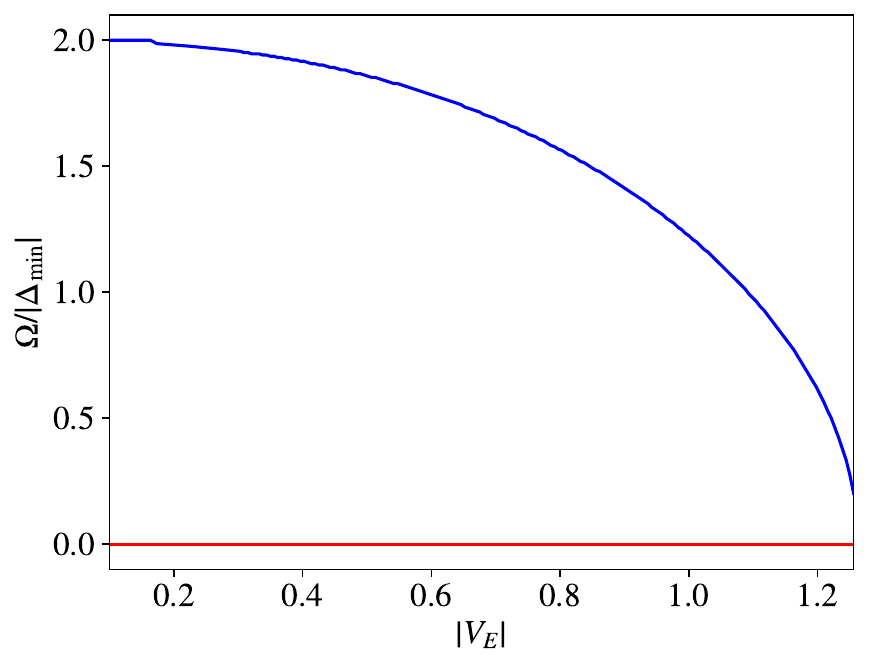}
            }
            \subfloat[
                \label{pwave_mode_spectrum}
            ]{
                \includegraphics[width=0.49\textwidth]{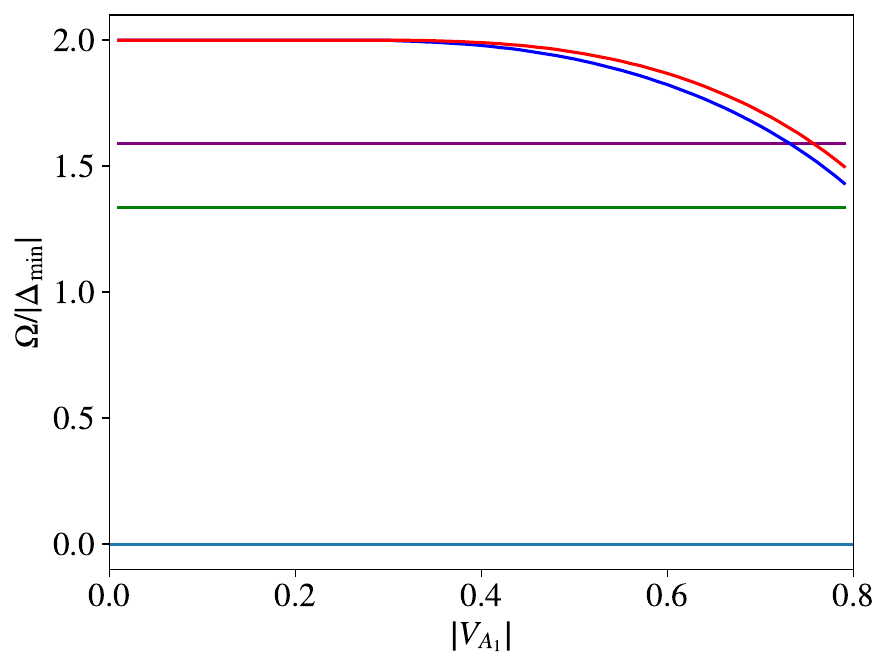}
            }
				\caption{
                Collective-mode spectra for the Rashba model with interband pairing (a) in the real in-plane phase and (b) in the chiral out-of-plane phase.
                Each line corresponds to a collective mode. The $A_1$ phase (a) features only one NG mode (red) and one BS mode (blue). The chiral phase (b) features two BS modes: The red line corresponds to a "left-winding" BS-mode of $e^{+i\omega t}A_1$-fluctuations, and the blue one to a "right-winding" mode with $e^{-i\omega t}A_1$-fluctuations. The other two modes in purple and green correspond to linear superpositions of relative phase oscillations and relative amplitude oscillations.
                Again, the parameters are $\mu=-3$, $t=1$, $\alpha=0.1$, $\bm{q}=0$, and $\omega_c=1$.
				}
				\label{rashbaModeSpectra}
		\end{figure*}
	The transition between the two phases is discontinuous, which means that the corresponding BS modes will not soften (become gapless) at the transition. This is confirmed by the collective mode spectra in the two phases (see Fig.~\ref{rashbaModeSpectra}). In the real phase (Fig.~\ref{swave_mode_spectrum}), the only sub-gap collective mode besides the NG mode is the two-fold degenerate BS mode characterized by fluctuations of the $E_x$ and $E_y$ order parameters. Even though it does not become gapless at the transition, just like in Sec.~\ref{s_vs_d_section}, the mode moves to lower energies with larger $|V_E|$. As the subdominant channel becomes more competitive, one can excite the corresponding fluctuations easier.
    
\begin{figure*}
		\centering
        \subfloat[
            $V_E=0$
            \label{swave_conductivity_nomode}
        ]{
            \includegraphics[width=0.49\textwidth]{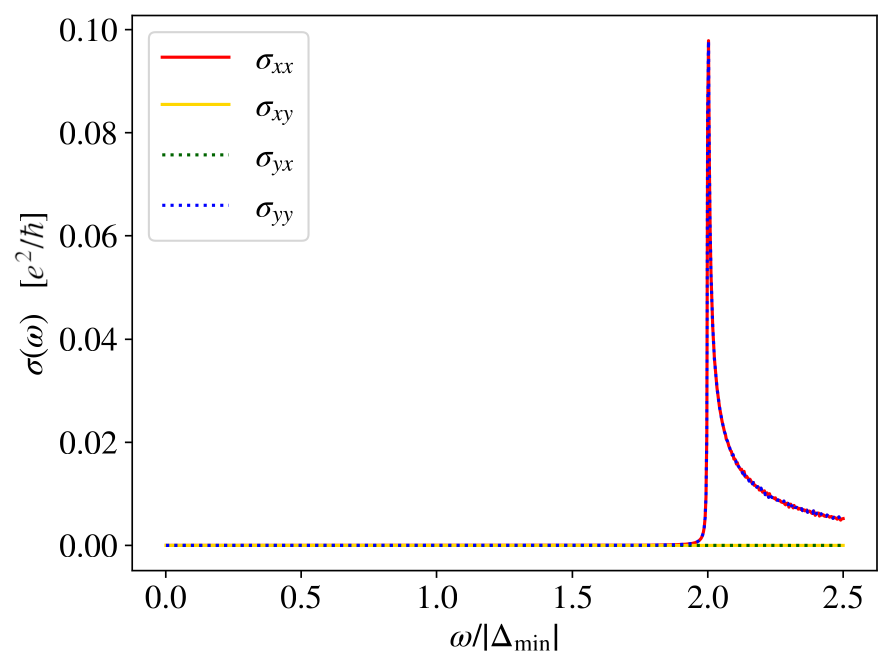}
        }
        \subfloat[
            $V_E=1$
            \label{swave_conductivity_mode}
        ]{
            \includegraphics[width=0.49\textwidth]{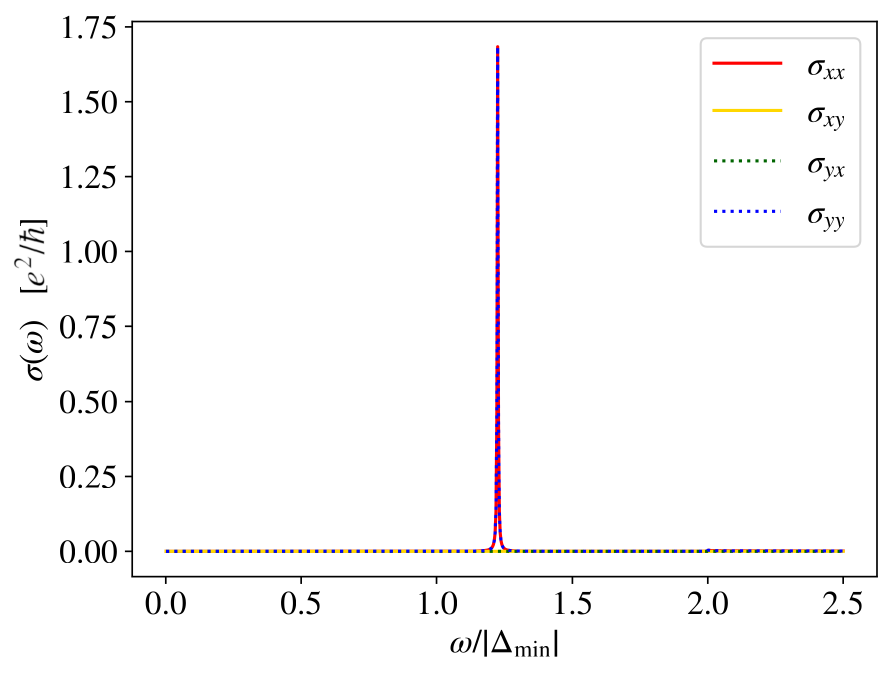}
        }
		\caption{
        Optical conductivity in the real phase of the Rashba model with interband pairing with (a) $V_E=0$ and (b) $V_E=1$.
        The parameters are $V_{A_1}=-1$, $\mu=-3$, $t=1$, $\alpha=0.1$, $\bm{q}=0$, and $\omega_c=1$.
        }
        \label{swave_conductivity}
	\end{figure*}

	The chiral phase (Fig.~\ref{pwave_mode_spectrum}) shows a more diverse spectrum: In addition to two non-degenerate BS modes (the red and blue lines) corresponding to fluctuations of the one-component $A_1$ order parameter, it shows two more collective modes due to relative oscillations of the $E_x$ and $E_y$ order parameters, the frequency of which is independent of the strength of the sub-dominant in-plane order. From the eigenvectors of the effective coupling matrix at the mode frequency, we can identify different fluctuating components and their relative phase shifts. This yields the following expressions for the 4 differently colored modes in Fig. \ref{pwave_mode_spectrum}, expressing $\delta\Delta(t)$ in real time in terms of $\widehat{\varphi}_{A_1}$, 
    $\widehat{\varphi}_{E_x}$ and
    $\widehat{\varphi}_{E_y}$:
    \begin{align}
        \text{Red: } \delta\Delta &\propto 
        \widehat{\varphi}_{A_1} e^{+i\omega t}\\
        \text{Blue: } \delta\Delta &\propto 
        \widehat{\varphi}_{A_1} e^{-i\omega t}\\
        \begin{split}
        \text{Purple: } \delta\Delta 
        &\propto
        2\left[\widehat{\varphi}_{E_x}
        -i\widehat{\varphi}_{E_y}\right]
        \cos(\omega t)\\
        &\quad-
        \left[\widehat{\varphi}_{E_x}
        +i\widehat{\varphi}_{E_y}\right]
        \sin(\omega t)
        \end{split}
        \\
        \begin{split}
        \text{Green: } \delta\Delta 
        &\propto
        \left[\widehat{\varphi}_{E_x}
        -i\widehat{\varphi}_{E_y}\right]
        \cos(\omega t)\\
        &\quad+2
        \left[\widehat{\varphi}_{E_x}
        +i\widehat{\varphi}_{E_y}\right]
        \sin(\omega t)
        \end{split},
    \end{align}
    where $\omega$ is the mode frequency which is of course different for each mode.
	Given that this model breaks inversion symmetry, these collective modes are in principle allowed to contribute to the optical conductivity. Again, the particular system is chosen because it has been shown to feature Lifshitz invariants in the GL free energy \cite{Lifshitz_Rashba}, which are related to the optical response of collective modes \cite{nagashima-san-leggett}.
    In fact, this turned out to be the case for the BS modes, as seen in the numerical data in Fig.~\ref{swave_conductivity}.
		 Fig.~\ref{swave_conductivity_nomode} shows that the model in the real in-plane phase without any subdominant pairing exhibits a peak in the optical conductivity due to quasiparticle excitations. This is already a signature of the Rashba SOC, which breaks inversion symmetry and allows for such a peak. In Fig.~\ref{swave_conductivity_mode}, the optical conductivity with the subdominant pairing $V_E=-1$ is shown. Just like in Sec.~\ref{s_vs_d_section}, the quasiparticle peak becomes suppressed and a new peak at the frequency of the BS mode appears, with significantly higher intensity than the original quasiparticle peak.
        In both Figs.~\ref{swave_conductivity_nomode} and \ref{swave_conductivity_mode}, the peaks appear in both the $x$- and $y$-components of the longitudinal conductivity, as opposed to just being visible in one direction. This is because here the inversion symmetry is broken due to the presence of the SOC, and therefore the system has no preferred direction, as opposed to the finite momentum pairing discussed before.
        
		\begin{figure*}
			\centering
            \subfloat[
                $V_{A_1}=0$
                \label{pwave_conductivity_nomode}
            ]{
                \includegraphics[width=0.47\textwidth]{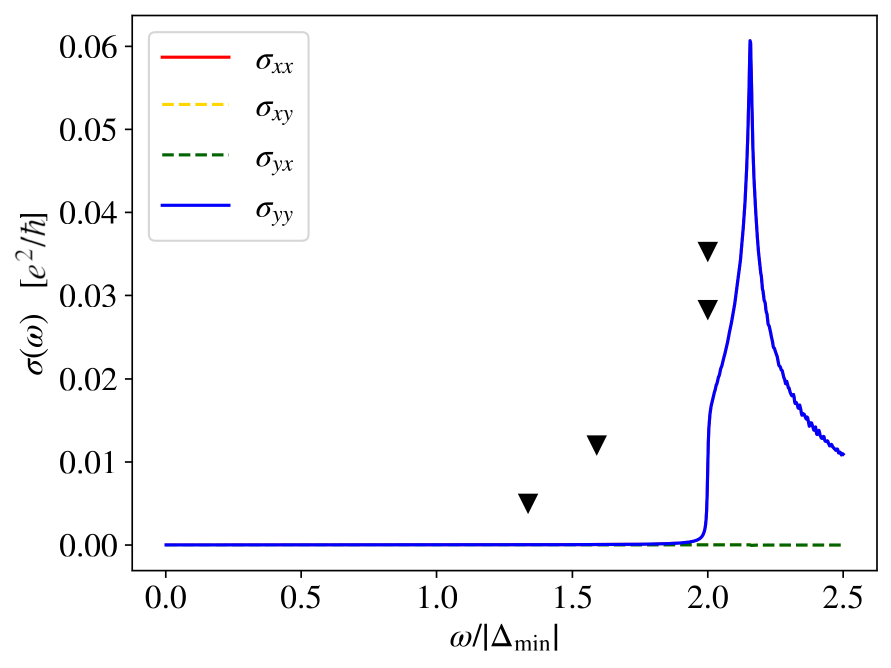}
            }
            \subfloat[
                $V_{A_1}=0.7$
                \label{pwave_conductivity_mode}
            ]{
                \includegraphics[width=0.5\textwidth]{
					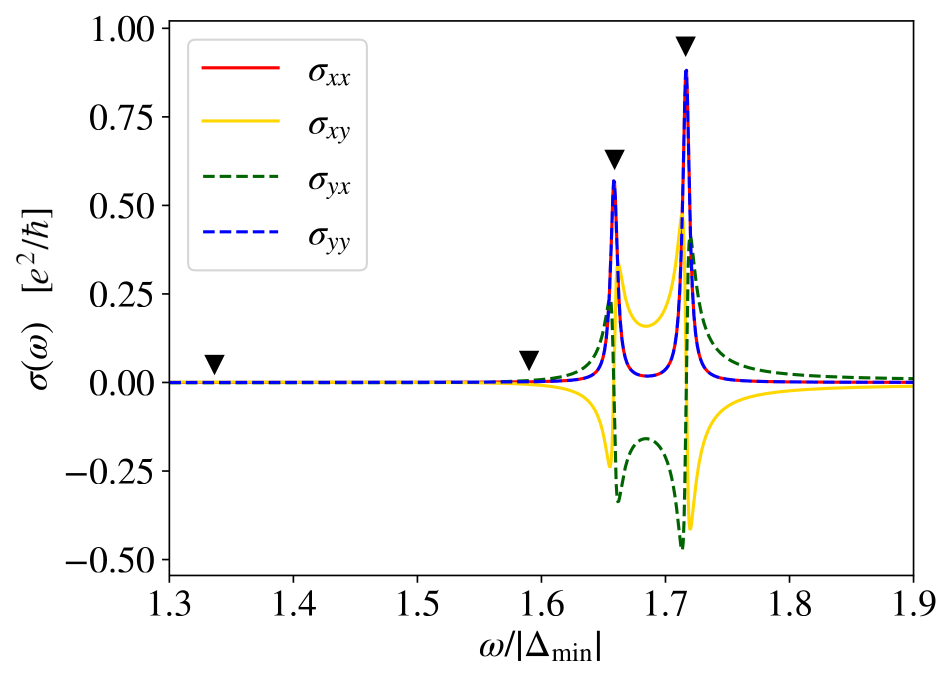}
            }
			\caption{
            Optical conductivity in the chiral phase of the Rashba model with interband pairing with (a) $V_{A_1}=0$ and (b) $V_{A_1}=0.7$. 
            The parameters are $V_{E}=1$, $\mu=-3$, $t=1$, $\alpha=0.1$, $\bm{q}=0$, and $\omega_c=1$. The black arrows denote the frequencies of the collective modes, as seen in Fig.~\ref{pwave_mode_spectrum}.
            }
			\label{pwave_conductivity}
		\end{figure*} 
		For the chiral phase, the optical conductivity is shown in Fig.~\ref{pwave_conductivity}. As $V_{A_1}$ is increased, the BS modes found in Fig.~\ref{pwave_mode_spectrum} generate peaks in the optical conductivity. The most notable difference with respect to all the other optical responses studied so far is that here, the modes also result in a nonzero transverse optical conductivity (i.e., the ac Hall conductivity). Satisfying $\sigma_{xy}=-\sigma_{yx}$, these parts of the optical conductivity have opposite signs on each side of the mode peak. Exactly at the mode frequency, the components vanish. Such peaks can be measured in experiments via the polar Kerr effect, where the polarization angle of a reflected ray of light becomes different from that of the incident light \cite{kerrAngle, kerrAngleMeasurement}.
        
		The lower two modes are not optically active, in spite of the Rashba SOC. This can be explained by considering the form of Eq.~(\ref{rashbaPhotonVertex}). For $\bm{q}=0$, this is an odd function in $\bm{k}$. Considering the form of Eq.~(\ref{qvecGeneralSimplified}), this means that only fluctuations of the irreps with different parity than the ground state may contribute to the optical conductivity. When finite-momentum pairing is considered, one can see in Eq.~(\ref{rashbaPhotonVertex}) that this gives rise to an additional term, which is even in $\bm{k}$. This means that it allows contributions to the optical conductivity by precisely those modes which have not contributed before. 
		\begin{figure*}
			\centering
            \subfloat[
                $\bm{q}=0.03\bm{e}_x$
            ]{
                \includegraphics[width=0.49\textwidth]
				{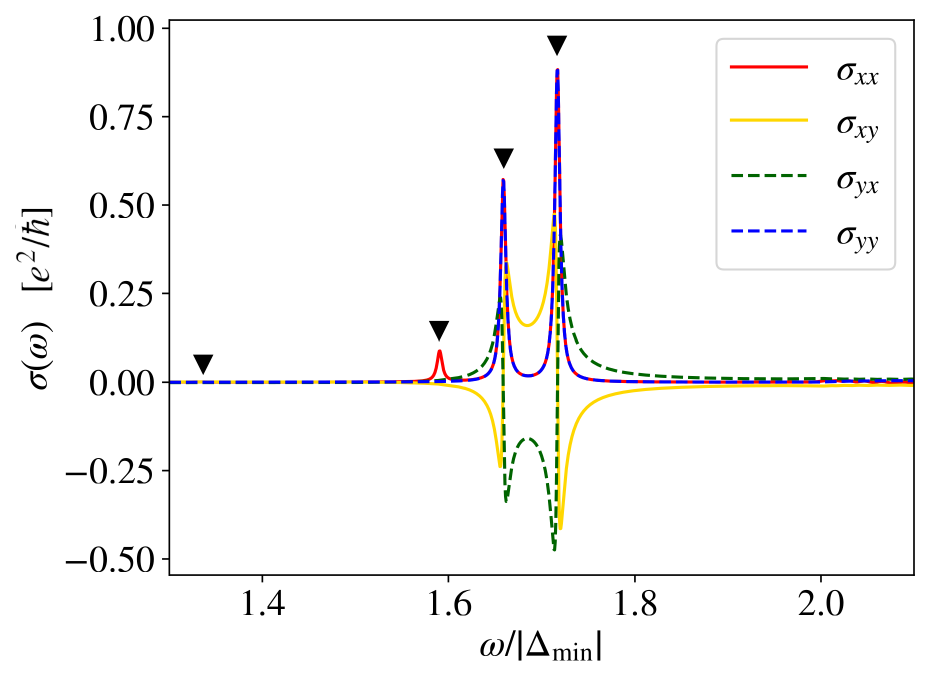}
            }
            \subfloat[
                $\bm{q}=0.03\bm{e}_y$
            ]{
                \includegraphics[width=0.49\textwidth]
				{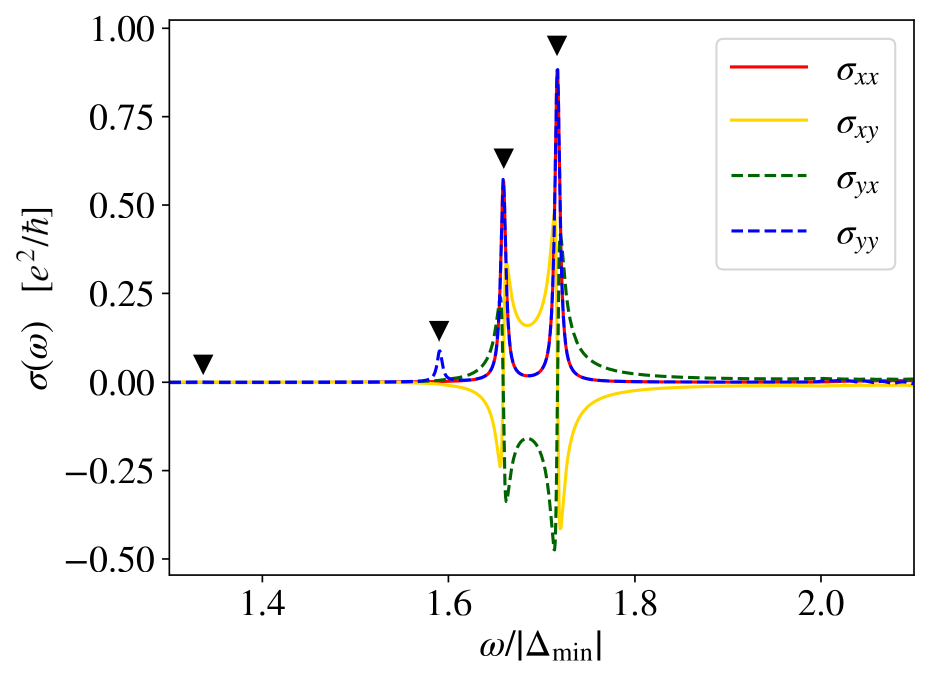}
            }
			\caption{
            Optical conductivity in the chiral phase of the Rashba model with interband pairing in the presence of supercurrent along (a) the $x$ direction and (b) $y$ direction.
            The parameters are $V_E=-1$, $V_{A_1}=-0.4$, $\mu=-3$, $t=1$, $\alpha=0.1$, and $\omega_c=1$. The current is chosen large enough to make the peak visible next to the pre-existing peaks, but small enough not to change the value of the gap.}
			\label{rashbaChiralConductivityCurrent}
		\end{figure*}
		Figure \ref{rashbaChiralConductivityCurrent} confirms this. Here, the same parameters are used as in Fig.~\ref{pwave_conductivity_mode}, and in addition a suppercurrent along (a) the $x$ direction and (b) $y$ direction is applied. As it turns out, it is possible to excite the upper one of the chiral $p$-wave modes, i.e., a relative phase oscillation of the $p_x$ and $p_y$ components. Just like in Sec.~\ref{s_vs_d_section}, this mode can then only be excited by fields along the direction of the supercurrent. However, regardless of the direction of the supercurrent, the out-of-phase amplitude mode does not show any optical response. This is similar to the findings for the symmetric bilayer model in Sec.~\ref{s_vs_d_twoband}, where the amplitude oscillations of the sub-dominant $d$-wave order parameter can only be optically excited when they are in phase, but not when they are out of phase. 
        
		Note that the additional peaks in Fig.~\ref{rashbaChiralConductivityCurrent} do not depend on the strength, or even the presence, of the Rashba SOC. They are simply caused by the relative phase mode of the chiral $p$-wave order parameter \cite{trsbModes}, which becomes optically active in the presence of a supercurrent.
        
\section{Discussion}

In conclusion, we have demonstrated that in the presence of a supercurrent, collective modes in unconventional superconductors that are usually only visible through nonlinear responses become visible as peaks in the linear optical conductivity, provided that they have the same parity as the gap in the ground state.
This was done using an effective action approach within the path integral formalism, deriving general formulas for the linear optical response of collective modes in unconventional superconductors with several pairing channels and bands.
By integrating out fluctuations of the superconducting gap, a gauge-invariant optical response was obtained, with a collective mode contribution given by an effective coupling matrix.

We first applied these to toy models with $s$-wave and $d$-wave pairing interactions. In these examples, various collective modes were found to become optically active in the presence of an injected supercurrent. A one-band model exhibits a peak associated with a BS mode in the $s$-wave phase, which softens at the transition and turns into an MSBS mode of the $s-id$-fluctuations in the $s+id$-phase. This peak stays visible throughout the entire phase, although the intensity changes depending on the mode frequency. This collective mode also strongly suppresses the known peak \cite{supercurrent_crowley} at the quasiparticle gap, which happened for all the models studied in this paper, including the Rashba system (where no supercurrent was needed to induce a non-trivial optical response). The presence of a BS mode always seems to suppress the quasiparticle contribution in the optical conductivity, an effect which becomes stronger as the BS mode lies further below the gap.

We then studied this situation in a simple two-band model. Neglecting the possibility of interband cooper pairs,  we found that interband scattering of intraband cooper pairs was sufficient to generate qualitative changes in the collective-mode spectrum, which now contained a Leggett mode and two BS modes: One which is in phase, one which is out of phase between the two bands. We found that the Leggett mode and in-phase BS mode always generate a peak in the optical conductivity, while the out-of-phase mode only appears when the gaps do not coinside between the two bands.

This could be applied to the study of iron-based superconductors (FeSC), which are thought to feature competing $s$-wave and $d$-wave pairing interactions \cite{recentFeSC-review}. The BS modes have been observed in such materials using Raman spectroscopy \cite{dwaveBSexperiment1, dwaveBSexperiment2}. 
The current-enabled linear optical conductivity will provide a new way of studying and better understanding the pairing mechanism. However, it should be mentioned that the optical conductivity does not provide information about the symmetry of the mode. This is in contrast to the Raman spectroscopy, where the signature of a collective mode can be separated in different channels.
Furthermore, iron-based superconductors are also thought to feature a greater variety of $s$-wave ground states such as $s_\pm$, $s_{++}$ or TRS-breaking combinations like $s_\pm+is_{++}$ can arise \cite{recentFeSC-review}. Whether such states also feature optically active collective modes in the presence of a supercurrent remains an open question.
Lastly, we considered a model for $p$-wave superconductors with the interband pairing and the Rashba-type SOC, again with several pairing channels. Here, we found that the model exhibits peaks due to BS modes in the linear optical response even without a supercurrent. In a chiral $p$-wave phase, these peaks are also seen in the ac Hall conductivity, suggesting possible experimental observation via the polar Kerr effect \cite{kerrAngle, kerrAngleMeasurement}.

Note that in this model, due to the previously calculated Lifshitz invariants \cite{Lifshitz_Rashba}, finite-momentum Cooper pairing could be realized without the need for considering a supercurrent. Regardless of the physical origin, we also found that in the chiral $p$-wave phase, finite-momentum Cooper pairing allows for a peak in the longitudinal optical conductivity due to relative phase oscillations of the two order parameter components \cite{trsbModes}. This is not associated with the Rashba system specifically, and remains for a chiral $p$-wave order parameter without any spin-orbit coupling. Therefore, the presence (or absence) of such a current-enabled peak provides a new experimental probe for confirming (or ruling out) the chiral $p$-wave symmetry of an order parameter. However, as mentioned before, the peaks contain no information about the symmetry of collective modes, so this would have to be combined with other experimental probes.
Sr$_2$RuO$_4$, for example, has often been theorized as an example for such states, but this has been difficult to verify experimentally \cite{chiralKallin}. Optical measurements in the presence of an injected supercurrent therefore present a promising avenue for exploring the ground state of this and related materials.
All of these suggests a promising way of exploring the properties of unconventional superconductors, in particular those suspected to feature sub-dominant pairing channels or multiple superconducting bands. The calculated formulas are fairly general, and can easily be applied to any band structures and pairing interactions, allowing comparison with experimental data.

There also remain several open questions for further theoretical investigation. Firstly, the mechanism behind the suppression of the quasiparticle peak by the BS modes should be investigated. Secondly, disorder in superconductors may have significant effects on the optical conductivity, such as introducing Mattis-Bardeen terms above the superconducting gap \cite{MattisBardeen}.\\ 
It has been shown that contributions from the Higgs mode to the optical conductivity exist in the presence of disorder, so we would expect that collective modes below the gap will also survive after introducing impurities. Further research is needed to verify this and to quantify the effect of disorder on our results. 
\begin{acknowledgements}
    G.N. thanks Manfred Sigrist and Raigo Nagashima for helpful discussions, and the Department of Physics at the University of Tokyo for its hospitality.
    G.N. received financial support from the Swiss-European Mobility Programme.
    This work was supported by JST FOREST (Grant No.~JPMJFR2131), JST PRESTO (Grant No.~JPMJPR2256), and JSPS KAKENHI (Grant Nos.~JP22K20350, JP23K17664, JP24H00191, JP25H01246, JP25H01251, and JP25K17312).
\end{acknowledgements}

\appendix

\section{Derivation of the gap equation from the path integral}\label{gapAppendix}

To derive the equilibrium gap equation for a general one-band superconductor, let us consider the full expression for the effective action [Eq.~(\ref{eq: action})] after integrating out the fermions:
	\begin{align}
		S_\text{eff}&=-\beta\sum_{\bm{kk'}}V_{\bm{kk'}}b^*_{\bm{k}}b_{\bm{k'}}
		-\text{Tr}\ln(G^{-1}).
	\end{align}
    An expansion of the logarithm in powers of $\mathcal{G}\Sigma$ would (up to a constant) yield the expression in Eq.~(\ref{effectiveActionFluctuations}). Here, the full expression may be used. Demanding that the effective action has a stationary point with respect to the mean field at the equilibrium value gives
    \begin{align}
		\frac{\delta S_\text{eff}}{\delta \Delta^*_{\bm{k}}}&=
		\beta b_{\bm{k}}-\text{Tr}\left(
		\frac{\delta\mathcal{G}^{-1}}{\delta \Delta^*_{\bm{k}}}\mathcal{G}
		\right) \notag
        \\
        &=\beta b_{\bm{k}}-\sum_{n}\text{Tr}
		\left[
		-\tau^-\mathcal{G}_{\bm{k,q}}
		\right]\overset{!}{=}0, \\
        \sum_{n}\text{Tr}
		\left[
		-\tau^-\mathcal{G}_{\bm{k,q}}
		\right]
        &=\frac12
        \sum_n\frac{\text{Tr}\left[(\tau^x-i\tau^y)
        (\tau^x\Delta^R_{\bm{k}}-i\tau^y\Delta^I_{\bm{k}}\right]}
        {(i\omega-\xi_{\bm{k}}')^2-\delta_{\bm{k}}^2}.
    \end{align}
    The sum over the Matsubara frequencies is carried out in the usual way using the residue theorem, yielding
    \begin{align}
        \sum_{n}\text{Tr}
		\left[
		-\tau^-\mathcal{G}_{\bm{k,q}}
		\right]&=\frac{\Delta_{\bm{k}}}{2\delta_{\bm{k}}}(n_F(E^+)-n_F(E^-))\\
		\Rightarrow0&=\beta b_{\bm{k}}+\frac{\Delta_{\bm{k}}}{2\delta_{\bm{k}}}(n_F(E^+)-n_F(E^-))\\
		\Rightarrow \Delta_{\bm{k}}&=\sum_{\bm{k'}}
		V_{\bm{kk'}}\frac{\Delta_{\bm{k'}}}{2\delta_{\bm{k'}}}(n_F(E^+)-n_F(E^-)),\label{gapequationgeneral}
	\end{align}
    where $n_F$ is the Fermi distribution. 
    In the case of multiband superconductors with interband scattering but only intraband pairing, this formula is generalized to
    \begin{align}
        \Delta_{\bm{k}}^\alpha=\sum_{\bm{k'}}
		V_{\bm{kk'}}^{\alpha\beta}\frac{\Delta_{\bm{k'}}^\beta}{2\delta_{\bm{k'}}^\beta}(n_F(E^+_\beta)-n_F(E^-_\beta)). \label{gapequationgeneralMultiband}
    \end{align}
    
\section{Simplifying the polarization bubbles}\label{simplifying}
In this appendix we give some further detail on how the formulas given in Eq. \ref{phiGeneral} - \ref{qGeneral} can be simplified, including how the contained sums over Matsubara frequencies are calculated. After explaining this in the general case, further simplifications are made for the specific case of competing $s$- and $d$-wave order discussed in Sec. \ref{s_vs_d_section} to obtain simple equations for the collective mode spectrum.

\subsection{In general}

The fluctuations $\Delta^{\mu,x/y}$ are coupled to the Green's function [Eq.~(\ref{explicitGreensfunction})] via a vertex of the form $\tau^{1/2}$. The photon vertex [Eq.~(\ref{bareVertex})] contains $\tau^{0,3}$, which means that the terms in Eqs.~(\ref{phiGeneral})-(\ref{qGeneral}) are all possible Matsubara sums of the form,
	 \begin{align}
	 	\begin{split}
	 		D_{ab}&=\frac{1}{\beta}
	 		\sum_n 
	 		\frac{1}
	 		{\left[(i\Omega+i\omega'_n)^2-\delta_{\bm{k}}^2\right]
	 			\left[(i\omega'_n)^2-\delta_{\bm{k}}^2\right]}\\
	 		\times\text{Tr}\biggm[
	 		&\tau_a
	 		[(i\omega'+i\Omega)\tau_0
	 		+\bar{\xi}_{\bm{k}}\tau_3
	 		+\Delta_{\bm{k}}\tau_1
	 		-\Delta^I_{\bm{k}}\tau_2
	 		]\\
	 		\times
	 		&\tau_b\left[
	 		i\omega'_n\tau_0
	 		+\bar{\xi}_{\bm{k}}\tau_3
	 		+\Delta^R_{\bm{k}}\tau_1-\Delta^I_{\bm{k}}\tau_2\right]
	 		\biggm],
	 	\end{split}\label{fullMatsubaraExpression}
	 \end{align}
	 where the trace is taken over the Nambu indices and $i\omega'_n=i\omega_n-\xi'_{\bm{k}}$. Since only $\tau_0$ has a non-zero trace, only the terms $\propto\tau_0$ contribute to the trace with a factor of 2. Furthermore, the following Matsubara sums can be calculated with the residue theorem:
	 \begin{align}
	 		\frac{1}{\beta}\sum_n\frac{2(i\omega_n')^2}
	 		{\left[(i\Omega+i\omega'_n)^2-\delta_{\bm{k}}^2\right]
	 			\left[(i\omega'_n)^2-\delta_{\bm{k}}^2\right]}
	 		&=
	 		\frac{(i\Omega)^2-2\delta_{\bm{k}}^2}
	 		{\delta_{\bm{k}}
	 			\left[(i\Omega)^2-4\delta_{\bm{k}}^2\right]
	 		},
	 	\\
	 		\frac{1}{\beta}\sum_n\frac{2i\omega_n'}
	 		{\left[(i\Omega+i\omega'_n)^2-\delta_{\bm{k}}^2\right]
	 			\left[(i\omega'_n)^2-\delta_{\bm{k}}^2\right]}
	 		&=
	 		\frac{-i\Omega}
	 		{\delta_{\bm{k}}
	 			\left[(i\Omega)^2-4\delta_{\bm{k}}^2\right]
	 		},
	 	\\
	 		\frac{1}{\beta}\sum_n\frac{2}
	 		{\left[(i\Omega+i\omega'_n)^2-\delta_{\bm{k}}^2\right]
	 			\left[(i\omega'_n)^2-\delta_{\bm{k}}^2\right]}
	 		&=
	 		\frac{2}
	 		{\delta_{\bm{k}}
	 			\left[(i\Omega)^2-4\delta_{\bm{k}}^2\right]
	 		}.
	 \end{align}
	 Applying this to Eqs.~(\ref{phiGeneral})-(\ref{qGeneral}) yields Eqs.~(\ref{phiGeneralSimplified})-(\ref{qvecGeneralSimplified}) in the main text.
     
    \subsection{For competing $s$- and $d$-wave pairing}
    
    With the $s$- and $d$-wave pairing channels, we can put Eq.~(\ref{piGeneralSimplified}) into an explicit matrix form, using the fact that the gap is of the $s+id$ type:
    \begin{widetext}
    \begin{align}
        \Pi &=-2
        \begin{pmatrix}
			2\Delta_d^2I[\varphi^4_d]+2I[\bar{\xi}^2\varphi_d^2]
			&2I[(\bar{\xi}^2+\Delta_d^2\varphi_d^2)\varphi_d]
			&-i(i\Omega)I[\bar{\xi}\varphi_d^2]+2\Delta_s\Delta_dI[\varphi_d]
			&+2\Delta_s\Delta_dI[\varphi_d^2]-i(i\Omega)I[\bar{\xi}\varphi_d]\\
			2I[(\bar{\xi}^2+\Delta_d^2\varphi_d^2)\varphi_d]
			&2\Delta_d^2I[\varphi^2_d]+2I[\bar{\xi}^2]
			&+2\Delta_s\Delta_dI[\varphi^2_d]-i(i\Omega)I[\bar{\xi}\varphi_d]
			&-i(i\Omega)I[\bar{\xi}]+2\Delta_s\Delta_dI[\varphi_d]
			\\
			i(i\Omega)I[\bar{\xi}\varphi_d^2]+2\Delta_s\Delta_dI[\varphi_d^3]
			&2\Delta_d\Delta_sI[\varphi^2_d]+i(i\Omega)I[\bar{\xi}\varphi_d]
			&2\Delta_s^2I[\varphi^2_d]+2I[\varphi^2_d\bar{\xi}^2]
			&2I[(\bar{\xi}^2+\eta_s^2)\varphi_d]\\
			+2\Delta_s\Delta_dI[\varphi^2_d]+i(i\Omega)I[\bar{\xi}\varphi_d]
			&i(i\Omega)I[\bar{\xi}]+2\Delta_s\Delta_dI[\varphi_d]
			&2I[(\bar{\xi}^2+\Delta_s^2)\varphi_d]
			&2\eta_s^2I[1]+2I[\bar{\xi}^2]
		\end{pmatrix}.
    \end{align}
    \end{widetext}
    This expression can be simplified when considering the square lattice model at half filling. 
    Neglecting also the effect of the supercurrent given by $\bm{q}$, any integral containing an odd power of $\bm{\bar{\xi}}$ or $\varphi_d$ vanishes.
    Furthermore, the self-consistent conditions [Eq.~(\ref{self-consistency})] can be used to simplify several terms:        
    \begin{align}
    \begin{split}
		I[(i\Omega)^2-4\delta_{\bm{k}}^2]&=\sum_{\bm{k}}\frac{n_F(E^+)-n_F(E^-)}{\delta_{\bm{k}}}
        \\&\overset{!}{=}\frac{2}{V_s},
    \end{split}
        \\
    \begin{split}
		I[\varphi^2_d((i\Omega)^2-4\delta_{\bm{k}}^2)]&=
		\sum_{\bm{k}}\varphi^2_d\frac{n_F(E^+)-n_F(E^-)}
		{\delta_{\bm{k}}}
        \\
        &\equiv L_d\overset{!}{=}\frac{2}{V_d}.
    \end{split}\label{d-wave-selfconsistency}
    \end{align}
    Since the self-consistency equation for each gap component only holds while the component is non-zero, the condition (\ref{d-wave-selfconsistency}) can only be used once the system is in the $s+id$ phase. Before that, one simply has to insert $L_d$ instead of $\frac2{V_d}$.
    \begin{align}
		\begin{split}
		I[\varphi^2_d(\bar{\xi}^2+\Delta_d^2\varphi^2_d)]
		&=
		I[\varphi^2_d(\delta_{\bm{k}}^2-\Delta_s^2)]
        \\=
		-&\Delta_s^2I[\varphi^2_d]
		+\frac{(i\Omega)^2}{4}I[\varphi^2_d]
		-\frac1{2V_d},
        \end{split}\\
        \begin{split}
		I[\bar{\xi}^2+\Delta_d^2\varphi^2_d]
		&=
		I[\delta_{\bm{k}}^2-\Delta_s^2]
        \\=-&\Delta_s^2I[1]
		+\frac{(i\Omega)^2}{4}I[1]
		-\frac1{2V_s},
        \end{split}\\
        \begin{split}
		I[\varphi^2_d(\bar{\xi}^2+\Delta_s^2)]
		&=
		I[\varphi_d^2(\delta_{\bm{k}}^2-\Delta_d^2\varphi_d^2)]
        \\
		=-&\Delta_d^2I[\varphi^4_d]
		+\frac{(i\Omega)^2}{4}I[\varphi^2_d]
		-\frac1{2V_d},
        \end{split}\\
        \begin{split}
		I[\bar{\xi}^2+\Delta_s^2]&=
		I[\delta_{\bm{k}}^2-\Delta_d^2\varphi_d^2]\\
		=-&\Delta_d^2I[\varphi_d^2]+\frac{(i\Omega)^2}{4}I[1]-\frac1{2V_s}.
		\end{split}
	\end{align}
    
    For the $s$-wave phase, this leads to a diagonal effective coupling matrix. Defining $I_n\equiv I[\varphi_d^n]$:
    \begin{widetext}
        
    \begin{align}
	 	V_\text{eff}^{-1}&=
	 	\begin{pmatrix}
	 		\frac{1}{V_d}-L_d+\left(\frac{(i\Omega)^2}{2}-2\Delta_s^2\right)I_2
	 		&0
	 		&0
	 		&0
	 		\\
	 		0
	 		&\left(\frac{(i\Omega)^2}{2}-2\Delta_s^2\right)I_0
	 		&0
	 		&0
	 		\\
	 		0
	 		&0
	 		&\frac{1}{V_d}-L_d+\frac{(i\Omega)^2}{2}I_2
	 		&0
	 		\\
	 		0
	 		&0
	 		&0
	 		&\frac{(i\Omega)^2}{2}I_0
	 	\end{pmatrix}.
	 \end{align}

This decouples the mode condition (\ref{modecondition}) into four separate equations, for each of the four components of the above matrix to be zero.
The last component, corresponding to $is$-fluctuations, is zero for $\Omega=0$, which yields the expected Nambu-Goldstone mode. The third component defines the equation for the BS mode:
\begin{align}
    \frac{1}{V_d}-L_d+\frac{(i\Omega)^2}{2}I_2=0,\label{swave_BSmode}
\end{align}
while the other two components have no sub-gap solution. Once $\Delta_d$ becomes non-zero, the coupling matrix takes the form,
\begin{align}
		V_\text{eff}^{-1}&=
		\begin{pmatrix}
			\left(\frac{(i\Omega)^2}{2}-2\Delta_s^2\right)I_2
			&0
			&0
			&2\Delta_s\Delta_dI_2
			\\
			0
			&\left(\frac{(i\Omega)^2}{2}-2\Delta_s^2\right)I_0
			&2\Delta_s\Delta_dI_2
			&0
			\\
			0
			&2\Delta_s\Delta_dI_2
			&\frac{(i\Omega)^2}{2}I_2-2\Delta_d^2I_4
			&0
			\\
			2\Delta_s\Delta_dI_2
			&0
			&0
			&\frac{(i\Omega)^2}{2}I_0-2\Delta_d^2I_2
		\end{pmatrix}.
	\end{align}
    \end{widetext}
    This time, the mode condition decouples into two equations. The first one, containing the $s$- and $id$-fluctuations, gives rise to the MSBS mode:
    \begin{align}
		I_0\left(\frac{(i\Omega)^2}{2}-2\Delta_s^2\right)
		\left(
		\frac{(i\Omega)^2}{2}I_0-2\Delta_d^2I_4
		\right)
		-4\eta_s^2\Delta_d^2(I_2)^2
		=0. \label{s+idwave_MSBSmode}
	\end{align}
    While the second equation, governing the $is$- and $d$-fluctuations, contains the NG mode, which can again be seen from the fact that $i\Omega=0$ is always a solution,
    \begin{align}
		I_2\left(\frac{(i\Omega)^2}{2}-2\Delta_s^2\right)
		\left(
		\frac{(i\Omega)^2}{2}I_0-2\Delta_d^2I_2
		\right)
		-4\Delta_s^2\Delta_d^2(I_2)^2&=0\\
		\Leftrightarrow
		\frac{(i\Omega)^4}{4}I_2I_0
		-(i\Omega)^2\left(\Delta_s^2I_2I_0+\Delta_d^2(I_2)^2\right)&=0.
	\end{align}

\section{Quench dynamics in the one-band model with $s$- and $d$-wave pairing interactions}
\label{dynamicsAppendix}

In this Appendix, we discuss the quench dynamics of collective modes in the one-band model with the $s$-wave and $d$-wave pairing interactions which has been studied in Sec.~\ref{sec: one-band model}.
As shown in Table \ref{collective-modes-characteristics}, the BS and MSBS modes show undamped oscillations with frequency below the energy gap, while the Higgs mode shows damped oscillations with frequency corresponding to the gap.
In order to understand these behaviors, we also perform the linearized analysis.

We adopt two methods for the quench simulation: 
The first one is a ``$\Delta$-quench'', in which a small fluctuation from equilibrium is introduced in the initial gap function, and then we let the system evolve from the corresponding initial state.
The second method is a ``$V$-quench'', in which the interaction parameters $V_s$ and $V_d$ are suddenly changed.
We use both methods here, since it seems impossible to induce the BS mode in the pure $s$-wave phase by the $V$-quench, and so is it to discuss the damping behavior based on the linearized analysis for the $\Delta$-quench.  
We confirm that the oscillation behavior of the Higgs and MSBS modes are consistent between these two quench methods. 
\begin{table}[h]
    \centering
    \caption{
    Characteristics of the collective modes in the one-band model with the $s$- and $d$-wave pairing interactions studied in Sec.~\ref{sec: one-band model}.
    }
    \begin{tabular}{ccccc} \hline \hline
        mode & ground state & fluctuation               & asymptotics & frequency     \\ \hline
        Higgs  & pure-$s$     & $\Delta_s$                & damped               & on the gap    \\
        BS     & pure-$s$     & $\Delta_d$                & undamped             & below the gap \\
        MSBS   & $s+id$       & $\Delta_s, \Delta_d$ & undamped             & below the gap \\ \hline \hline
    \end{tabular}
    \label{collective-modes-characteristics}
\end{table}

\subsection{Equilibrium phase diagram}

\begin{figure}[h]
    \centering
    \includegraphics[width=\linewidth]{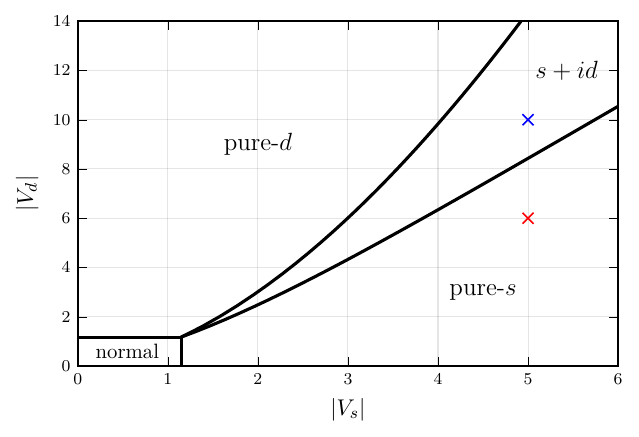}
    \caption{
    Equilibrium phase diagram of the one-band model with the $s$- and $d$-wave pairing interactions studied in Sec.~\ref{sec: one-band model}. There are three superconducting phases (pure-$s$, pure-$d$, and $s+id$) as well as the normal state. The red and blue markers represent $(V_s, V_d) = (-5, -6)$ and $(-5, -10)$, used in the simulation later. The parameters are $\mu = 0$, $t = 1$, and $T = 0.05$.
    }
    \label{equilibrium-phase-diagram}
\end{figure}

We briefly review the equilibrium phase diagram of the one-band model. 
In the mean-field approximation, the gap functions,
\begin{align}
    \Delta_s^{\textrm{eq}} &= - \frac{V_s}{N} \sum_{\bm{k}} \bigl\langle c_{\bm{k}, \uparrow}^{\dag} c_{-\bm{k}, \downarrow}^{\dag} \bigr\rangle, \\
    \Delta_d^{\textrm{eq}} &= - \frac{V_d}{N} \sum_{\bm{k}} \varphi_d(\bm{k}) \bigl\langle c_{\bm{k}, \uparrow}^{\dag} c_{-\bm{k}, \downarrow}^{\dag} \bigr\rangle,
\end{align}
satisfy the self consistent relation,
\begin{equation}
    \begin{pmatrix}
        \Delta_s^{\textrm{eq}} \\ \Delta_d^{\textrm{eq}}
    \end{pmatrix} 
    = \begin{pmatrix}
        - \frac{V_s}{2N} \sum_{\bm{k}} \frac{t_{\bm{k}}^{\textrm{eq}}}{E_{\bm{k}}^{\textrm{eq}}} & 
        - \frac{V_s}{2N} \sum_{\bm{k}} \frac{\varphi_d(\bm{k}) t_{\bm{k}}^{\textrm{eq}}}{E_{\bm{k}}^{\textrm{eq}}} \\
        - \frac{V_d}{2N} \sum_{\bm{k}} \frac{\varphi_d(\bm{k}) t_{\bm{k}}^{\textrm{eq}}}{E_{\bm{k}}^{\textrm{eq}}} & 
        - \frac{V_d}{2N} \sum_{\bm{k}} \frac{\varphi_d(\bm{k})^2 t_{\bm{k}}^{\textrm{eq}}}{E_{\bm{k}}^{\textrm{eq}}} 
    \end{pmatrix} 
    \begin{pmatrix}
        \Delta_s^{\textrm{eq}} \\ \Delta_d^{\textrm{eq}}
    \end{pmatrix},
    \label{equilibrium-gap-equation}
\end{equation}
where $E_{\bm{k}}^{\textrm{eq}}$ and $t_{\bm{k}}^{\textrm{eq}}$ are defined as 
\begin{align}
    E_{\bm{k}}^{\textrm{eq}} &\coloneqq \sqrt{\xi_{\bm{k}}^2 + |\Delta_s^{\textrm{eq}}|^2 + \varphi_d(\bm{k})^2 |\Delta_d^{\textrm{eq}}|^2}, \\
    t_{\bm{k}}^{\textrm{eq}} &\coloneqq \tanh\bigl(\beta E_{\bm{k}}^{\textrm{eq}} / 2\bigr). 
\end{align}
We solve Eq.~(\ref{equilibrium-gap-equation}) under the assumption that the $s+d$ solution is unstable \cite{s-vs-d-mixedSymmetry}. 
There are three types of non-trivial solutions: pure-$s$, pure-$d$, and $s+id$. 
Figure \ref{equilibrium-phase-diagram} shows the equilibrium phase diagram. 
We will later show the results of the quench dynamics for the set of parameters, $(V_s, V_d) = (-5, -6)$, $(-5, -10)$ (marked by red and blue crosses in Fig.~\ref{equilibrium-phase-diagram}), which correspond to the pure-$s$ and $s+id$ phases, respectively. 

\subsection{The dynamics of collective modes}
\subsubsection{$\Delta$-quench}

\begin{figure*}[t]
    \centering
    \subfloat[
        \label{fig-Delta-quench-Higgs}
    ]{
        \includegraphics[width=0.33\textwidth]{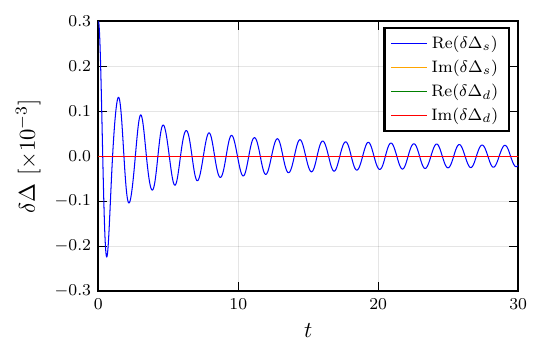}
    }
    \subfloat[
        \label{fig-Delta-quench-BS}
    ]{
        \includegraphics[width=0.33\textwidth]{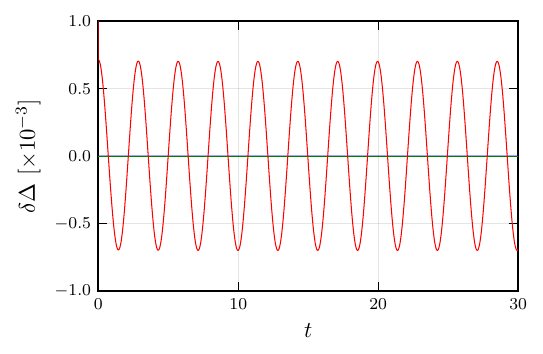}
    }
    \subfloat[
        \label{fig-Delta-quench-MSBS}
    ]{
        \includegraphics[width=0.33\textwidth]{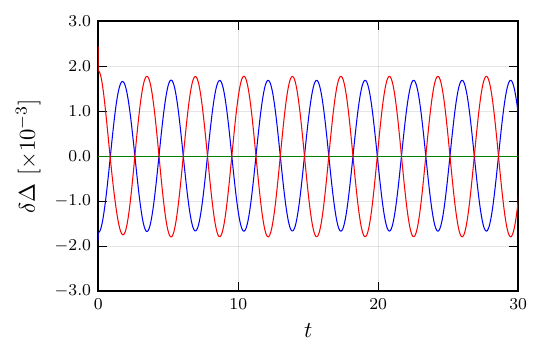}
    }
    \caption{%
        Order-parameter variations after the $\Delta$-quench in the one-band model with the $s$- and $d$-wave pairing interactions studied in Sec.~\ref{sec: one-band model}. 
        Each panel shows the real and imaginary parts of $\delta \Delta_s$ and $\delta \Delta_d$.
        [(a), (b)] The pure $s$-wave phase with $V_s=-5, V_d=-6$
        and the quench parameters given by (a) $\delta\Delta_s^{\rm i}=0.001, \delta\Delta_d^{\rm i}=0$ and (b) $\delta\Delta_s^{\rm i}=0, \delta\Delta_d^{\rm i}=0.001 i$.
        (c) The $s+id$-wave phase with $V_s=-5, V_d=-10$ and the quench parameters given by $\delta\Delta_s^{\rm i}=0, \delta\Delta_d^{\rm i}=0.001 i$.
        The other parameters are $\mu = 0$, $t = 1$, and $T = 0.05$. Note that the curves that do not oscillate are overlapped with the horizontal axis ($\delta\Delta=0$).
    }
    \label{fig-Delta-quench}
\end{figure*}

We first show the numerical results for the $\Delta$-quench,
which consists of three steps: (i) The equilibrium gap functions are calculated for a set of given interaction parameters $V_s$ and $V_d$. (ii) Small shifts ($\Delta_s\to \Delta_s+\delta\Delta_s^{\rm i}, \Delta_d \to \Delta_d+\delta\Delta_d^{\rm i}$) are introduced in the initial values of the gap functions. (iii) The time evolution of the gap functions is numerically calculated from those initial conditions by the fourth-order Runge-Kutta (RK4) method.

Figure \ref{fig-Delta-quench} shows typical oscillation dynamics of the gap functions induced by the $\Delta$-quench.
In order to focus on the fluctuations, the offset (the center of the oscillations) is subtracted. 
The panels (a)-(b) and (c) correspond to the results for the pure $s$-wave phase ($V_s=-5, V_d=-6$) and the $s+id$-wave phase ($V_s=-5, V_d=-10$), respectively.
The initial displacements are $\delta \Delta_s^{\textrm{i}} = 0.001, \delta \Delta_d^{\textrm{i}} = 0$ in (a), and $\delta \Delta_s^{\textrm{i}} = 0, \delta \Delta_d^{\textrm{i}} = 0.001 i$ in (b) and (c). 
In (a), the dominant component is $\operatorname{Re}(\delta \Delta_s)$, which exhibits a damped oscillation corresponding to the Higgs mode. In (b), the dominant component is $\operatorname{Im}(\delta \Delta_d)$, which oscillates permanently and corresponds to the BS mode. In (c), $\operatorname{Re}(\delta \Delta_s)$ and $\operatorname{Im}(\delta \Delta_d)$ oscillate in opposite phases, corresponding to the MSBS mode.
These results suggest that the Higgs mode is damped, while the BS and MSBS modes are undamped. 

\subsubsection{$V$-quench}

We next show the results for the $V$-quench,
which consists of three steps: (i) The equilibrium gap functions are calculated for a set of given interaction parameters $V_s, V_d$. (ii) The interaction parameters $V_s, V_d$ are suddenly changed at $t=0$. (iii) The time evolution of the gap functions is numerically calculated by the RK4 method.

\begin{figure*}
    \centering
    \subfloat[
        \label{fig-V-quench-Higgs}
    ]{
        \includegraphics[width=0.49\textwidth]{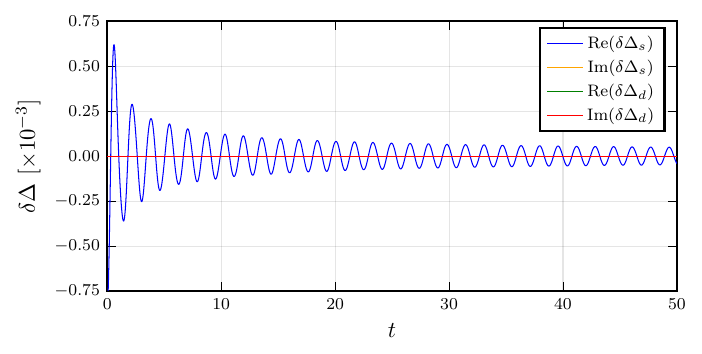}
    }
    \subfloat[
        \label{fig-V-quench-MSBS}
    ]{
        \includegraphics[width=0.49\textwidth]{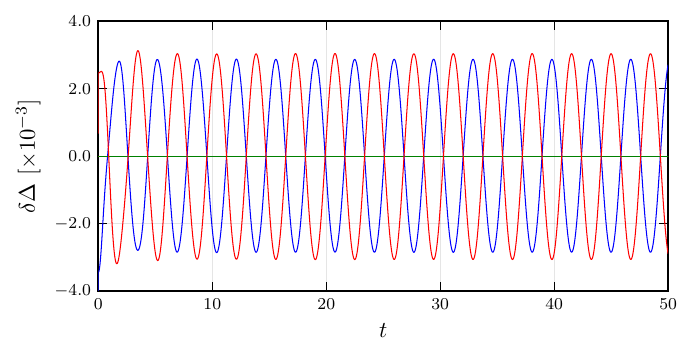}
    }
    \caption{
        Order-parameter variations after the $V$-quench in the one-band model with the $s$- and $d$-wave pairing interactions studied in Sec.~\ref{sec: one-band model}. 
        Each graph shows the real and imaginary parts of $\delta \Delta_s$ and $\delta \Delta_d$. 
        (a) The pure $s$-wave phase with $V_s=-5, V_d=-6$.
        (b) The $s+id$-wave phase with $V_s=-5, V_d=-10$.
        The interaction parameters are quenched as $V_s\to 1.001 \times V_s$ and $V_d\to 1.001 \times V_d$.
        The other parameters are $\mu = 0$, $t = 1$, and $T = 0.05$.
        Note that the curves that do not oscillate are overlapped with the horizontal axis ($\delta\Delta=0$).
    }
    \label{fig-V-quench}
\end{figure*}

Figure \ref{fig-V-quench} shows typical oscillation dynamics of the gap functions induced by the $V$-quench.
The panels (a) and (b) correspond to the pure $s$-wave phase ($V_s = -5, V_d = -6$) and the $s+id$-wave phase ($V_s = -5, V_d = -10$), respectively. 
The interaction parameters are quenched by a factor of $1.001$. 
Each graph shows the real and imaginary parts of $\delta \Delta_s$ and $\delta \Delta_d$. In (a), the dominant component is $\operatorname{Re}(\delta \Delta_s)$, which shows a damped oscillation and represents the Higgs mode. 
In (b), $\operatorname{Re}(\delta \Delta_s)$ and $\operatorname{Im}(\delta \Delta_d)$ oscillate in opposite phases without damping,  corresponding to the MSBS mode. Again, these results are consistent with the previous observation that the Higgs mode is damped and the MSBS modes is undamped.

We can also extract a more precise form of the oscillations.
The results in Fig.~\ref{fig-V-quench-Higgs} indicate that the Higgs mode shows a $1/\sqrt{t}$ decay with frequency $\omega = 3.9 \pm 0.1$ ($\approx 2\Delta_s^{\rm eq}$), which is consistent with Fig.~\ref{fig-Delta-quench-Higgs}. 
The results in Fig.~\ref{fig-V-quench-MSBS} show that the MSBS mode shows the undamped oscillation with frequency $\omega = 1.8 \pm 0.1$, which is also consistent with Fig.~\ref{fig-Delta-quench-MSBS}.
The frequency of each mode is summarized in Fig.~\ref{collective-modes-frequency}.

\subsection{Analytical evaluation of the collective-mode frequencies}
\label{appendix-linear-response-analysis}

The numerical simulation suggests that the Higgs mode shows the damped oscillation with the frequency being equal to the superconducting gap, and the BS and MSBS modes show the undamped oscillations with the frequency below the gap. 
This behavior can be understood from the linearized equation of motion analytically \cite{Tsuji2015}.

Here, we define Anderson's pseudospins \cite{Anderson1958} as 
\begin{equation}
    \sigma_{\bm{k}}^{\alpha}(t) \coloneqq \frac{1}{2} \bigl\langle \psi_{\bm{k}}^{\dag} \tau_{\alpha} \psi_{\bm{k}} \bigr\rangle. 
\end{equation}
Then, the gap functions can be written as 
\begin{align}
    \Delta_s(t) &= - \frac{V_s}{N} \sum_{\bm{k}} \bigl(\sigma_{\bm{k}}^x(t) + i \sigma_{\bm{k}}^y(t)\bigr), \\
    \Delta_d(t) &= - \frac{V_d}{N} \sum_{\bm{k}} \varphi_d(\bm{k}) \bigl(\sigma_{\bm{k}}^x(t) + i \sigma_{\bm{k}}^y(t)\bigr). 
    \label{gap-equation}
\end{align}
In order to calculate the frequency of the collective modes, we assume that the displacement from the equilibrium solution is small enough so that we can linearize the equation of motion with respect to the variation for each quantity from the equilibrium one.
The pseudospin $\sigma_{\bm{k}}^{\alpha}(t)$ and the gap function $\Delta_{\mu}(t)$ are expanded as 
\begin{align}
    \sigma_{\bm{k}}^{\alpha}(t) &= \sigma_{\bm{k}, \textrm{eq}}^{\alpha} + \delta \sigma_{\bm{k}}^{\alpha}(t) \quad (\alpha = x, y, z), \label{pseudospin-expansion} \\
    \Delta_{\mu}(t) &= \Delta_{\mu}^{\textrm{eq}} + \delta \Delta_{\mu}(t) \quad (\mu = s, d). \label{gap-func-expansion}
\end{align}
Here, $\sigma_{\bm{k}, \textrm{eq}}^{\alpha}$ and $\Delta_{\mu}^{\textrm{eq}}$ are the equilibrium solutions. 
We substitute Eqs.~(\ref{pseudospin-expansion}) and (\ref{gap-func-expansion}) into the Bloch equation for the pseudospins \cite{Tsuji2015},
\begin{equation}
    \partial_t \begin{pmatrix}
        \sigma_{\bm{k}}^x(t) \\ \sigma_{\bm{k}}^y(t) \\ \sigma_{\bm{k}}^z(t) 
    \end{pmatrix} = 2 \begin{pmatrix}
        - \xi_{\bm{k}} \sigma_{\bm{k}}^y(t) - \Delta_{\bm{k}}''(t) \sigma_{\bm{k}}^z(t) \\
        \xi_{\bm{k}} \sigma_{\bm{k}}^x(t) + \Delta_{\bm{k}}'(t) \sigma_{\bm{k}}^z(t) \\
        - \Delta_{\bm{k}}'(t) \sigma_{\bm{k}}^y(t) + \Delta_{\bm{k}}''(t) \sigma_{\bm{k}}^x(t) 
    \end{pmatrix}. 
\end{equation}
Ignoring the second-order terms in displacement and performing the Fourier transformation, the Bloch equation becomes 
\begin{widetext}
    \begin{equation}
        \begin{pmatrix}
            \delta \widetilde{\sigma}_{\bm{k}}^x(\omega) \\
            \delta \widetilde{\sigma}_{\bm{k}}^y(\omega) \\
            \delta \widetilde{\sigma}_{\bm{k}}^z(\omega) 
        \end{pmatrix} 
        = \frac{t_{\bm{k}}^{\textrm{eq}} / E_{\bm{k}}^{\textrm{eq}}}{4 (E_{\bm{k}}^{\textrm{eq}})^2 - \omega^2} 
        \begin{pmatrix}
            2 \bigl[\xi_{\bm{k}}^2 + \Delta_{\bm{k}, \textrm{eq}}''^2\bigr] \delta\widetilde{\Delta}_{\bm{k}}'(\omega) - \bigl[i\omega \xi_{\bm{k}} + 2\Delta_{\bm{k}, \textrm{eq}}' \Delta_{\bm{k}, \textrm{eq}}''\bigr] \delta\widetilde{\Delta}_{\bm{k}}''(\omega) \\
            \bigl[i\omega \xi_{\bm{k}} - 2\Delta_{\bm{k}, \textrm{eq}}' \Delta_{\bm{k}, \textrm{eq}}''\bigr] \delta\widetilde{\Delta}_{\bm{k}}'(\omega) + 2 \bigl[\xi_{\bm{k}}^2 + \Delta_{\bm{k}, \textrm{eq}}'^2\bigr] \delta\widetilde{\Delta}_{\bm{k}}''(\omega) \\
            \bigl[i\omega \Delta_{\bm{k}, \textrm{eq}}'' - 2\xi_{\bm{k}} \Delta_{\bm{k}, \textrm{eq}}'\bigr] \delta\widetilde{\Delta}_{\bm{k}}'(\omega) - \bigl[i\omega \Delta_{\bm{k}, \textrm{eq}}' - 2\xi_{\bm{k}} \Delta_{\bm{k}, \textrm{eq}}''\bigr] \delta\widetilde{\Delta}_{\bm{k}}''(\omega) 
        \end{pmatrix}. 
    \end{equation}

    In order to explain the behavior observed in the numerical simulations, we assume that the relative phase between $\Delta_s(t)$ and $\Delta_d(t)$ is always $\pi/2$. We can choose a gauge such that the $s$-wave components ($\Delta_s^{\textrm{eq}}$ and $\delta \Delta_s(t)$) are real, and the $d$-wave components are imaginary. 
    In the following, we redefine $\Delta_d^{\textrm{eq}}$ and $\delta \Delta_d(t)$ as the imaginary part of the corresponding quantities.
    In other words, $\Delta_s$, $\Delta_d$ and $\Delta_{\bm{k}}$ satisfy the following relations: 
    \begin{align}
        \Delta_{\bm{k}, \textrm{eq}}' = \Delta_s^{\textrm{eq}}, \quad 
        &\Delta_{\bm{k}, \textrm{eq}}'' = \varphi_d(\bm{k}) \Delta_d^{\textrm{eq}}, \\
        \delta \Delta_{\bm{k}}'(t) = \delta \Delta_s(t), \quad 
        &\delta \Delta_{\bm{k}}''(t) = \varphi_d(\bm{k}) \delta \Delta_d(t) .
    \end{align}
    Using these relations, the linearized Bloch equation becomes 
    
    \begin{equation}
        \begin{pmatrix}
            \delta \widetilde{\sigma}_{\bm{k}}^x(\omega) \\
            \delta \widetilde{\sigma}_{\bm{k}}^y(\omega) \\
            \delta \widetilde{\sigma}_{\bm{k}}^z(\omega) 
        \end{pmatrix} 
        = \frac{t_{\bm{k}}^{\textrm{eq}} / E_{\bm{k}}^{\textrm{eq}}}{4 (E_{\bm{k}}^{\textrm{eq}})^2 - \omega^2} 
        \begin{pmatrix}
            2 \bigl[\xi_{\bm{k}}^2 + \varphi_d(\bm{k})^2 (\Delta_d^{\textrm{eq}})^2\bigr] \delta\widetilde{\Delta}_s(\omega) - \bigl[i\omega \xi_{\bm{k}} + 2\varphi_d(\bm{k}) \Delta_s^{\textrm{eq}} \Delta_d^{\textrm{eq}}\bigr] \varphi_d(\bm{k}) \delta\widetilde{\Delta}_d(\omega) \\
            \bigl[i\omega \xi_{\bm{k}} - 2\varphi_d(\bm{k}) \Delta_s^{\textrm{eq}} \Delta_d^{\textrm{eq}}\bigr] \delta\widetilde{\Delta}_s(\omega) + 2 \bigl[\xi_{\bm{k}}^2 + (\Delta_s^{\textrm{eq}})^2\bigr] \varphi_d(\bm{k}) \delta\widetilde{\Delta}_d(\omega) \\
            \bigl[i\omega \varphi_d(\bm{k}) \Delta_d^{\textrm{eq}} - 2\xi_{\bm{k}} \Delta_s^{\textrm{eq}}\bigr] \delta\widetilde{\Delta}_s(\omega) - \bigl[i\omega \Delta_s^{\textrm{eq}} - 2\xi_{\bm{k}} \varphi_d(\bm{k}) \Delta_d^{\textrm{eq}}\bigr] \varphi_d(\bm{k}) \delta\widetilde{\Delta}_d(\omega) 
        \end{pmatrix}. 
    \end{equation}
    Using the gap equation (i.e., the Fourier transformed version of Eq.~(\ref{gap-equation})), we finally obtain
    \begin{equation}
        \begin{pmatrix}
            0 \\ 0
        \end{pmatrix} = \begin{pmatrix}
            1 + \frac{V_s}{N} \sum_{\bm{k}} \frac{2 t_{\bm{k}}^{\textrm{eq}} / E_{\bm{k}}^{\textrm{eq}}}{4 \left(E_{\bm{k}}^{\textrm{eq}}\right)^2 - \omega^2} \left[\xi_{\bm{k}}^2 + \varphi_d(\bm{k})^2 \left(\Delta_d^{\textrm{eq}}\right)^2\right] & 
            - \frac{V_s}{N} \sum_{\bm{k}} \frac{2 \varphi_d(\bm{k})^2 t_{\bm{k}}^{\textrm{eq}} / E_{\bm{k}}^{\textrm{eq}}}{4 \left(E_{\bm{k}}^{\textrm{eq}}\right)^2 - \omega^2} \Delta_s^{\textrm{eq}} \Delta_d^{\textrm{eq}} \\
            - \frac{V_d}{N} \sum_{\bm{k}} \frac{2 \varphi_d(\bm{k})^2 t_{\bm{k}}^{\textrm{eq}} / E_{\bm{k}}^{\textrm{eq}}}{4 \left(E_{\bm{k}}^{\textrm{eq}}\right)^2 - \omega^2} \Delta_s^{\textrm{eq}} \Delta_d^{\textrm{eq}} & 
            1 + \frac{V_d}{N} \sum_{\bm{k}} \frac{2 \varphi_d(\bm{k})^2 t_{\bm{k}}^{\textrm{eq}} / E_{\bm{k}}^{\textrm{eq}}}{4 \left(E_{\bm{k}}^{\textrm{eq}}\right)^2 - \omega^2} \left[\xi_{\bm{k}}^2 + \left(\Delta_s^{\textrm{eq}}\right)^2\right] 
        \end{pmatrix} \begin{pmatrix}
            \delta\widetilde{\Delta}_s(\omega) \\ \delta\widetilde{\Delta}_d(\omega)
        \end{pmatrix}. 
        \label{fourier-transformed-gap-equation}
    \end{equation}
\end{widetext}

From Eq.~(\ref{fourier-transformed-gap-equation}), we can calculate each mode's frequency. 
The following calculation is classified into two cases, the pure $s$-wave phase and $s+id$-wave phase. 

\subsubsection{Pure $s$-wave phase}

In the pure $s$-wave phase, $\Delta_d^{\textrm{eq}} = 0$. 
From the equilibrium gap equation (\ref{equilibrium-gap-equation}), we have $1 = -\frac{V_s}{2N} \sum_{\bm{k}} \frac{t_{\bm{k}}^{\textrm{eq}}}{E_{\bm{k}}^{\textrm{eq}}}$, with which Eq.~(\ref{fourier-transformed-gap-equation}) becomes
\begin{equation}
	\begin{pmatrix}
		0 \\ 0
	\end{pmatrix} = \begin{pmatrix}
        - V_s C(\omega) & 0 \\
		0 & 1 + V_d D(\omega) 
	\end{pmatrix} \begin{pmatrix}
		\delta\widetilde{\Delta}_s(\omega) \\ \delta\widetilde{\Delta}_d(\omega)
	\end{pmatrix}. 
    \label{pure-s-gap-eq}
\end{equation}
Here, the functions $C(\omega)$ and $D(\omega)$ are defined as 
\begin{align}
    C(\omega) &\coloneqq \frac{1}{2N} \sum_{\bm{k}} \frac{t_{\bm{k}}^{\textrm{eq}} / E_{\bm{k}}^{\textrm{eq}}}{4 \left(E_{\bm{k}}^{\textrm{eq}}\right)^2 - \omega^2} \left[4 \left(\Delta_s^{\textrm{eq}}\right)^2 - \omega^2\right],  
    \label{C(omega)} \\
    D(\omega) &\coloneqq \frac{1}{N} \sum_{\bm{k}} \frac{2\varphi_d(\bm{k})^2 t_{\bm{k}}^{\textrm{eq}} E_{\bm{k}}^{\textrm{eq}}}{4 \left(E_{\bm{k}}^{\textrm{eq}}\right)^2 - \omega^2}. 
    \label{D(omega)}
\end{align}
In order to have non-trivial solutions in Eq.~(\ref{pure-s-gap-eq}), at least one of the diagonal components must be zero. 
These two solutions correspond to the Higgs and BS modes. 

From the upper left component, an equation
\begin{equation}
    C(\omega) = 0
\end{equation}
is obtained. This equation has a solution $\omega = 2 \Delta_s^{\textrm{eq}}$, where the derivative $\frac{d C(\omega)}{d\omega}$ diverges. There is no other solution in the range of $0 < \omega < 2 \Delta_s^{\textrm{eq}}$. 
This frequency corresponds to that of the Higgs mode.

From the lower right component, an equation
\begin{equation}
    D(\omega) = - 1 / V_d
    \label{D(omega)=1/Vd}
\end{equation}
is obtained. In the range of $0 < \omega < 2\Delta_s^{\textrm{eq}}$, the derivative $\frac{d D}{d \omega}$ is positive. It is not difficult to show that $D(0) < -1/V_d$ and $\lim_{\omega \to 2\Delta_s^{\textrm{eq}}} D(\omega) = +\infty$. Therefore, Eq.~(\ref{D(omega)=1/Vd}) has a unique solution in $0 < \omega < 2 \Delta_s^{\textrm{eq}}$. Here, $D(\omega)$ crosses $-1/V_d$ with a finite gradient. 
This frequency corresponds to that of the BS mode.

\subsubsection{$s+id$-wave phase}

In the $s+id$-wave phase, we obtain
$1 = -\frac{V_s}{2N} \sum_{\bm{k}} \frac{t_{\bm{k}}^{\textrm{eq}}}{E_{\bm{k}}^{\textrm{eq}}}$ and $1 = -\frac{V_d}{2N} \sum_{\bm{k}} \frac{\varphi_d(\bm{k})^2 t_{\bm{k}}^{\textrm{eq}}}{E_{\bm{k}}^{\textrm{eq}}}$
from the equilibrium gap equation (\ref{equilibrium-gap-equation}), with which Eq.~(\ref{fourier-transformed-gap-equation}) becomes
\begin{equation}
	\begin{pmatrix}
		0 \\ 0
	\end{pmatrix} = A(-i \omega) \begin{pmatrix}
		\delta\widetilde{\Delta}_s(\omega) \\ \delta\widetilde{\Delta}_d(\omega)
	\end{pmatrix}. 
\end{equation}
Here, the matrix $A(s)$ is defined as 
\begin{align}
    A(s) &= \begin{pmatrix}
        A_{11}(s) & A_{12}(s) \\ A_{21}(s) & A_{22}(s) 
    \end{pmatrix}, \\
    A_{11}(s) &= \frac{-V_s}{2N} \sum_{\bm{k}} \frac{t_{\bm{k}}^{\textrm{eq}} / E_{\bm{k}}^{\textrm{eq}}}{4 \left(E_{\bm{k}}^{\textrm{eq}}\right)^2 + s^2} \left[4 \left(\Delta_s^{\textrm{eq}}\right)^2 + s^2\right], \\
    A_{12}(s) &= \frac{-V_s}{2N} \sum_{\bm{k}} \frac{4\varphi_d(\bm{k})^2 t_{\bm{k}}^{\textrm{eq}} / E_{\bm{k}}^{\textrm{eq}}}{4 \left(E_{\bm{k}}^{\textrm{eq}}\right)^2 + s^2} \Delta_s^{\textrm{eq}} \Delta_d^{\textrm{eq}}, \\
    A_{21}(s) &= \frac{-V_d}{2N} \sum_{\bm{k}} \frac{4\varphi_d(\bm{k})^2 t_{\bm{k}}^{\textrm{eq}} / E_{\bm{k}}^{\textrm{eq}}}{4 \left(E_{\bm{k}}^{\textrm{eq}}\right)^2 + s^2} \Delta_s^{\textrm{eq}} \Delta_d^{\textrm{eq}}, \\
    A_{22}(s) &= \frac{-V_d}{2N} \sum_{\bm{k}} \frac{\varphi_d(\bm{k})^2 t_{\bm{k}}^{\textrm{eq}} / E_{\bm{k}}^{\textrm{eq}}}{4 \left(E_{\bm{k}}^{\textrm{eq}}\right)^2 + s^2} \left[4\varphi_d(\bm{k})^2 \left(\Delta_d^{\textrm{eq}}\right)^2 + s^2\right].
\end{align}
In order for this equation to have non-trivial solutions, 
\begin{equation}
    \det A(-i \omega) = 0
    \label{detA=0}
\end{equation}
must be satisfied. 
In the range of $0 < \omega < 2\Delta_s^{\textrm{eq}}$, the derivative $\frac{d}{d \omega} \bigl[\det A(-i \omega)\bigr]$ is negative. 
We can show $\det A(0) > 0$ and $\det A\bigl(-2i \Delta_s^{\textrm{eq}}\bigr) < 0$,
using the Cauchy-Schwarz inequality. 
Therefore, Eq.~(\ref{detA=0}) has a unique solution in $0 < \omega < 2 \Delta_s^{\textrm{eq}}$. 
Here, $\det A(-i \omega)$ crosses zero with a finite gradient. 
This frequency corresponds to that of the MSBS mode.

\begin{figure}
	\centering
	\includegraphics[width=\linewidth]{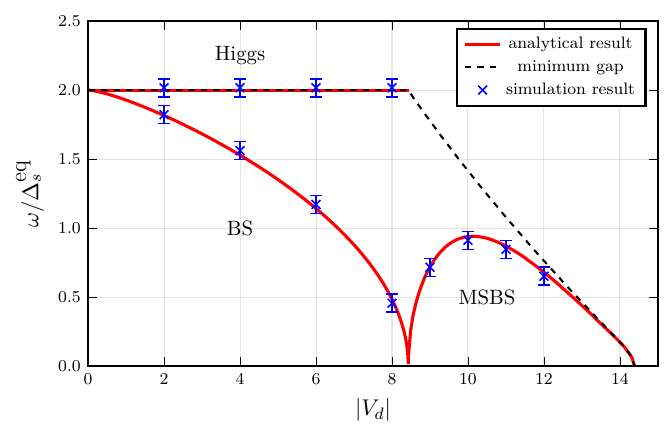}
	\caption{
    Frequency of the collective modes in the one-band model with the $s$- and $d$-wave pairing interactions studied in Sec.~\ref{sec: one-band model} for $V_s = -5$. The red lines represent the analytical results, and the blue markers show the numerical results estimated from the quench simulations. The errors come from Fourier transformation of the finite-time data.
    The vertical axis is normalized by the asymptotic value of $\Delta_s^{\textrm{eq}}$ in the limit $V_d \to 0$. The parameters are $\mu = 0$, $t = 1$, and $T = 0.05$.
    }
	\label{collective-modes-frequency}
\end{figure}

Figure \ref{collective-modes-frequency} shows the frequency of the Higgs, BS, and MSBS modes for $V_s = -5$. The red lines represent the analytical results, and the blue markers show the numerical results extracted from the quench simulations for several value of $V_d$. 
The dashed black line represents the minimum gap ($\min_{\bm{k}} \Delta_{\bm{k}}^{\textrm{eq}} = \Delta_s^{\textrm{eq}}$). 
We find that the analytical and numerical results are consistent with each other.

\subsection{Undamped behavior of the collective modes}

Both the $\Delta$-quench and $V$-quench results suggest that the MSBS mode is undamped. This behavior can also be understood analytically in the infinitesimal $V$-quench picture. 
Here, we use the Laplace transformation instead of the Fourier transformation to discuss the oscillation dynamics (following Ref.~\cite{VolkovKogan1973}).
The linearized Bloch equation is Laplace transformed as 
\begin{align}
    \begin{pmatrix}
        \delta \widetilde{\Delta}_s(s) \\ \delta \widetilde{\Delta}_d(s)
    \end{pmatrix} &= \bigl[A(s)\bigr]^{-1} \begin{pmatrix}
        \frac{\Delta_s^{\textrm{eq}}}{s} \frac{\delta V_s}{V_s} \\ \frac{\Delta_d^{\textrm{eq}}}{s} \frac{\delta V_d}{V_d}
    \end{pmatrix} \\
    &= \frac{1}{\det A(s)} \begin{pmatrix}
        A_{22}(s) & -A_{12}(s) \\ -A_{21}(s) & A_{11}(s)
    \end{pmatrix} \begin{pmatrix}
        \frac{\Delta_s^{\textrm{eq}}}{s} \frac{\delta V_s}{V_s} \\ \frac{\Delta_d^{\textrm{eq}}}{s} \frac{\delta V_d}{V_d}
    \end{pmatrix}.
\end{align}
The real-time gap function can be obtained by the inverse Laplace transformation,
in which singular points in the complex $s$-plane have an important contribution. 
There are two sources of singularities: $1/s$ and $1/\det A(s)$. 
We already know a pair of singular points $s = \pm i \omega_0$ such that $\det A(\pm i \omega_0) = 0$ with the derivative $\left.\frac{d}{d\omega} \bigl[\det A(-i\omega)\bigr]\right|_{\omega = \pm\omega_0}$ being finite. 
Since $\det A(s)$ is analytic in the vicinity of $s = \pm i \omega_0$, we can Taylor-expand it as 
\begin{equation}
    \det A(s) = \sum_{n = 0}^{\infty} \left.\frac{d^n}{ds^n}\bigl[\det A(s)\bigr]\right|_{s = \pm i \omega_0} \frac{(s \mp i \omega_0)^n}{n!}. 
\end{equation}
The finite derivative $\left.\frac{d}{d\omega} \bigl[\det A(-i\omega)\bigr]\right|_{\omega = \pm\omega_0}$ suggests that the first term in the Taylor expansion is non-zero. 
Therefore, using some complex constants $a_{\pm}$, we can write $\det A(s)$ as
\begin{equation}
    \det A(s) = a_{\pm} (s \mp i \omega_0) + O\bigl((s \mp i \omega_0)^2\bigr). 
\end{equation}
Therefore, the Laplace-transformed gap function is approximated to 
\begin{align}
    \begin{pmatrix}
        \delta \widetilde{\Delta}_s(s) \\ \delta \widetilde{\Delta}_d(s)
    \end{pmatrix} &= \frac{1}{a_{\pm} (s \mp i \omega_0)} \begin{pmatrix}
        A_{22}(\mp i\omega_0) & -A_{12}(\mp i\omega_0) \\ -A_{21}(\mp i\omega_0) & A_{11}(\mp i\omega_0)
    \end{pmatrix} 
    \notag
    \\
    &\quad\times
    \begin{pmatrix}
        \frac{\Delta_s^{\textrm{eq}}}{\mp i \omega_0} \frac{\delta V_s}{V_s} \\ \frac{\Delta_d^{\textrm{eq}}}{\mp i \omega_0} \frac{\delta V_d}{V_d}
    \end{pmatrix}. 
\end{align}
The singularity in this equation is $1 / (s \mp i \omega_0)$, whose inverse Laplace transformation becomes $e^{\pm i \omega_0 t}$.
This calculation suggests that the MSBS mode shows the undamped oscillation. 

The undamped behavior of the BS mode can also be understood in a similar way. 
The important point in the case of the MSBS mode is that the derivative $\left.\frac{d}{d\omega} \bigl[\det A(-i\omega)\bigr]\right|_{\omega = \pm i \omega_0}$ is finite. 
A similar property holds for the BS mode: The function $D(\omega)$ defined in Eq.~(\ref{D(omega)}) also has a finite derivative ($D(\omega)$ crosses $1/V_d$ with a finite gradient). 

\section{The Rashba gap in the spin basis}\label{rashbaGapTransformation}
In this Appendix, the derivation of Eqs. \ref{rashbaSingletSpinBasis}-\ref{rashbaTripletInterband} is shown, explaining how one obtains the usual gap in terms of the singlet component $\psi$ and the triplet components $\bm{d}$ from the gap matrix in terms of the gap matrix in the basis of the Rashba bands.
		The Nambu spinors in the Rashba-band basis take the form,
		\begin{align}
			\Psi_{\bm{k}}&=
			\begin{pmatrix}
				c_{\bm{k}+}\\
				c_{\bm{k}-}\\
				c^\dagger_{\bm{-k}+}\\
				c^\dagger_{\bm{-k}-}
			\end{pmatrix},
		\end{align}
		written here with the creation and annihilation operators, rather than Grassmann variables. The transformation from the $c$-operators in the Rashba basis, to the $a$-operators in the spin basis is defined in Eq.~(\ref{rashbaTransformation}). This can be extended to the four-dimensional Nambu space:
		\begin{align}
			\begin{pmatrix}
				c_{\bm{k}+}\\
				c_{\bm{k}-}\\
				c^\dagger_{\bm{-k}+}\\
				c^\dagger_{\bm{-k}-}
			\end{pmatrix}
			&=
			\begin{pmatrix}
				U_{\bm{k}}&0\\
				0&U_{\bm{-k}}^T
			\end{pmatrix}
			\begin{pmatrix}
				a_{\bm{k}+}\\
				a_{\bm{k}-}\\
				a^\dagger_{\bm{-k}+}\\
				a^\dagger_{\bm{-k}-}
			\end{pmatrix}.
		\end{align}
		From this, the transformation property of the gap function can be deduced:
		\begin{align}
			\mathcal{H}_\text{BdG}^\pm
			&=
			\begin{pmatrix}
				\xi^\pm_{\bm{k}}&\Delta_{\bm{k}}\\
				\Delta^\dagger_{\bm{k}}&-\xi^\pm_{\bm{-k}}
			\end{pmatrix},\\
			\mathcal{H}^{\uparrow\downarrow}_\text{BdG}
			&=
			\begin{pmatrix}
				U^\dagger_{\bm{k}}&0\\
				0&U_{\bm{-k}}^\text{T}
			\end{pmatrix}
			\mathcal{H}_\text{BdG}^\pm
			\begin{pmatrix}
				U_{\bm{k}}&0\\
				0&U^*_{\bm{-k}}
			\end{pmatrix}\\
			\Rightarrow 
			\widehat{\Delta}_{\bm{k}}^{\uparrow\downarrow}
			&= 
			U^\dagger_{\bm{k}}\widehat{\Delta}^\pm_{\bm{k}}U^*_{\bm{-k}}.
		\end{align}
		This may now be explicitly calculated, using the definition of $U_{\bm{k}}$ in Eq.~(\ref{rashbaTransformation}). Dropping for notational convenience the $\bm{k}$ subscripts, and writing $t\equiv t_{+}=-t_{-}$, this yields:
        \begin{widetext}
		\begin{align}
			\widehat{\Delta}_{\bm{k}}^{\uparrow\downarrow}
			&=\frac12
			\begin{pmatrix}
				1&1\\t^*&-t^*
			\end{pmatrix}
			\begin{pmatrix}
				t\widetilde{\Delta}_{++}&-t\widetilde{\Delta}_{+-}\\
				t\widetilde{\Delta}_{-+}&-t\widetilde{\Delta}_{--}
			\end{pmatrix}
			\begin{pmatrix}
				1&-t^*\\1&t^*
			\end{pmatrix}
            \notag
            \\
			&=\frac{1}{2}
			\begin{pmatrix}
				t(\widetilde{\Delta}_{++}
				+\widetilde{\Delta}_{-+}
				-\widetilde{\Delta}_{+-}
				-\widetilde{\Delta}_{--})
				&
				-\widetilde{\Delta}_{++}
				+\widetilde{\Delta}_{-+}
				+\widetilde{\Delta}_{+-}
				-\widetilde{\Delta}_{--}
				\\
				+\widetilde{\Delta}_{++}
				-\widetilde{\Delta}_{-+}
				-\widetilde{\Delta}_{+-}
				+\widetilde{\Delta}_{--})
				&
				t^*(
				-\widetilde{\Delta}_{++}
				+\widetilde{\Delta}_{-+}
				-\widetilde{\Delta}_{+-}
				+\widetilde{\Delta}_{--})
			\end{pmatrix}.
		\end{align}
        \end{widetext}
		Note that due to the symmetry property [Eq.~(\ref{rashbaGapSymmetry})], the intraband components of $\widetilde{\Delta}$ necessarily have even parity, while the interband components satisfy 
		$\widetilde{\Delta}_{\bm{k}}^{+-}\equiv\widetilde{\Delta}^\text{inter}_{\bm{k}}=-\widetilde{\Delta}_{\bm{-k}}^{-+}$, so they are odd under simultaneous transposition and inversion. This means that, using the usual parametrization of an unconventional gap function,
        \begin{align}
        \begin{split}
			\widehat{\Delta}^{\uparrow\downarrow} 
			&= \psi(\bm{k})i\sigma_y + i(\bm{d(k)\cdot\sigma})\sigma_y\\
            &=
			\begin{pmatrix}
				-d_x(\bm{k})+id_y(\bm{k})&d_z(\bm{k})+\psi(\bm{k})\\
				d_z(\bm{k})-\psi(\bm{k})&d_x(\bm{k})+id_y(\bm{k})
			\end{pmatrix}\label{d-vector-parametrization},
        \end{split}
		\end{align}
         one obtains Eq.~(\ref{rashbaSingletSpinBasis}) for the singlet components of the gap function. The $\bm{d}$-vector takes the form,
		\begin{align}
			{\bm{d(k)}}
			&=
			-\frac12
			\begin{pmatrix}
				\hat{\gamma}_x(\widetilde{\Delta}^{++}_{\bm{k}}
				-\widetilde{\Delta}^{--}_{\bm{k}})
				-i\hat{\gamma}_y(
				\widetilde{\Delta}^{\text{inter}}_{\bm{k}}
				+\widetilde{\Delta}^{\text{inter}}_{\bm{-k}})
				\\
				\hat{\gamma}_y(\widetilde{\Delta}^{++}_{\bm{k}}
				-\widetilde{\Delta}^{--}_{\bm{k}})
				+i\hat{\gamma}_x(
				\widetilde{\Delta}^{\text{inter}}_{\bm{k}}
				+\widetilde{\Delta}^{\text{inter}}_{\bm{-k}})
				\\
				\widetilde{\Delta}^{\text{inter}}_{\bm{k}}
				-\widetilde{\Delta}^{\text{inter}}_{\bm{-k}}
			\end{pmatrix},
		\end{align}
        which yields Eqs.~(\ref{rashbaTripletSpinBasis})-(\ref{rashbaTripletInterband}).
\section{Free energy calculations}\label{freeEnergy}
\begin{figure}
    \centering
    \includegraphics[width=\linewidth]{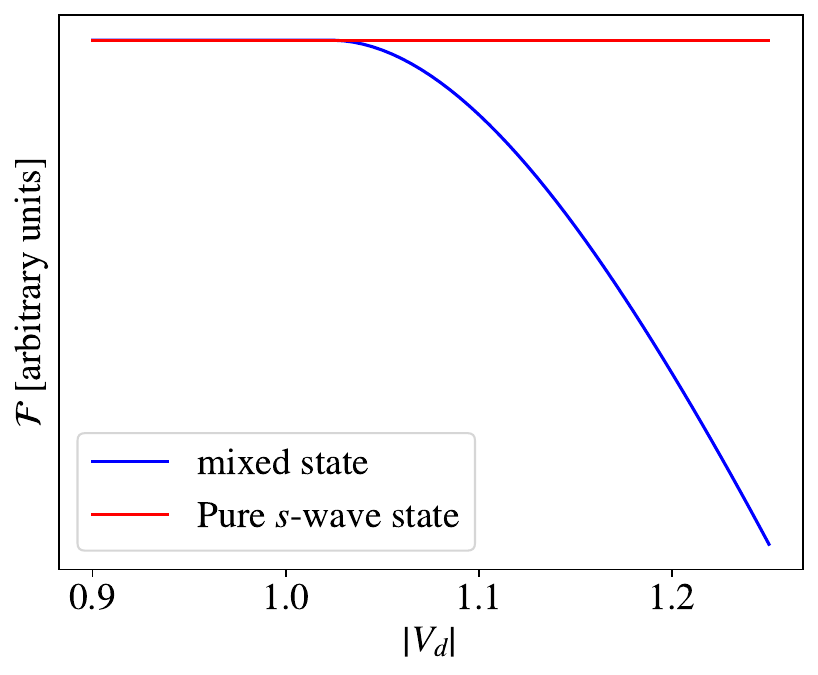}
    \caption{Free energy of the multiband system with asymmetrical bands (Sec. \ref{asymmMultiband}), which is analogous to the symmetrical two-band case (Sec. \ref{symmMultiBand}) and the one-band case (Sec \ref{sec: one-band model}). Parameters are again $\mu=-0.2$, $V_s=-1$, $t=1$, and $\bm{q}=0.001\bm{\hat{e}}_x$.}
    \label{freeEnergy_sd}
\end{figure}

\begin{figure}
    \centering
    \includegraphics[width=\linewidth]{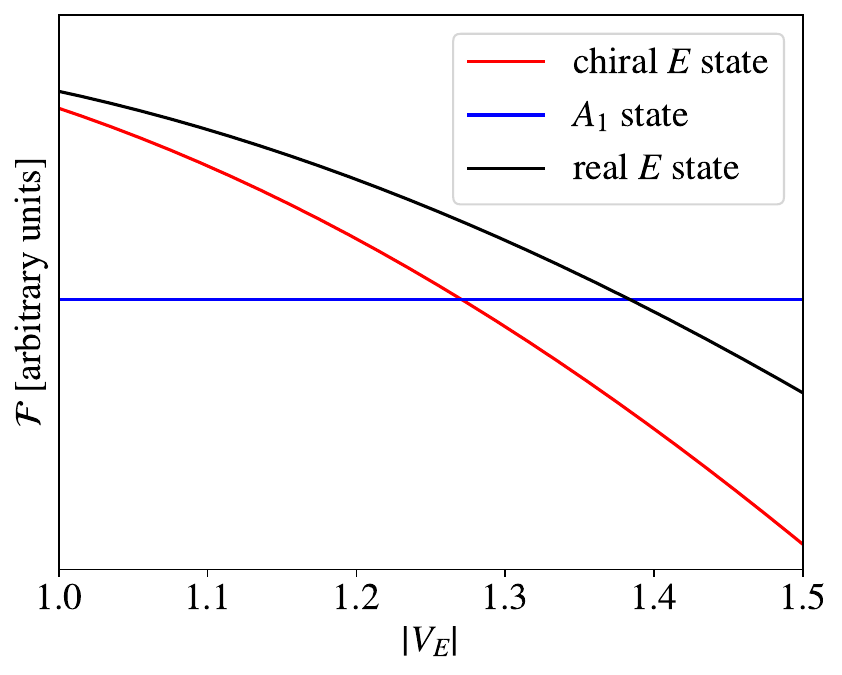}
    \caption{Free energies of possible ground states belonging to the $E$- or $A_1$-irrep of the Rashba system (Sec. \ref{rashba}), for increasing $V_E$. Parameters are $\mu=-3$, $t=1$, $\alpha=0.1$, $\bm{q}=0$, and $\omega_c=1$ and $V_{A_1}=1$.}
    \label{freeEnergy_rs}
\end{figure}

In cases where several ground states fulfill the gap equation (see Appendix \ref{gapAppendix}), a free energy calculation is necessary to find the state which will be realized in the system. This can be done using the usual formula for the free energy of non-interacting fermions\cite{AltlandSimons}
\begin{align}
    \mathcal{F} &= -\frac1\beta \sum_{\bm{k}j}\ln\left(1+e^{-\beta E^j_{\bm{k}}}\right),
\end{align}
where $\beta$ is the inverse temperature and $j$ are any quantum numbers that need to be summed over in addition to momentum $\bm{k}$. At zero temperature this simplifies to
\begin{align}
    \mathcal{F}&= \sum_{\bm{k},j: E^j_{\bm{k}}<0}E^j_{\bm{k}}
\end{align}
which is simply the total energy of all occupied states. For the systems with $s$- and $d$-wave interactions, the main competition lies in between the pure $s$-wave ground state and the mixed-symmetry $s+id$-wave ground state. As the interaction in the $d$-wave channel is increased, the pure $s$-wave state always remains a solution of the gap equation, so the free energies of the two states should be compared. For the example of the multiband system with asymmetrical bands, this comparison is shown in Fig. \ref{freeEnergy_sd}.
This graph shows how the free energies of the mixed and pure symmetry state initially coincide, when the "mixed" symmetry state has no $d$
-wave component, and they are identical. As soon as the states become different, one can observe that the free energy of the mixed state always lies below that of the pure $s$-wave solution to the gap equation. The plot is shown for the most complicated case of two asymmetrical bands, but is identical for the other cases which were discussed in the article.\\
For the Rashba system (see Sec. \ref{RashbaResults}), the transition we considered was between an $A_1$ ground state, corresponding to an in-plane $p$-wave $\bm{d}$-vector, and an out-of-plane, chiral $p_x+ip_y$-wave $\bm{d}$-vector. Additionally, a real out-of-plane $p_x+p_y$ state would have solved the relevant gap equation. The comparison of the relevant free energies is shown in Fig. \ref{freeEnergy_rs}.
This shows how the free energy of the chiral ground state in the $E$-irrep crosses that of the $A_1$ ground state at the transition. It can also be seen how the free energy of the real $p_x+p_y$ state in the $E$ irrep eventually falls below that of the $A_1$ state, but remains above that of the chiral state. This was used to confirm the phase diagram in Fig. \ref{s-p-transition}.
\bibliography{main}
\end{document}